\documentclass{aa}  

\usepackage{graphicx}
\usepackage{txfonts}

\usepackage{natbib}
\newcommand{\mygi}{MyGIsFOS}

\newcommand{\logg}{\ensuremath{\log\,g}}
\newcommand{\glog}{\ensuremath{\log\,g}}
\def\teff{$T\rm_{eff}$}
\newcommand{\kms}{$\rm km s ^{-1}$}

\begin{document}

\title{High-speed stars II: An unbound star, young stars, bulge metal-poor stars, and Aurora candidates. 
\thanks{Based on observations made at VLT with FORS2  under programmes 0105.D-0213,0106.D-0435,  
and UVES programme
0108.D-0372.}
}
\titlerunning{High-Speed stars II}

\author{
P.~Bonifacio \inst{1} \and
E.~Caffau    \inst{1} \and
L.~Monaco \inst{2} \and
L.~Sbordone \inst{3} \and      
M.~Spite \inst{1} \and
A.~Mucciarelli \inst{4,5} 
\and
P.~Fran\c{c}ois \inst{6,7} 
\and
L.~Lombardo \inst{1,8}
\and
A.d.M.~Matas Pinto\inst{1}
}

\institute{GEPI, Observatoire de Paris, Universit\'{e} PSL, CNRS,  5 Place Jules Janssen, 92190 Meudon, France
\and
Universidad Andres Bello, Facultad de Ciencias Exactas, 
Departamento de Ciencias F{\'\i}sicas - Instituto de Astrof{\'\i}sica,
Autopista Concepci{\'o}n-Talcahuano, 7100, Talcahuano, Chile
\and
European Southern Observatory, Casilla 19001, Santiago, Chile
\and
Dipartimento di Fisica e Astronomia, Universit\`a degli Studi di Bologna, Via Gobetti 93/2, I-40129 Bologna, Italy
\and
INAF -- Osservatorio di Astrofisica e Scienza dello Spazio di Bologna, Via Gobetti 93/3, I-40129 Bologna, Italy
\and
 GEPI, Observatoire de Paris, Universit\'{e} PSL, CNRS,  77 Av. Denfert-Rochereau, 75014 Paris, France
\and
UPJV, Universit\'e de Picardie Jules Verne, 33 rue St Leu, 80080 Amiens, France
\and
Goethe University Frankfurt, Institute for Applied Physics
(IAP), Max-von-Laue-Str. 12, 60438, Frankfurt am Main
}

   \date{Received September 15, 1996; accepted February 11, 1997}

  \abstract
{The data from the Gaia satellite led us to revise our conception of the Galaxy structure
and history. Hitherto unknown components have been discovered and a deep re-thinking
of what the Galactic halo is is in progress.}
{We selected from the Gaia catalogue stars with extreme transverse velocities with respect
to the Sun  ($|V_T| > 500 $ \kms) and observed them with FORS2 at the ESO VLT,
to classify them  using both their chemical and dynamical properties. Two apparently
young stars, identified in paper\,I, were observed with UVES.}
{We derived abundances for Na, Mg, Ca, Ti, Mn, and Fe, analysing the spectra with \mygi,\ while for
Ba we used line profile fitting.  We computed actions from parallaxes and kinematical data.}
{The stars span the metallicity range $\rm -3.5 \le [Fe/H] \le -0.5$ with 
$\rm \langle [Fe/H] \rangle = -1.6$. Star GHS143 has a total speed of about 1440 \kms, which is
almost three times faster than the local escape velocity of 522\,\kms , strongly implying this 
star is unbound to the Galaxy. Remarkably, this star is not escaping from the Galaxy, but it is falling into
it.
Ten stars are apparently young  with masses in excess of 1.3 $M\odot$. 
Their interpretation as evolved blue stragglers is doubtful. The existence of 
a young metal-poor population is possible. The two stars observed with UVES show no lithium,
suggesting they are blue stragglers. We detected a metal-poor  population, confined to the bulge, that we call SpiteF, and argue that
it is the result of a recent accretion event. We detect 102 candidates of the Aurora population that
should have formed prior to the formation of the disc.
}
{Our sample is non-homogeneous and mainly retrograde. The stars are
metal poor, and 23\% have [Fe/H] $\le -2.0$. Our selection 
is efficient at finding very metal-poor stars, but it selects peculiar populations.}

   \maketitle

%
\section{Introduction\label{intro}}

The stars belonging to the Galactic disc, such as the Sun, move around the Galactic centre 
with a velocity dependent on their distance from the Galactic centre. 
At the Galactic radius of about 8\,kpc, where the Sun dwells, the velocity of the stars around 
the Galactic centre is 220\,\kms\ \citep[the recommended IAU value;][]{1986MNRAS.221.1023K}. 
The Galactic disc moves as a rigid body up to about the Sun, while further out the speed 
of the stars is more or less constant. For this reason, disc stars
do not show a high speed when observed from the Earth. 
Halo stars instead can have any orbit around the centre
of the Galaxy. 
If the Milky Way has formed in a hierarchical scenario, we expect the halo to contain stars that
were formed 
in accreted dwarf galaxies as well as, possibly, stars that formed
during the collapse of the halo.
According to \citet{haywood18},  what are usually called `halo stars' 
are a mixture of the following: (1) stars 
that were formed in the massive satellite galaxy that was accreted 8 to  10\,Ga ago,
which was initially discovered by \citet{belokurov18} using Gaia data release 1
\citep{gaiadr1} and later confirmed by \citet{haywood18} and also by
\citet{helmi18} using Gaia data release 2 \citet{gaiadr2}; and (2) disc stars that were scattered
in halo-like orbits as a result of the collision.
The Galactic structure resulting from this accretion was called Gaia-Sausage by
\citet{belokurov18} and Gaia-Enceladus by \citet{helmi18}, following a now well -established convention in the literature we shall refer to as Gaia-Sausage-Enceladus (GSE). Other accretion events have since been discovered and some of them
are discussed in detail in Sec.\,\ref{kin}.
Recently, \citet{BV2022} claimed the discovery of a population
of stars that formed in the last turbulent phases of the Galactic collapse, prior to the formation of the disc.
This population should represent the dawn of star formation,  for this reason they call it Aurora\footnote{Latin for dawn.}. 
\citet{dimatteo19},  based on Gaia 
data release 2 \citep{gaiadr2} kinematic data and Apache Point Observatory Galactic Evolution Experiment  
\citep[APOGEE][]{apogee} chemical abundances,
found no evidence for the existence of stars that formed during the collapse
of the Galactic halo. The halo, as explored by the data they took into consideration, 
seems to be formed exclusively by thick disc stars, heated by the collision
that gave rise to GSE and stars that were formed in the colliding galaxy itself.
The issue is clearly still open to debate.

By looking at the Toomre diagrams in Fig.6 of \citet{haywood18}, it is clear that there are groups 
of stars that have a high retrograde rotation and high speed. 
These stars may indeed be the result of other minor accretion events. 
Since dwarf galaxies 
have a shallow gravitational potential, they cannot retain large amounts of chemically enriched material 
and the stars that formed in dwarf galaxies are generally more metal poor than stars in the Galactic disc
(exceptions are e.g. Fornax and Sgr, the most massive  dwarf spheroidal galaxies).

Among the halo stars, we are especially interested in metal-poor and extremely metal-poor 
(MP [Fe/H]$\leq -1.0$ and EMP [Fe/H]$\leq -3.0$) stars. 
These stars formed from a gas cloud that  was hardly enriched in metals by supernovae explosions, 
the last stage of the evolution of the massive stars. 
This low metal content of the gas was the typical chemical composition of a pristine phase of the Universe. 
The massive stars have a short life; if formed in the pristine Universe, they are long gone at present and are no longer observable. 
The pristine low-mass stars are still observable today because their 
lifetime is longer than the age of the Universe, but their chemical composition is a witness of a long gone 
chemical composition that characterised the pristine Universe. 
The knowledge on the detailed chemical composition of these old stars is providing us information on the early Universe. 
In fact, the chemical content of the primordial stars puts constraints on the masses of the first generations 
of massive stars that exploded as supernovae and enriched the primordial interstellar medium and also on 
the physics of the supernovae explosions.

In the classical picture, the halo is populated only by old metal-poor stars since no high-density
molecular clouds, suitable to sustain a recent star formation, are present in the halo.
Most apparently, young stars found in the halo are near the turnoff and can be interpreted as blue straggler stars.
Yet, since the availability of Gaia parallaxes, stars that are compatible with young metal-poor stars
among giants become apparent in colour-magnitude diagrams. An excellent example is
figure 2 of \citet{hattori18}, although these stars are not further discussed in that paper. 
Even though a recent star formation could not have taken place in the halo, it could have taken place
in Local Group dwarf galaxies, which then lost these stars due to a tidal interaction with the Milky Way.
For instance, Fornax is known to have had a star formation ongoing up
to until 100 Ma ago \citep{deboer12}.
If many of the dwarf galaxies contain these metal-poor young stars, then it would not be 
surprising to find some of them in the Galactic halo.

The mass of the Galactic halo is much smaller than the mass of the Galactic disc. 
This means that halo stars are a minor component of the Galactic stars (about 1\%) 
and the majority of MP and EMP stars belong to the halo. 
It must be however noted that a minority of MP and EMP stars can also be found
on disc orbits \citep{PristineX,dimatteo2020}.
In general, MP stars are not common objects. 
In order to find them, large amounts of data have to be gathered and analysed. In the past years, 
several projects focussed  on the search and chemical investigations of MP and EMP stars, based on photometry 
or on the analysis of low-resolution spectra obtained by large surveys. 
However, there are other ways to select halo stars. 
A way to select halo stars is to rummage in the kinematic database and select stars whose speed is faster than  the speed of disc stars. 
In this way we do not really select halo stars, but we cut out disc stars.

We began this research in \citet[][hereafter Paper\,I]{ghs104}, by selecting stars with a high transverse
speed from the Gaia \citep{gaiacol16} second data release \citep{gaiadr2} and observing them at a low spectral
resolution with the FOcal Reducer/low dispersion Spectrograph 2 (FORS2) at the ESO  8.2\,m Very Large Telescope in ESO period 104.
Given the very promising results of the first 72 stars presented in Paper\,I, 
the programme was continued in ESO periods 105 and 106.
In this paper we describe these observations, their analysis, and our conclusions
from this chemo-dynamical study.

\section{Target selection\label{sel}} 
\begin{figure}
\centering
\resizebox{7.5cm}{!}{\includegraphics{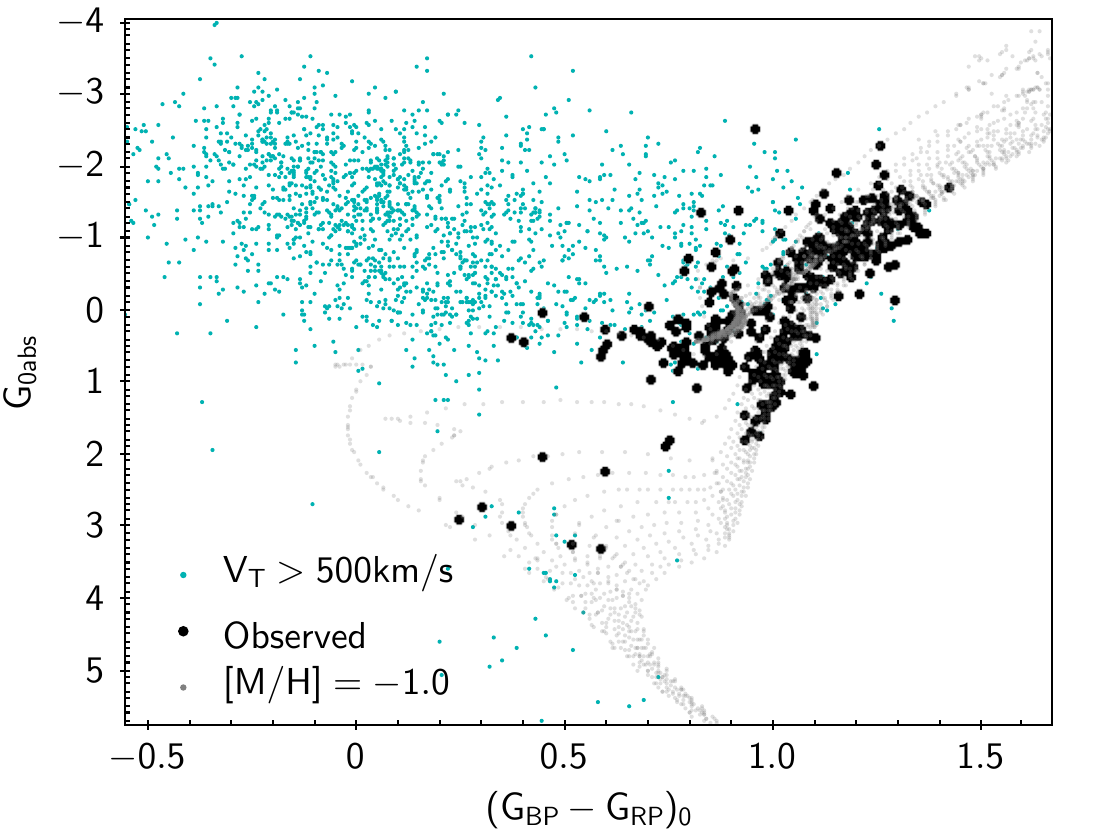}}
\caption{Colour magnitude diagram of the   high transverse speed stars (cyan) selected
from the Gaia DR3 catalogue using the query in appendix \ref{query} and those
observed (black). PARSEC isochrones \citep{bressan12} of metallicity --1.0 (grey) are over plotted. 
}
\label{cmd_hs}
\end{figure}

\begin{figure}
\centering
\resizebox{7.8cm}{!}{\includegraphics[clip=true]{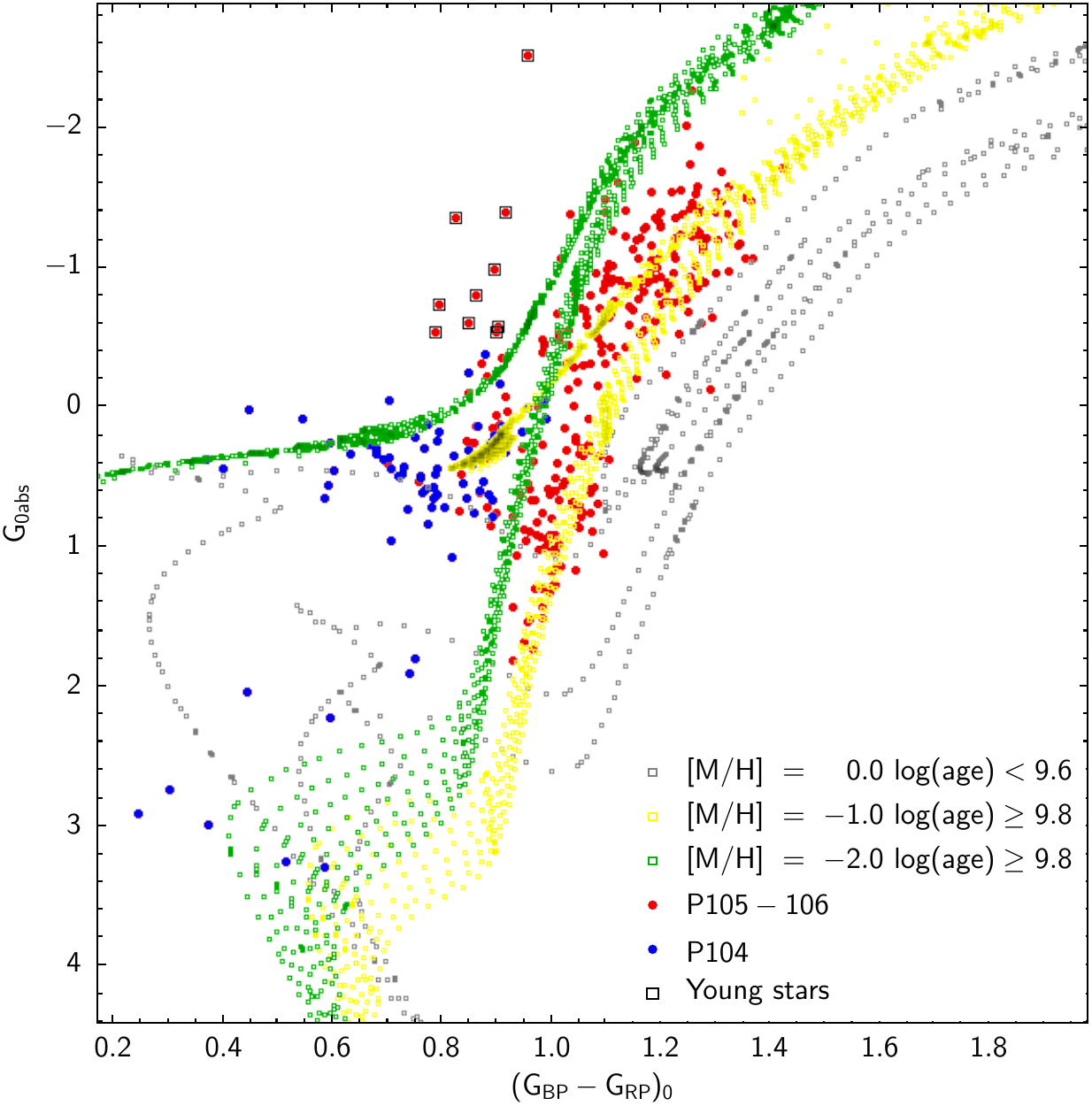}}
\caption{Colour magnitude diagram of the observed high-speed stars. The
stars discussed in Paper\,I are in blue, which were observed in ESO period 104, and those observed
in period 105 and 106 are in red. We provide PARSEC isochrones \citep{bressan12} of metallicity 0.0 (grey),\,--1.0 (yellow), 
and -2.0 (green) of different age ranges. The age difference between consecutive isochrones
of any given metallicity is  1\,Ga.  The apparently young stars observed
in periods 105 and 106 are identifiable with an open square around the symbol.}
\label{cmd}
\end{figure}

In Paper\,I we selected from Gaia DR2 \citep{gaiadr2} stars with transverse 
velocity\footnote{ $V_t = 4.740470446\sqrt{(\mu_\alpha/\varpi)^2 +(\mu_\delta/\varpi)^2}$, where $\varpi$
is the parallax and $\mu_\alpha,\mu_\delta$ are the proper motions in right ascension and declination. All these
quantities have been taken from the Gaia catalogue, where they are provided in units of mas. }
in excess of 500 \kms\ and we privileged `blue' stars because
we were aiming at selecting the most metal-poor population
of these high-speed stars. For the following observations,
during ESO period 105 and 106, we used Gaia EDR3, we kept the
condition of transverse velocity in excess of 500 \kms\ 
but observed also redder stars with the aim
of obtaining a representative metallicity distribution
function for high-speed stars.  We excluded in any case stars 
with $(G_{BP}-G_{RP})> 1.5 $ because of the complexity
of modelling the spectra of stars cooler than this limit.
The Gaia DR3 identifiers of all our programme stars (including those
in Paper\,I) can be found in appendix\,\ref{starnam}.
 The input catalogues used for the three ESO periods are different,
also because they had folded in the range of right ascension suitable for the given period,
colour and magnitude cuts are also different, and the Gaia catalogues used were
different. We stress that in this investigation we have no claim of
completeness, therefore the lack of homogeneity among the observations
in the different ESO periods does not affect any of our results. 
To provide an idea of what the general population of high transverse speed
stars is, we show in Fig.\,\ref{cmd_hs} our observed stars together with a
selection obtained on the Gaia DR3 catalogue with
the adql query  in appendix \ref{query}. We focussed
on giant stars for two reasons: {\em i)} the cooler temperatures
imply stronger lines at any given metallicity, making the chemical
analysis easier with the low resolution spectra we used; {\em ii)}
the age-metallicity degeneracy is weaker for giant stars. 
The magnitudes were limited to $G\le 14.3$
in order to observe a large sample in a relatively short amount of time.

In Fig.\,\ref{cmd} we show the colour-magnitude diagram in Gaia DR3 \citep{gaiadr3}
broad band observed colours\footnote{ The absolute magnitudes where computed as
$G_{abs} = G +5 +5\log(\varpi/1000) $, where $\varpi$ is the parallax in mas
as provided by the Gaia catalogue.}. We note that this is the case also for the stars in Paper\,I 
shown as blue dots, while in figure 1 of Paper\,I the Gaia DR2 colours were used. 
To guide the eye we show also three sets of PARSEC isochrones \citep{bressan12} 
of metallicity
--2.0, --1.0 and 0.0. For the two most metal-poor sets only old isochrones
are shown with ages between 6 and 14  Ga, at steps of 1  Ga. For the solar
metallicity ones only young isochrones are shown, with ages between 1 and 3  Ga.
It is clear from the figure how the stars observed in periods 105 and 106 are
redder than those observed in period 104. It is also clear, from
the comparison with the isochrones, that the colours imply that the 
majority  of the stars should have metallicities in the range --2.0 to --1.0,
with only few more metal-poor or metal-rich. 
In a colour-magnitude diagram the metallicity, to a first approximation, decreases
along a vector orthogonal to the Red Giant Branch (RGB), the most metal-poor stars
being the bluest. This colour dependence saturates around metallicity --2.5,
when line blanketing has no further effect on the broad band colours.
This behaviour is clear in Fig.\,\ref{cmd} where isochrones of several metallicities are
shown. From Fig.\,\ref{cmd_hs} it appears that we have sampled
the blue-edge (most metal-poor stars) of the RGB and have observed several stars
on the red-edge (most metal-rich).
Similar to what done in Paper\,I we assign to each star a working name GHSXXX
where GHS stands for Gaia High Speed and XXX is a three digit integer.
In Table \,\ref{tablenam} we provide Gaia DR3 source IDs for each star.

One feature that stands out from  Fig.\,\ref{cmd} is that there are ten stars
that are far too blue and luminous to lie on old isochrones of whatever
metallicity. We refer to these stars as `young' and we derived  their ages and masses
using isochrones as described in Sect.\ref{sec:param}.

\section{Observations and radial velocities\label{obs}} 

\subsection{FORS}
In periods 105 and 106,  we observed 287 FORS  spectra  of 278 stars.
The instrumental setup was the same as used for Paper\,I.
We discuss in this paper together the stars observed in periods 105, 106 and 104 (Paper\,I)
for a total of 350 stars. Two of the stars observed are not discussed in the main text: star GHS36 of
Paper\,I that was not analysed chemically, due to the low S/N ratio in the spectrum
and star GHS076, observed in period 105, a reddened B-type star discussed  in  appendix\,\ref{remarks}.
Variable stars are also discussed in   in  appendix\,\ref{remarks}.

\begin{figure}
\centering
\includegraphics[width=\hsize,clip=true]{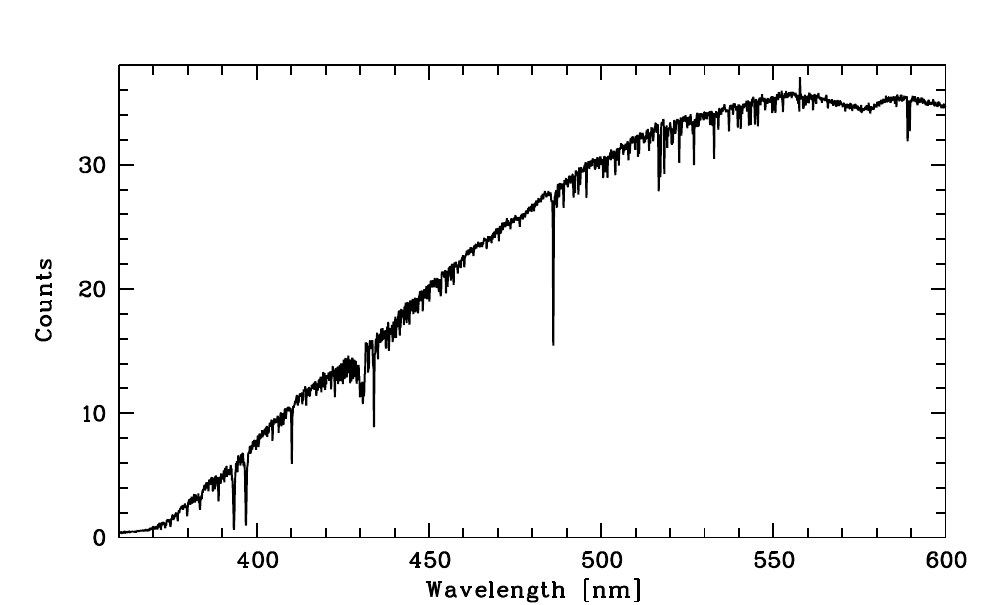}
\caption{Observed FORS spectrum of GHS337.}
\label{fig:obs337}
\end{figure}

We used GRISM 600B+22 and a 0\farcs{28} slit. The CCD was read with 
a binning $1\times 1$.  
This configuration provides a resolving power of about\ 2800\ in the spectral range 330–621 nm.
In Fig.\,\ref{fig:obs337} we show an example for the very metal-poor star GHS337.
The spectra were reduced using the ESO FORS pipeline.

\begin{figure}
\centering
\includegraphics[width=\hsize,clip=true]{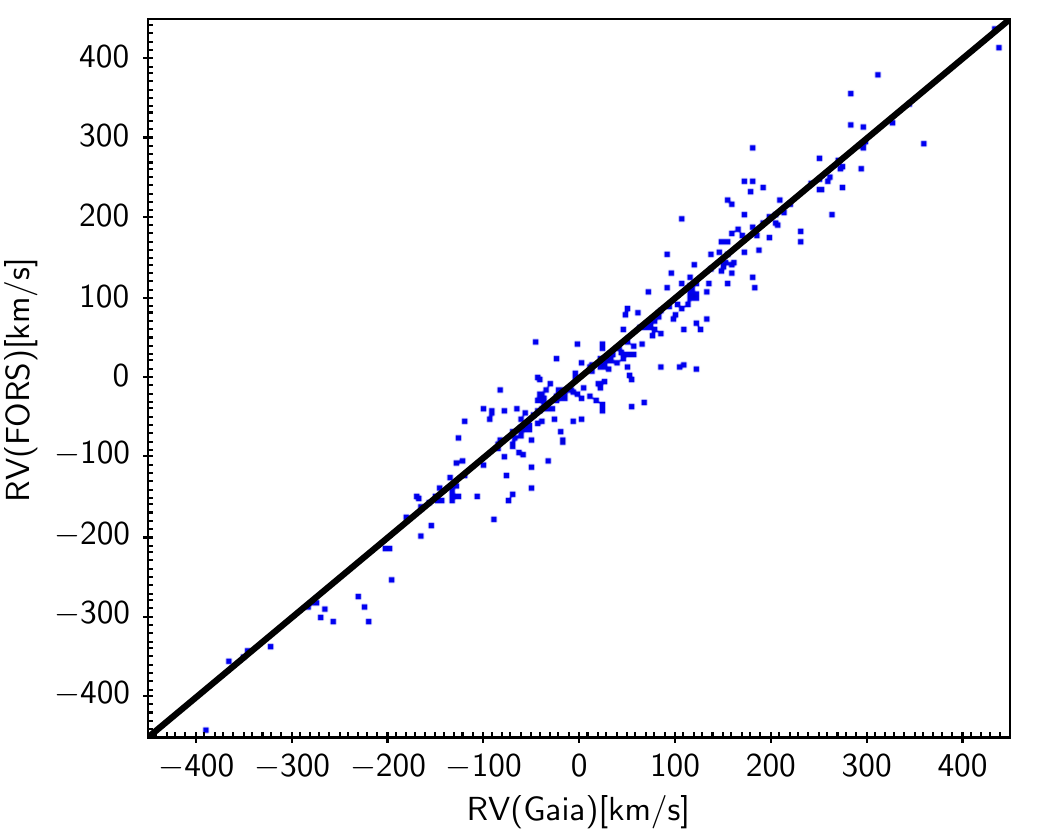}
\caption{Comparison of the radial velocities measured with FORS with those of Gaia RVS. The black line is the bisector, to guide the eye.}
\label{fig:vrad}
\end{figure}

For each spectrum a first estimate of the radial velocity was determined, like
in Paper\,I by template matching. This was used to shift the spectrum 
to rest wavelength and fed to \mygi\ \citep{mygi14} for the chemical analysis.
Among the \mygi\ output there is also the radial velocity shift derived for each feature.
For each star the shifts derived for the features retained for the abundance analysis
were averaged and a correction was applied to the radial velocity derived
from template matching.
In Fig.\,\ref{fig:vrad} we show the comparison of our radial velocities
with those of Gaia RVS.  The errors on the Gaia radial velocities
range from 0.8\,\kms to 10.5\,\kms with a median value of 2.6\,\kms and a standard deviation 
of 1.6\,\kms . On average there is a small offset of about 9\,\kms ,
in the sense Gaia--FORS, with a standard deviation of about 34\,\kms .
Since the FORS radial velocities are affected by systematic errors
caused by instrument flexures and
centring of the star on the slit \citep{caffau18,PristineXI}, we prefer to use the Gaia radial velocities, whenever available,
yet we consider the good correlation between the two sets of radial velocities, satisfying.

\subsection{UVES observations}

A possible diagnostic to distinguish blue straggler stars from truly
young stars, is the Li
abundance, since blue stragglers generally do not have measurable lithium
\citep{1991PASP..103..431H,1991ApJ...373..105P,1994AJ....108..271G,2001ApJ...547..231R}.
With the FORS spectra it is not possible to use the Li diagnostic
because the only available Li abundance indicator, the \ion{Li}{i} resonance doublet
 at 670.7\,nm,
is not within the observed wavelength range and in any case the resolution is too low. 
For this reason, we requested a high resolution follow-up of two of the sub-giant
candidate blue stragglers from Paper\,I with UVES in the ESO period 108, in order to measure the
\ion{Li}{i} resonance doublet.
We were allocated nine hours of observation, but only three were executed on two targets. 
The stars GHS69 ans GHS70 were observed between December 2021 and March 2022\ in service mode 
using the standard setting DIC1 390+580 (326-454\,nm  in the blue arm and 476-684\,nm  in the red arm).
With a slit of 1\farcs{0} and a $1\times1$ binning this setting provides a resolving power of 
$\sim$\, 40\,000. 
With one hour observing blocks (corresponding to 3000\,s integration) 
we expected to obtain 60\,$\leq$\,S/N\,$\leq$\,80 at 671\,nm and 30\,$\leq$\,S/N\,$\leq$\,40 at 400\,nm. 
The constraint on the  seeing was better than 1\farcs{2}, all observations
were taken  at airmass $\sim$\,2. 
The star GHS70 was observed twice, 
since the mean seeing of the first observation was above the requested one. 
With these constraints, we achieved a mean S/N 
lower than expected around the \ion{Li}{i} doublet, 
with S/N\,$\sim$\,42 for GHS69, and S/N\,$\sim$\,55 for GHS70 (by summing the two exposures).

\section{Kinematics}\label{kin}

In order to characterise the stellar kinematics, we used the {\tt galpy} code together with its default Galactic potential \citep[MWPotential2014,][]{bovy15}. We further adopted the \citet[][]{sch10} solar peculiar motions, 8\,kpc as solar distance and 220\,\kms\ as circular velocity at the solar distance \citep[][]{1986MNRAS.221.1023K}, that
is consistent with the recent determination by \citet[][$218\pm 6$\,\kms]{bovy12}. 

For all the stars, we fed {\tt galpy} with Gaia\,DR3 coordinates, proper motions and radial velocities. Distances were obtained from Gaia\,DR3 parallaxes, corrected for the zero-point according to the prescriptions of \citet[][]{lindegren21}. Similarly to \citet[][]{topos6}, uncertainties on the derived quantities were evaluated by extracting random realisations of the input parameters (positions, proper motions, distances and radial velocities) considering the error in the input parameters and the astrometric covariance matrix using the {\tt Pyia} code \citep[][]{pyia18}. For each stars, 1000 realisations were fed to {\tt galpy} and we adopted as uncertainties the standard deviations of the calculated quantities.

There is always some concern in deriving distances by inverting parallaxes
\citep[see e.g.][]{luri18} and some prefer to use bayesian estimates \citep[see e.g.][]{bj21}.
One should be aware, however that most of the troubles arise when the
parallax error is large, the example
in \citet{luri18} has been obtained assuming 0.3 mas errors on the parallaxes. 
Our sample has a mean parallax error of 0.016 mas and the relative
errors range from 2\% to 21\% with a median of 8\%.
In fact, when we compare the distances obtained from inverting the parallax
with the bayesian photogeometric estimates of \citet{bj21}, we obtain an excellent
correlation. A linear fit of the two distance estimates yields a slope of 0.87 and 
a root-mean-square deviation of 0.3\,kpc.
None of our main results would change, had we adopted the bayesian distance estimates. 
Finally we should also note that the use of bayesian distance estimates
introduces a bias, that is due to the assumed prior. 
We chose not to use bayesian estimates to avoid this bias.
A similar approach was used also by \citet{marchetti22}.

Gaia\,DR3 radial velocities (RV) are not available for three stars, namely GHS91, GHS248 and GHS277. For these stars we adopted the RVs measured from the FORS spectra and a formal error of 30\,\kms, consistent with the
standard deviation discussed in Sect.\,\ref{obs}. 

We decided to perform the same analysis also on the stars from Paper\,I. In this case 5 stars do not have RVs from Gaia\,DR3 (GHS29, GHS65, GHS66, GHS69, GHS70). For these stars we used the RVs and errors derived from the FORS spectra in Paper\,I. Stars GHS22, GHS33, GHS37, GHS58, GHS64 were indicated as unbound in Paper\,I, while they are found to be bound in the present analysis. This difference is due to the zero-point corrected parallaxes adopted here and entirely consistent with the discussion on the parallax zero point presented in section 4.2 of Paper\,I. On the other hand, with a total galactocentric space velocity of 1439.8\,kms, GHS143 results unbound to the Milky Way according to the current analysis (see Sect. \ref{ghs143}). 

Out of the 348 stars in our combined sample, 346 belong to the halo and two to the thick disc according to the criteria introduced by \citet[][]{bensby14}. Ninety of the halo stars belong to the GSE 
and 17 to the Sequoia (Seq,\,\citealt{barba19,myeong19,villanova19}, but see also \citealt{myeong18,koppelman18}) structures, following the criteria introduced by \citet[][]{feuillet21}.

We noticed among our targets, a group of stars located in the inner part of the Galaxy. In order to assess the presence among our targets of stars confined to the bulge, we repeated the analysis described above but this time adding to the MWPotential2014 Galactic potential a rotating bar \citep[][]{dehnen00} generalised to three dimensions as in \citet[][]{monari16}. We identify 16 stars having an apocentric radius r$_{ap}$ lower than 3.5\,kpc and for which more than 50\% of the random realisation of the input parameters fed to {\tt galpy} returned a r$_{ap}<$3.5\,kpc. We classify them as bulge stars. It is worth notice that 14 out of 16 stars are in retrograde motion (L$_Z<$0). Star GHS247 belongs both to the bulge and Sequoia,
according to our adopted definitions. 

With 222 over 348 analysed stars (64\%, we exclude GHS076) having L$_Z<$0, our sample is dominated by stars in retrograde motion. Excluding GSE (90 stars, 51 with L$_Z>$0), Seq (17 stars, all with L$_Z<$0), the two thick disc stars (L$_Z>$0) and the stars confined to the bulge ( 16 of which 14 with 
L$_Z<$0), the remaining 223 halo stars are divided into 141 retrograde  (63\%) and 82 prograde (37\%)  motions.

Of the group of ten young stars, as identified from the CMD in Sect\,\ref{sel}, 
GHS108 and GHS110 are bulge stars, GHS212 belongs to GSE and GHS120 to Seq. Star GHS143, the only star classified as unbound within our sample, is also a young star. 
Star GHS110 is photometrically variable according to  Gaia, however 
its classification is uncertain
(see section \ref{var}).

\begin{figure*}
\centering
\resizebox{\hsize}{!}{
\includegraphics[clip=true]{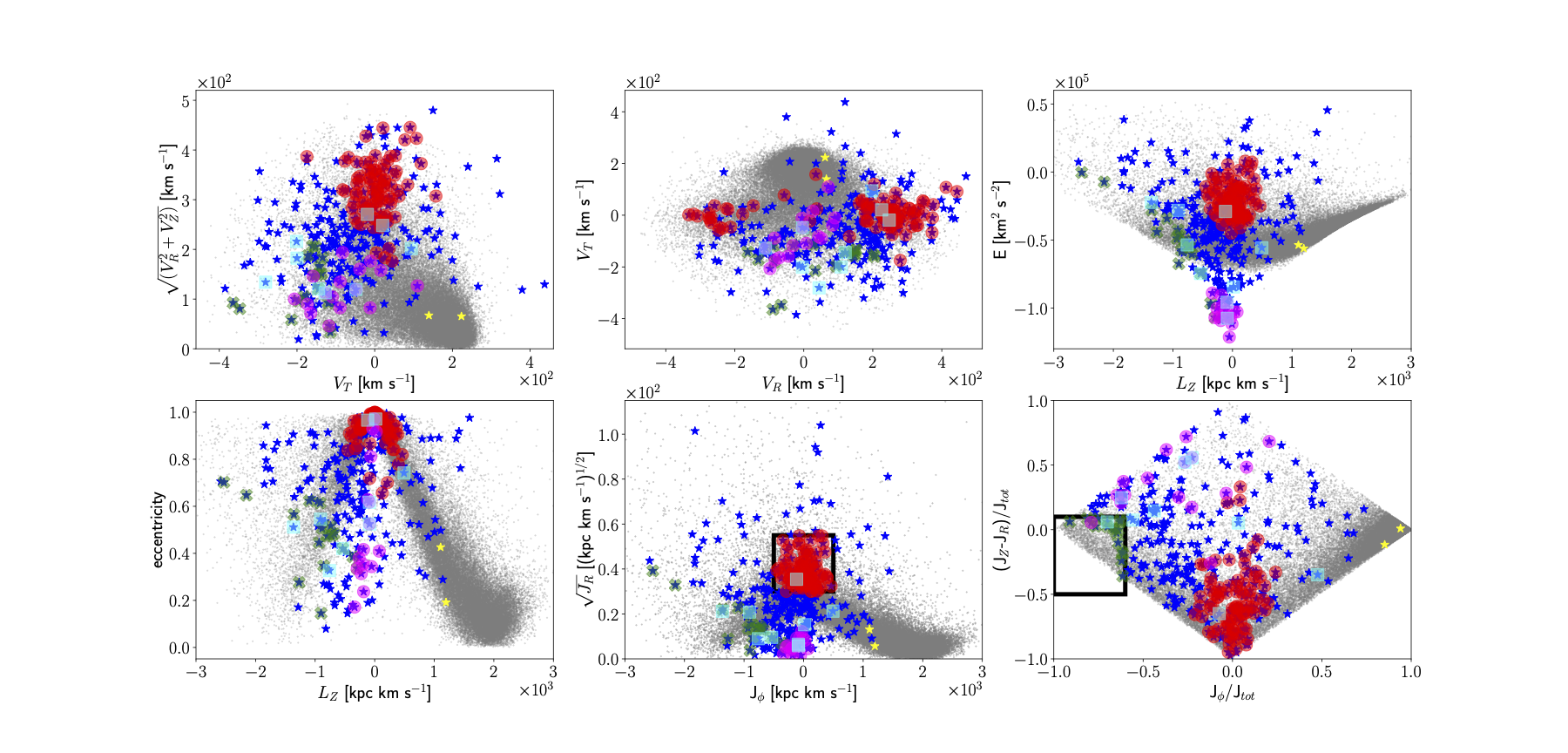}}
\caption{Target stars (coloured filled symbols) are presented in several planes. Top panels:  E  vs L$_Z$ (right, orbital energy versus the vertical component of the angular momentum), V$_R$  vs V$_T$ (middle, radial versus transversal velocity component in galactocentric cylindrical coordinates) and V$_T$  vs $\sqrt(V_R^2+V_Z^2)$ (left, Toomre diagram: transversal velocity component, versus a combination of the radial and vertical velocity components in galactocentric cylindrical coordinates). Bottom panels: (J$_Z$-J$_R$)/J$_{tot}$  vs J$_\phi$/J$_{tot}$ (right, action diamond, where: J$_{tot}$=|J$_\phi$|+J$_R$+J$_Z$, the box is the region used to select Seq candidates), J$_\phi$=L$_Z$  vs  $\sqrt{J_R}$ (middle, square root of the radial action versus the vertical component of the angular momentum, the box indicates the region of the plane used to select GSE candidates),  orbital
eccentricity  vs  L$_Z$ (left). Grey dots are stars of the  "good parallax sub-sample" of \citet[][]{topos6} and are plotted for reference. Filled blue and yellow stars indicate targets classified as halo and thick disc, respectively. Filled circles are GSE (red),  bulge (magenta) stars. Green crosses are Seq star.} Filled cyan squares are young stars.   
\label{char1}
\end{figure*}

We present in figure\,\ref{char1} the target stars in several, commonly used, kinematic planes \citep[][]{lane22}, using combinations of orbital Energy (E) and angular momentum L$_Z$ (top-right panel), eccentricity (bottom-left panel), actions (J$_\phi$=L$_Z$, J$_R$, J$_Z$, bottom middle and right panels), and velocity components in galactocentric cylindrical coordinates (V$_T$, V$_R$, V$_Z$, top left and middle panels). In the action diamond in the bottom-right panel, quantities are normalised to the total action J$_{tot}$=|J$_\phi$|+J$_R$+J$_Z$.

Halo and thick disc stars are shown as blue and yellow stars in the figure, respectively. 
Stars belonging to GSE, Seq or to the bulge are shown as big filled, red, green and magenta circles, respectively. Cyan filled squares are stars classified as young. Grey points are stars from the "good parallax sub-sample" of \citet[][]{topos6} and are plotted only for reference. The shaded red and green boxes in the figure, correspond to the \citet[][]{feuillet21} criteria employed to select likely GSE and Seq candidates, respectively.
The selection of GSE can be found in the lower row, middle panel, that of Seq on the bottom
row right panel.

\begin{figure*}
\centering
\resizebox{\hsize}{!}{
\includegraphics[clip=true]{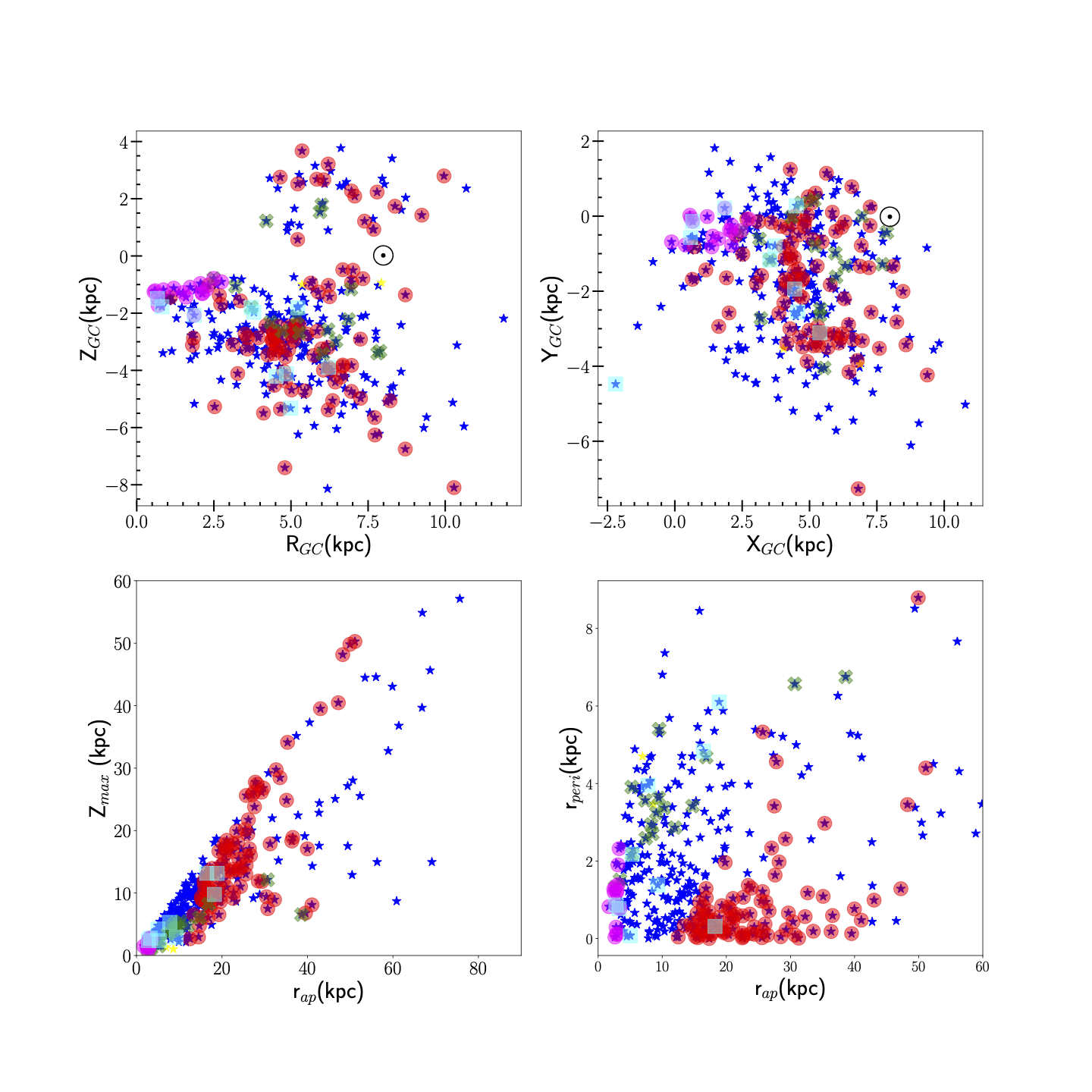}}
\caption{The top panels show Y  vs  X (right, galactocentric Cartesian coordinates), Z  vs  R (left, galactocentric height over the galactic plane versus galactocentric cylindrical distance). The bottom panels show r$_{peri}$  vs  r$_{ap}$ (right, pericentric versus apocentric distances), Z$_{max}$ (maximum height over the galactic plane)  vs  r$_{ap}$ (left). Symbols are the same as in fig.\,\ref{char1}. 
The position of the Sun is shown in the top two panels with the solar symbol.}
\label{char2}
\end{figure*}

Figure \ref{char2} shows the target stars in the   Y  vs  X  plane (top-right panel, Galactocentric Cartesian coordinates), the height over the galactic plane Z  vs  the cylindrical radius R ($ R=\sqrt{(X^2+Y^2)}$   top-left panel), maximum height over the galactic plane Z$_{max}$  vs  r$_{apo}$ plane 
(bottom-left  panel) and apocentric  vs  pericentric  distances (r$_{peri}$  vs  r$_{apo}$, bottom-right panel).  
The position of the Sun is $(X,Y,Z)=(8.0,0,0.0208)$ and $R=8.0$, everything expressed in kpc.

Several investigations 
have introduced chemical and dynamical criteria to select stars likely sharing common origins 
\citep[e.g.][]{helmi18,Naidu,feuillet21,limberg22,buder22}.  
Such criteria are based on extensive databases, like Gaia, APOGEE, GALAH, and H3 to name a few. 
If one adopts  some of these criteria for our sample of stars,
one is  lead to tag  a few stars as belonging to some of these structures. 
However, 
this does not take into account the existence of any
significant signal in our sample,  that points towards 
the presence of a prominent feature. 
Such an outcome can 
be clearly seen from our analysis of the stars selected from the Gaia Universe Model (GUM, \citealt{robin03}\footnote{Available at
\url{https://gea.esac.esa.int/archive/}, see also
the Gaia DR3 documentation
\url{https://gea.esac.esa.int/archive/documentation/GDR3/Data_processing/chap_cu2sim/sec_cu2UM/}. }
see Sec.\,\ref{sec:spitef}). 
We thus decided to perform a clustering analysis among the stars of our small sample, based on the derived integrals of motion only, in order to have some objective
insight into the groups of stars that can be found. 
The details of this analysis can be found in  appendix \ref{clustering}. 

The main conclusion is that two main structures are present among the stars of our sample. One is connected to GSE and, indeed, the clustering analysis recovers a selection similar to that introduced in \citet{feuillet21}.  
In the [Mg/Fe] vs [Fe/H] plane, 
when we consider stars selected according to our clustering analysis and 
to the \citet{feuillet21}  selection criteria, we detect qualitatively similar trends. 
However, since these latter are now widely used in the literature, and are 
based on a much more extensive database, we decided to use them in the end. 
Similarly, we decided to still tag stars as Seq according to \citet{feuillet21}, although our clustering analysis does not reveal the presence of a structure compatible with the \citet{feuillet21} criteria. Indeed, our GUM analysis and the lack of a clear pattern in Fig.\,\ref{char3} cast 
doubts on the reality of the association of these stars between them and with the Seq accretion event. 

The second structure revealed by our clustering analysis, is composed of stars at low energy and belonging to the inner part of the Galaxy. The kinematics of these stars is, however, clearly affected by the presence of the bar. Therefore, we decided to repeat our analysis but this time adding a rotating bar to the standard {\tt galpy} potential. 
We ended up with 16 stars only that respect the criterion $r_{ap}<3.5$\,kpc,  these are the stars 
we tagged as SpiteF (see Sec.\,\ref{sec:spitef}). 
Comparing in the [Mg/Fe] vs [Fe/H] plane the stars selected in this second structure by our clustering analysis (34 stars, see Appenidx\,\ref{clustering})  with the 16 SpiteF stars, we believe that a much clearer signal is apparent considering this latter selection. 

In summary, the clustering analysis we performed convinced us that GSE and SpiteF stars represent prominent groups among the stars of our sample and not just overdensities corresponding to random fluctuations of parameters. 
It is interesting to note that an analysis similar to the one described in the appendix, 
performed on the GUM sample does not reveal any significant 
cluster. This is expected, since GUM does not contain any substructure, but is a sanity
check, for the clustering analysis.
We describe this analysis in appendix\,\ref{clustering}.
 
\subsection{GHS143 an unbound star \label{ghs143}}

Star GHS143 has a total galactocentric space velocity of $\sim$1440\,\kms which, compared to a local escape velocity of 521.7\,\kms, suggests that the orbit of this stars is unbound to the Galaxy\footnote{The total orbital energy E of this star is positive. However for the MWPotential2014 this does not always indicate that one star is unbound to the Galaxy, as clearly specified in the {\tt galpy} documentation.} and make this an hyper-velocity star. To our knowledge, four stars only have larger space velocities \citep[see Table A.1 in][and references therein]{li22a}, GHS143 being the only late-type star among them. \citet[][]{boubert18} identified one late-type star only (LAMOST J115209.12+120258.0) unbound to the galaxy. \citet[][]{li22a} identified 52 marginally hyper-velocity stars candidates which, similarly to GHS143, are metal-poor late type halo stars. Their galactocentric space velocities are, however, significantly lower ($<$750\,\kms).

The large galactocentric space velocity of GHS143 is dominated by the Y component (in cartesian coordinates), which is also affected by the largest uncertainty: v$_{X,Y,Z}$=(535$\pm$149, -1318$\pm$408, 222$\pm$59\,\kms). The total uncertainty on the space velocity is 438\,\kms, while the difference between the space velocity of GHS143 and the local escape velocity is 918\,\kms. The minimum value of this difference obtained over 1000 random realisation of the input parameters is 227\,\kms, while the standard deviation is 465\,\kms. Notice that GHS143 (Gaia DR3 6632370485122299776) has {\tt RUWE}=1.33, lower than the recommended threshold of 1.4, and {\tt parallax\_over\_error}=2.7.

\section{Chemical analysis}
\begin{figure}
\centering
\resizebox{7.7cm}{!}{
\includegraphics[clip=true]{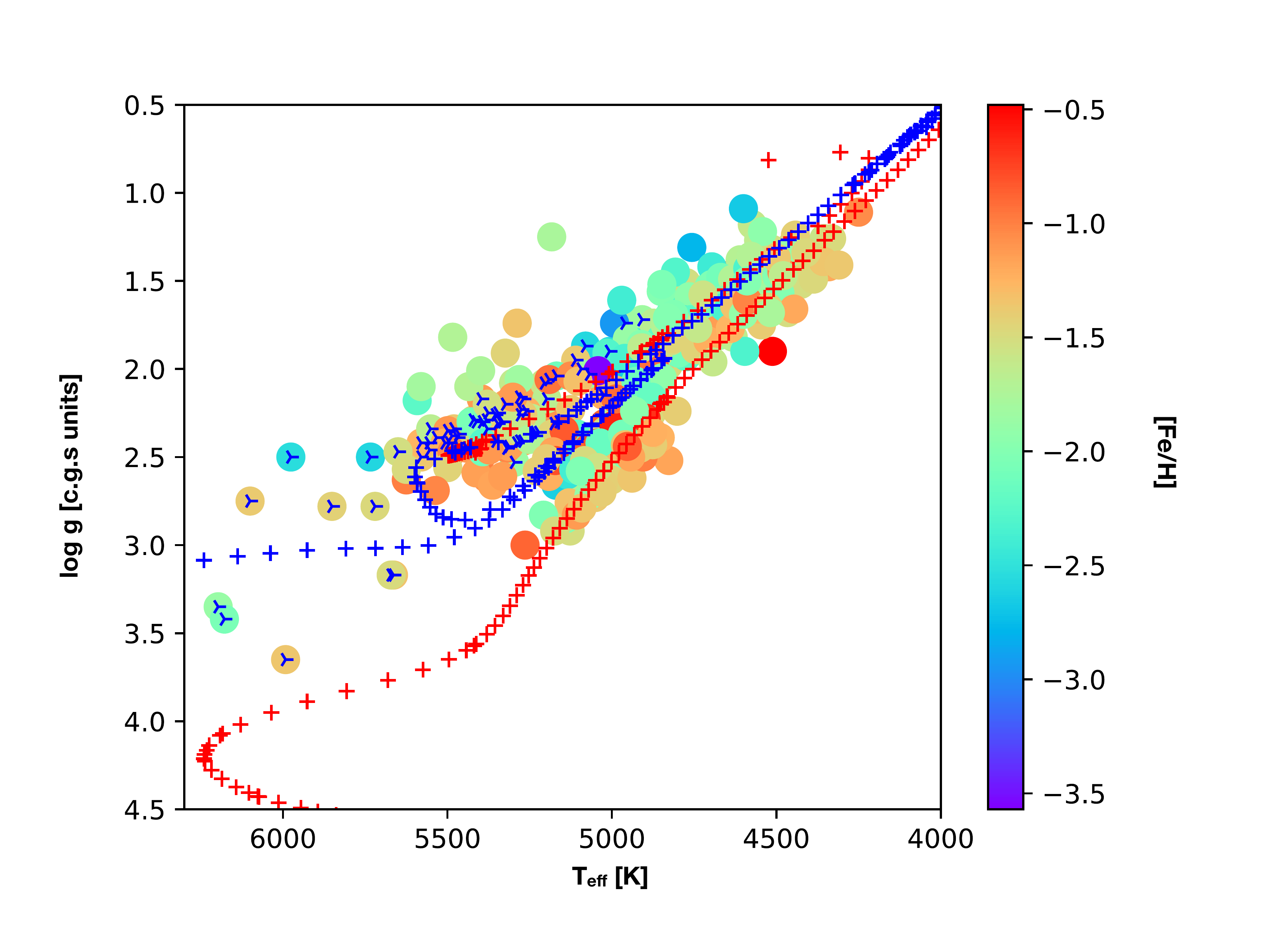}}
\caption{Observed stars in the (\teff,\logg) diagram, compared to PARSEC isochrones
of metallicity --1.0 and age of 0.2 (blue crosses) and  10\,Ga (red crosses), to guide the eyes.
Filled circles with a cross are from \citet{ghs104}.
}
\label{fig:plotiso}
\end{figure}

\begin{figure}
\centering
\resizebox{8.5cm}{!}{
\includegraphics[clip=true]{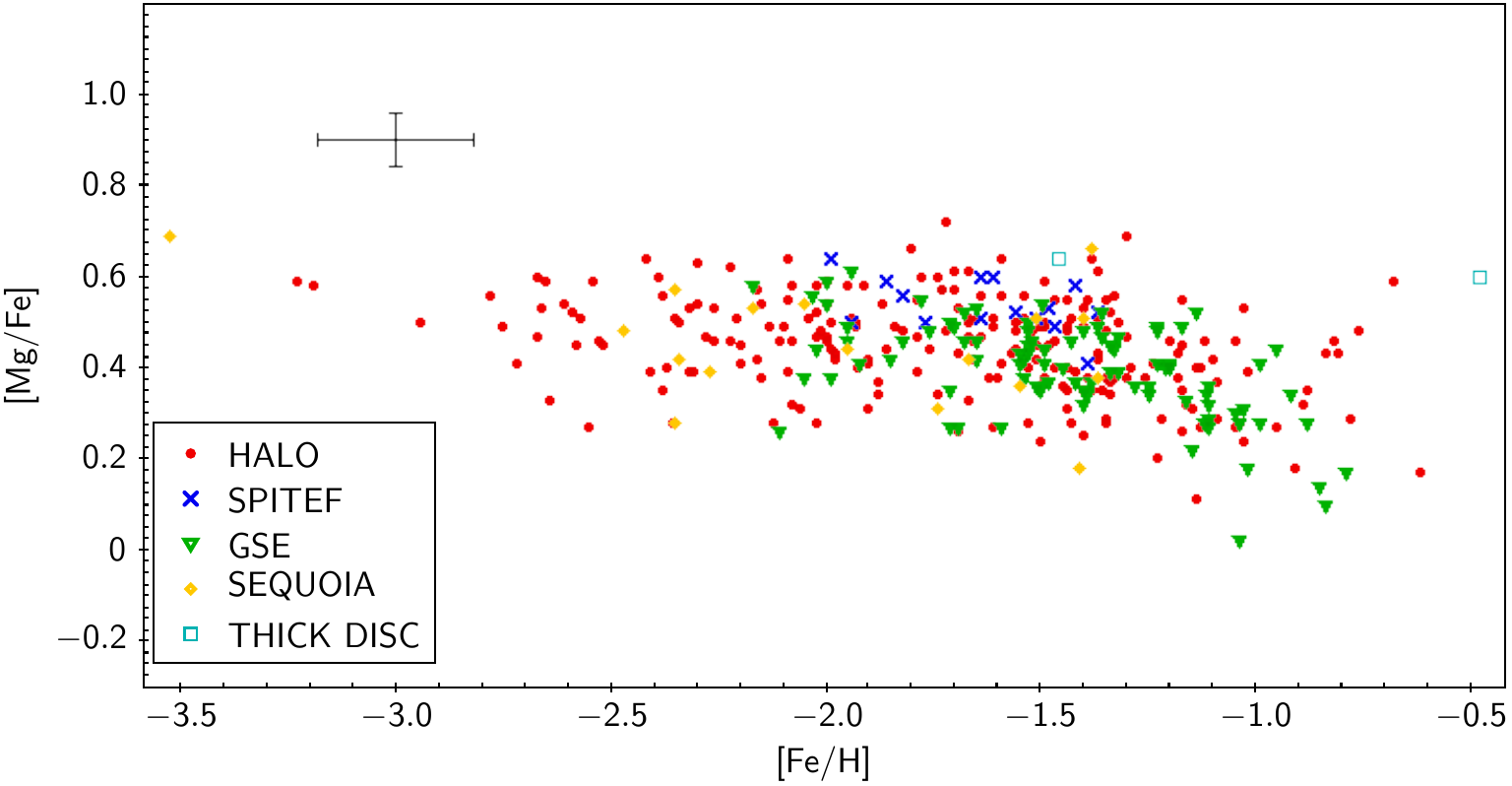}}
\caption{[Mg/Fe] versus [Fe/H] for the observed sample.
The different Galactic components identified kinematically
are shown by different symbols and colours, as detailed in the legend.
}

\label{fig:mgfe}
\end{figure}

\begin{figure}
\centering
\resizebox{8.5cm}{!}{
\includegraphics[clip=true]{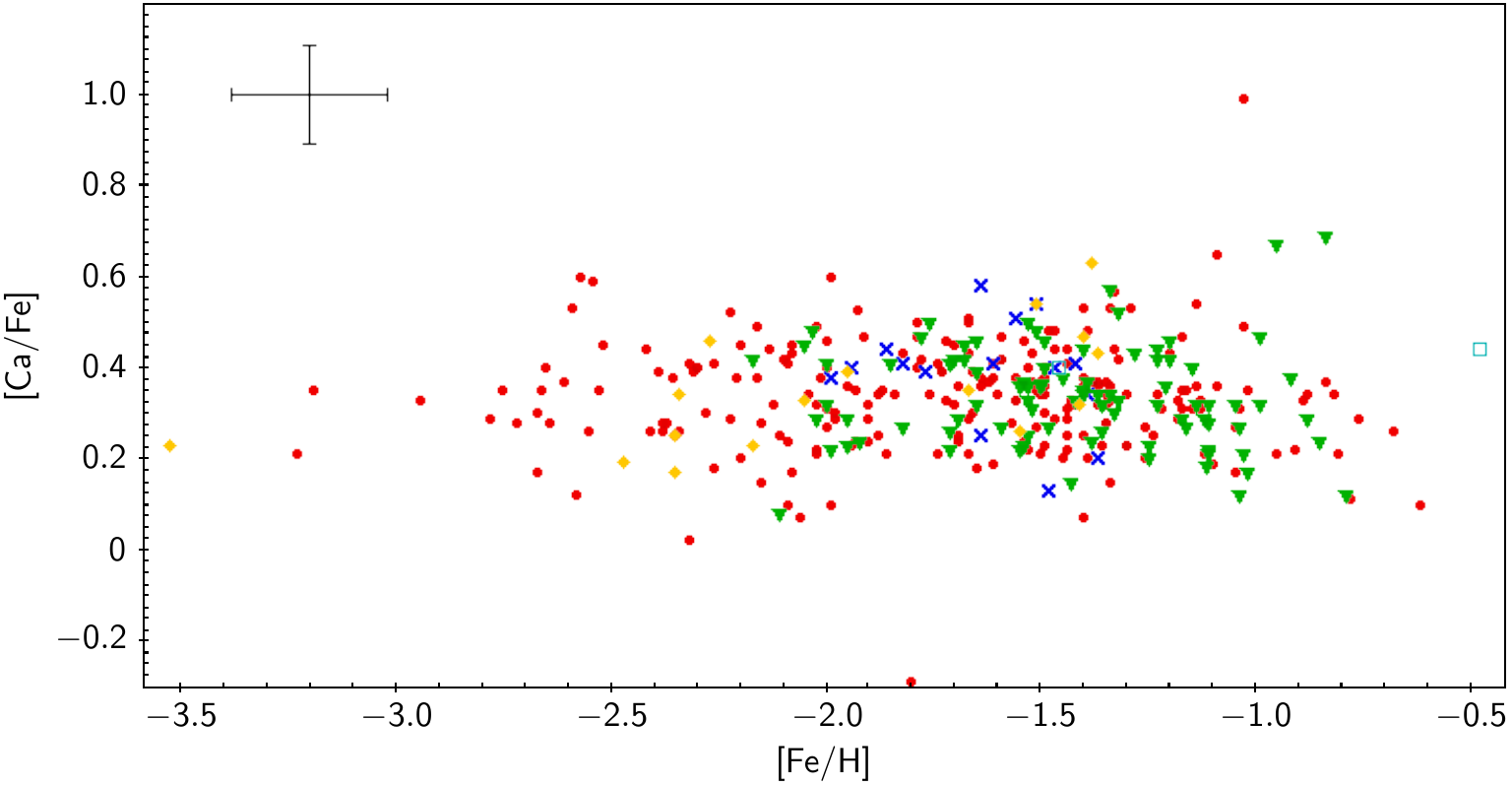}}
\caption{[Ca/Fe] versus [Fe/H] for the observed sample.
Symbols are the same as in Fig.\,\ref{fig:mgfe}.
}
\label{fig:cafe}
\end{figure}
\begin{table}
\caption{Masses,  ages, and metallicities for the young stars.\label{masses}}
\centering
\begin{tabular}{crrc}
\hline
Star & Mass & Age & [Fe/H]\\
     & M\sun &  Ma & dex \\
\hline
GHS108 & 1.9 & 941 & --1.39\\
GHS110 & 2.2 -- 2.8 & 398 -- 628  & --1.32 \\
GHS114 & 1.4 -- 2.1 & 646 -- 2024 & --1.56\\
GHS117 & 2.2 -- 2.8 & 315 -- 698  & --1.59\\
GHS120 & 1.4 -- 2.1 & 571 -- 1845 & --2.17\\
GHS143 & 3.1 -- 3.8 & 156 -- 279  & --1.74\\ 
GHS145 & 1.3 -- 2.1 & 631 -- 2545 & --1.79 \\
GHS189 & 1.7 -- 2.3 & 485 -- 1388 & --1.69 \\
GHS209 & 1.4 -- 2.3 & 498 -- 1999 & --1.67\\
GHS212 & 1.7 -- 2.2 & 574 -- 1270 & --1.71\\
\hline
\end{tabular}
\end{table}

\subsection{Stellar parameters} \label{sec:param}

We adopted the reddening from the maps of \citet{SF11}.
The procedure to determine \teff\ and \glog\ is iterative
and is described in detail in \citet{2021A&A...656A.155L}.
In a nutshell: the dereddened 
$G_{BP}-G_{RP}$ Gaia\,DR3 colour is compared to the synthetic photometry  
to derive \teff\ and bolometric correction, the \glog\ is then determined from the Stefan-Boltzman
equation, the extinction coefficients are updated and the whole procedure is iterated
to convergence.  The synthetic colours were computed
from the ATLAS\,9 fluxes of the grid of Mucciarelli  et al. (in preparation). 
For most stars we used a subset of the whole grid suitable for giant stars,
it covers effective temperatures from 3500\,K to 5625\,K 
in steps  of 125\,K, surface gravities from 0.0 to 3.0
(c.g.s. units) in steps of 0.5\,dex, and metallicities from 
--5.0 to 0.5, the metallicity steps are of 0.5\,dex between --5.0 and --2.5
and 0.25\,dex for metallicities above --2.5.
Two stars, GHS292 and GHS312, are warm horizontal branch stars 
and for these two we used a smaller grid, covering the same metallicity range,
but effective temperatures from 4875\,K to 6000\,K in 
 steps of 125\,K and surface gravities from 2.0 to 4.0 (c.g.s. units) in steps of 0.5\,dex. 
As convergence criteria we adopted $\Delta T < 50$ and $\Delta$\glog $< 0.05$, where
$\Delta$ denotes the difference between the current parameters and those of the previous iteration.
The required iterations were typically two in\relax\ \teff\ and one in \glog. 
The stellar parameters were then  used to derive the metallicity from the spectrum by using \mygi\ \citep{mygi14}.
The metallicity so derived was then used to  update the 
stellar parameters and the process was iterated again.
A single iteration at this stage was sufficient to achieve convergence in the above-defined sense.
The microturbulent velocities ($\xi$) are derived by using the calibration of \citet{mashonkina17}. 
Our estimate of the uncertainties (statistical plus systematic) is of 100\,K in effective
temperature and 0.2\,dex in surface gravity.

In a first run we assumed for all stars a mass of 0.8M\,\sun, appropriate for an old stellar population, then for
the 10 young stars we used the derived metallicity to interpolate in the  PARSEC isochrones \citep{bressan12,marigo17}
to derive their masses and ages. We then re-derived their atmospheric parameters with the new masses.
As expected the largest change was to \glog\ while the temperature hardly changed, neither did the
metallicity (see Sect.\,\ref{secabbo}). 
The region of the colour-magnitude diagram where the young stars are found
is characterised by the presence of loops and different evolutionary stages from
sub-giant to core helium burning. This implies an ambiguity in the derived
masses and ages, depending on the assumed evolutionary stage. We therefore provide
for each star a range of masses and ages when different solutions are possible.
The results on masses and ages for the young stars are summarised in Table\,\ref{masses}.
We considered the effect of errors on the photometry. The largest effects come
from the uncertainty on the reddening and on the parallax, however the combined error
of both is negligible compared to the uncertainty on the evolutionary stage of the star.
We stress that we only use the photometry to derive stellar masses and ages, thus
our adopted effective temperatures and surface gravities do not enter into this derivation.
In Fig.\,\ref{fig:plotiso} the adopted stellar parameters, compared to PARSEC isochrones, are shown.


\subsection{FORS spectra\label{secabbo}}

The derived atmospheric parameters were used as input to \mygi\ \citep{mygi14}
to derive the chemical abundances from the spectra, as done in Paper\,I.
In spite of the low resolution of the FORS spectra, the high S/N ratios (always above 100
at 500\,nm) and the large spectral coverage allowed us to determine the
abundances of Na, Mg, Ca, Ti, Mn, Fe and Ba.  
The results are summarised in Figs.\,\ref{fig:mgfe} to \ref{fig:bafe}.
The full table with all abundances for each star is available in electronic form at CDS
via anonymous ftp to cdsarc.u-strasbg.fr (130.79.128.5) or via http://cdsweb.u-strasbg.fr/cgi-bin/qcat?J/A+A/.
For star GHS110, that is a variable star, we only provide
\teff , \glog , and [Fe/H] but refrain from doing a multi-element
abundance analysis given the additional uncertainty on its effective temperature
and surface gravity, as discussed in Sect.\,\ref{var}.
For nine stars we have two spectra, mostly observed both in periods 105 and 106,
in some cases one spectrum was observed, but assigned C quality and then re-observed
with A quality. In these cases we ran \mygi\ independently on the two spectra,
in order to be able to estimate the errors on the abundances derived from different
spectra. 
The comparison is generally quite good, with abundances derived from the different
spectra showing a standard deviation of the order of 0.1\,dex or less, with the exception
of the cases of spectra of different quality, like C and A, in this case the standard deviation
can be as large as 0.3\,dex.
For these nine stars we averaged the different abundances and use the spectrum-to-spectrum scatter
as error estimate on the abundances. 

\subsubsection{Carbon abundances}

Although we let \mygi\ fit the G-band to provide the abundance of C
we do not provide these values since we estimate it to be very uncertain.
The uncertainty on these values arises from two facts: i) the width of the band makes
the continuum placement highly undertain; ii) \mygi\  interpolates between synthetic
spectra of different metallicity, that have different ionisation structure 
(see appendix \ref{mygi_ion}), and this affects the band formation.
Nevertheless we believe these abundances output by \mygi\ are
useful to alert us in case of large carbon overabundances.
In this way we flagged six stars for which we performed a traditional
fitting of the G-band to determine their carbon abundances that are reported
in Table\,\ref{cabun}.
This allowed us to identifiy six stars that are enhanced in carbon,
although only one has a large enough enhancement
to be classified as Carbon Enhanced Metal Poor star: GHS162.
Remarkably all six stars are also clearly enhanced in Ba,
which suggests pollution from an AGB companion. 
For all these stars the error on the Gaia  radial velocity is $\ga 1.4$\,\kms,
being as large as 4.6\,\kms for GHS341. This  makes it possible that at least some
of them are radial velocity variables.
Further scrutiny of these stars is encouraged.  

\begin{table}
\caption{Carbon enhanced stars.}
\label{cabun}
\begin{tabular}{cccc}
\hline
Star & [C/H] & [Fe/H] & [Ba/Fe]$_{NLTE}$\\
\hline
GHS151 & --3.08 & --3.53 & +1.2\\
GHS162 & --0.79 & --1.80 & +1.5 \\
GHS230 & --1.29 & --1.74 & +1.3 \\
GHS268 & --1.23 & --1.48 & +1.0 \\
GHS284 & --1.19 & --1.34 & +1.3 \\
GHS341 & --1.47 & --2.09 & +0.5 \\ 
\hline
\end{tabular}
\end{table}

\subsubsection{$\alpha$ elements abundances}

\begin{figure}
\centering
\resizebox{8.5cm}{!}{
\includegraphics[clip=true]{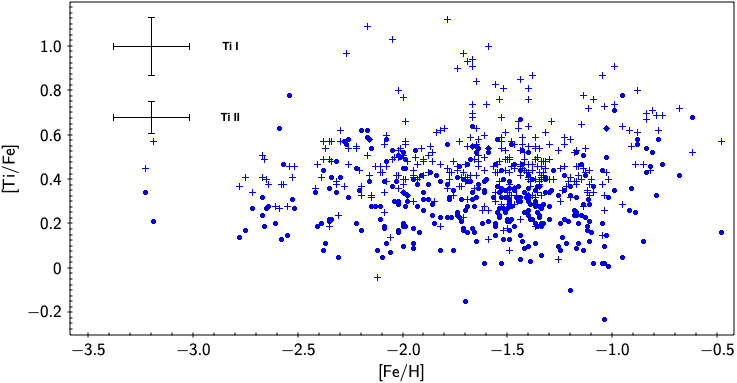}}
\caption{[Ti/Fe] versus [Fe/H] for the observed sample.
Filled circles A(Ti) from \ion{Ti}{i} and crosses from \ion{Ti}{ii} lines,
but A(Fe) is always from \ion{Fe}{i} lines.
}
\label{fig:tife}
\end{figure}

The  abundance ratios of the $\alpha$ elements Mg, Ca and Ti to iron
are shown in figures \ref{fig:mgfe}, \ref{fig:cafe} and 
\ref{fig:tife}.
While all three elements show a plateau below [Fe/H]=--1.5,
it is obvious that at any metallicity there is a large scatter, 
and that  many stars have low $\alpha$-to-iron ratios.
Part of the scatter is due to the observational uncertainties,
yet, given the size of uncertainties,  it is likely partly intrinsic.
As discussed in Sect. \ref{halo_aurora}, it is tempting to identify stars with low  $\alpha$-to-iron ratios
with stars accreted from dwarf galaxies, that have experienced
a slow or bursting star formation.
There is no obvious distinction between the different Galactic components that
are identified in the plots. The case of the Sequoia star GHS151 is intriguing, 
it is the most metal-poor star in our sample, [Fe/H]=--3.5, and 
it has [Mg/Fe]=+0.7, [Ca/Fe]=+0.2, [\ion{Ti}{i}/Fe]=--1.0 and [\ion{Ti}{ii}/Fe]=--0.8.
It is theoretically expected that not all $\alpha$ elements vary in lockstep,
especially at the lowest metallicities, because the nucleosynthesis sites are not the same.
The resolution of our spectra is too low to make a strong claim about this, 
however this star certainly deserves a closer scrutiny and analysis
with higher resolution spectra.
There is a clear offset of about 0.15\,dex in the [Ti/Fe] ratios derived from
\ion{Ti}{i} and \ion{Ti}{ii} lines, and we attribute most of this to NLTE 
effects on \ion{Ti}{i} lines \citep{sitnova_ti}. For the readers interested
in Galactic chemical evolution we recommend to use the abundances derived from the
\ion{Ti}{ii} lines.

\subsubsection{Sodium and manganese abundances}
\begin{figure}
\centering
\resizebox{8.5cm}{!}{
\includegraphics[clip=true]{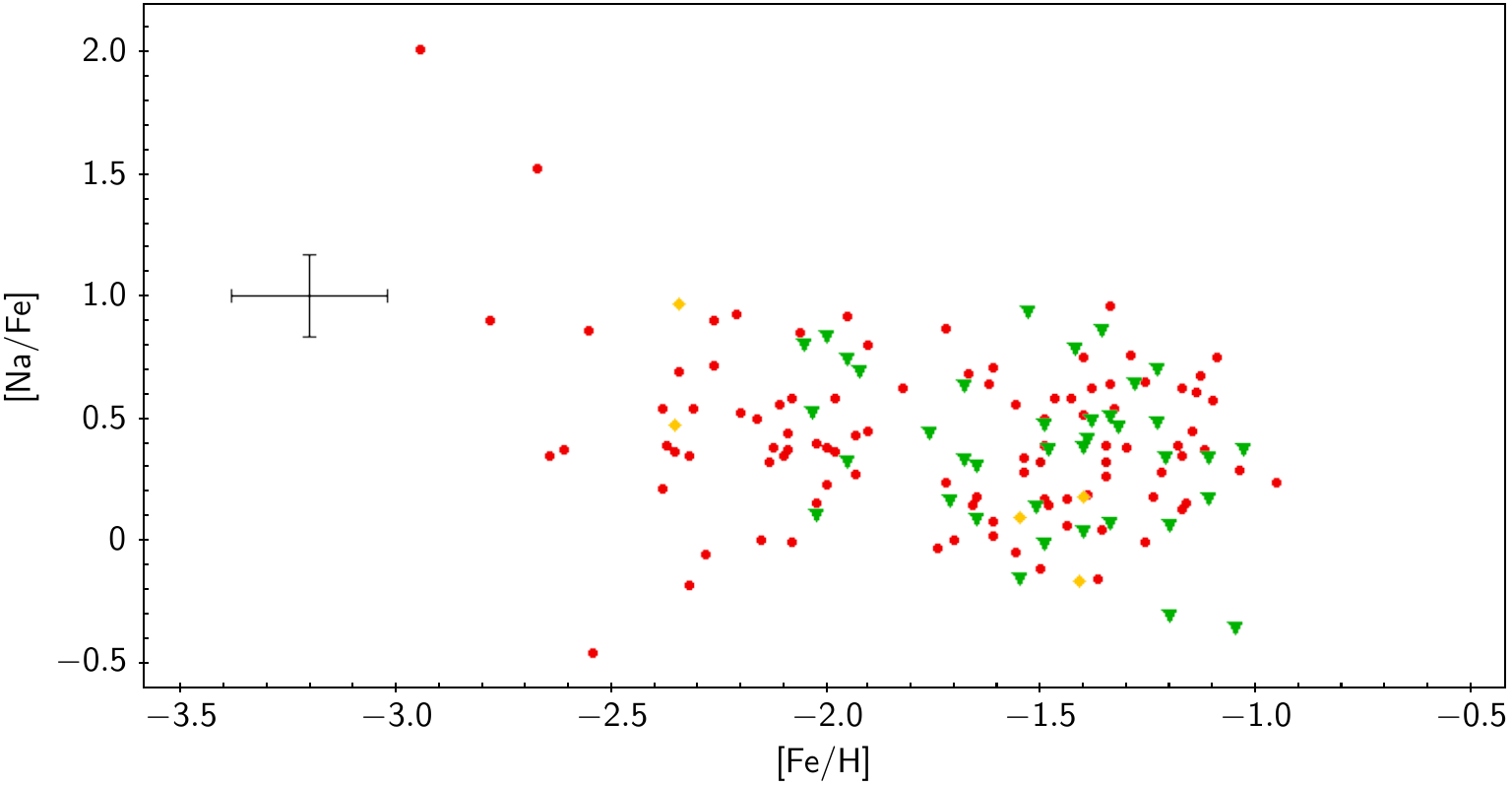}}
\caption{[Na/Fe] versus [Fe/H] for the observed sample.
}
\label{fig:nafe}
\end{figure}

 On average the stars appear enhanced in Na, with no clear distinction
among the different Galactic components as shown in Fig.\,\ref{fig:nafe}. This is not expected
for metal-poor stars \citep[see e.g. figure 8 of ][]{andrievsky07}.
This is very likely due to the fact that in our analysis we have neglected the deviations
from Local Thermodynamic Equilibrium that are important for \ion{Na}{i} D lines
\citep[see e.g.][figure 1]{andrievsky07}. Another possible cause of concern
is contamination with interstellar (IS)  \ion{Na}{i} D lines, that at the resolution of our spectra,
can contaminate the stellar lines, even if the star has a high radial velocity. 
We provide no Na measure of any of the Bulge stars, since
they are all clearly contaminated by  IS lines, in some cases a wide structure
of these lines is even visible. In order to obtain reliable Na abundances in these
stars higher resolution spectra are necessary. These will allow to disentagle
IS from stellar lines in many cases and in many cases other non-saturated \ion{Na}{i} 
lines will be usable. Corrections for NLTE effects should also be considered.
Two stars, among the most metal-poor, stand out for having a very high
[Na/Fe] ratio: GHS080 and GHS144. While GHS080 has essentially a solar
[Ba/Fe] ratio, GHS144 is strongly enhanced in barium, [Ba/Fe]=+1.14. For the latter star
one may suspect a pollution from an AGB companion. The error on the Gaia radial velocity
is 2.1\,\kms\ a little bit large for a 13th magnitude star, leaving margin
for possible radial velocity variations. Higher resolution observations and radial
velocity momitoring for this star are encouraged.  

\begin{figure}
\centering
\resizebox{8.5cm}{!}{
\includegraphics[clip=true]{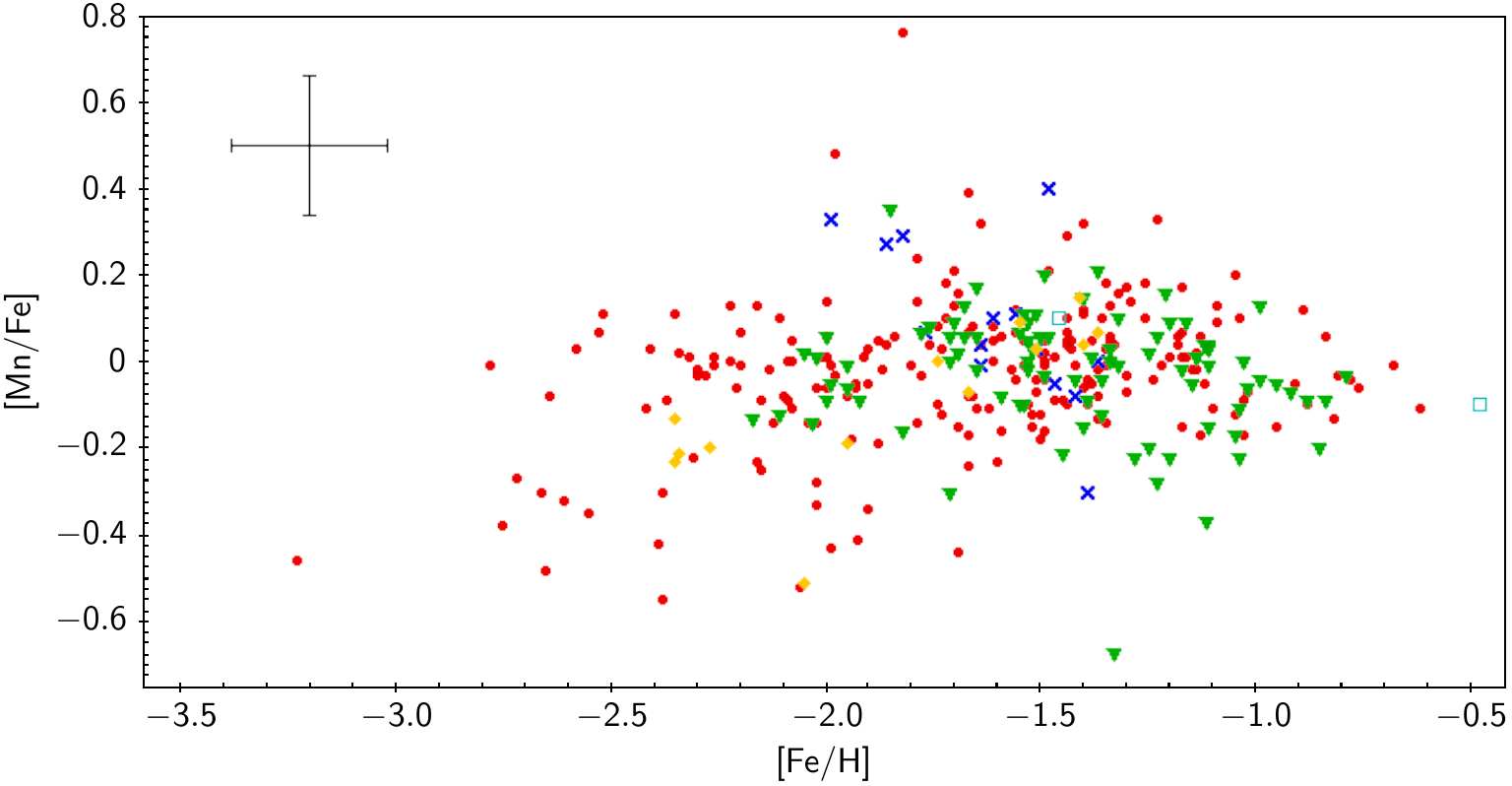}}
\caption{[Mn/Fe] versus [Fe/H] for the observed sample.
}
\label{fig:mnfe}
\end{figure}

The [Mn/Fe] ratios in our sample of stars are on average solar as shown
in Fig.\,\ref{fig:mnfe}, however there is a clear
tendency to subsolar values at metallicties below --2.0.
This is certainly due to NLTE effects \citep{2008A&A...492..823B}, that for stars
of these parameters should be in the range +0.2 to +0.4\,dex and would bring 
[Mn/Fe] to a very flat behaviour. This can be theoretically expected since both
Mn and Fe are formed in nuclear statistical equilibrium.

\subsubsection{Ba abundances}

\begin{figure}
\centering
\resizebox{8.5cm}{!}{
\includegraphics[clip=true]{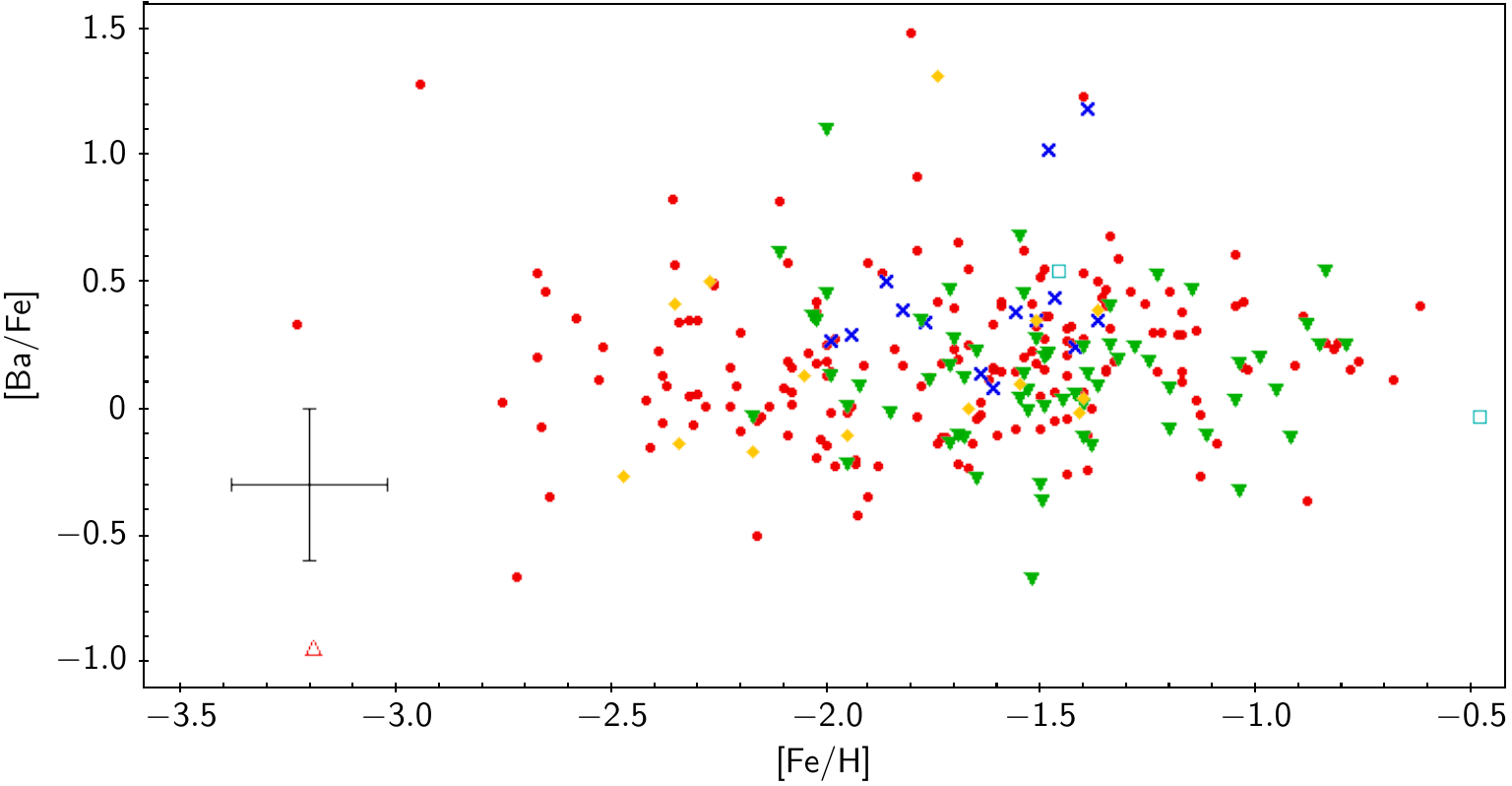}}
\caption{[Ba/Fe] versus [Fe/H] for the observed sample.
NLTE corrections have been applied to the Ba abundances.
Symbols in Fig.\,\ref{fig:mgfe}, the red open triangle is an upper limit
for star GHS118, belonging to the halo. 
}
\label{fig:bafe}
\end{figure}

Given the limitations of \mygi\ for ionised species, characterised by large over or under abundances, with respect to iron, as explained in appendix \ref{mygi_ion}, we determined the Ba abundances with line profile fitting. We used the \ion{Ba}{ii} 455.4\,nm resonance line, taking into account the full hyperfine and isotopic structure, for an assumed solar isotopic ratio. For each star we computed an ATLAS 9 model atmosphere using the new set of opacity distribution functions of Mucciarelli \& Bonifacio (in preparation). The adopted microturbulence in the ODFs was 1 \kms\ and in the model computation we assumed a mixing length parameter 1.25.
To compute the line profiles we used the {\tt turbospectrum} \citep{1998A&A...330.1109A,2012ascl.soft05004P}\ spectrum synthesis code.
We interpolated the NLTE corrections of \cite{korotin15} and applied them to the derived LTE Ba abundances.
 The [Ba/Fe] ratios are displayed in Fig.\,\ref{fig:bafe}

\subsection{UVES spectra}
With the stellar parameters in Paper\,I and the new spectra obtained with UVES, we derived again the metallicities for the 
two blue straggler candidates
GHS69 and GHS70 using \mygi\ \citep{mygi14}, and we obtained [Fe/H]=--2.23$\pm$0.26 for GHS69 and [Fe/H]=--1.89$\pm$0.15 for GHS70. 
The new metallicities are slightly lower than the one found in Paper\,I 
([Fe/H]=--1.94 for GHS69, and [Fe/H]=--1.59 for GHS70), but compatible within errors. 
The abundances for \ion{Fe}{i}, \ion{Fe}{ii}, \ion{Mg}{i},
\ion{Ca}{i}, \ion{Mn}{i}, \ion{Co}{i} and \ion{Ni}{i} are available
as an on-line table at CDS. They are unremarkable except perhaps, that
the $\alpha$ elements are slightly low, for this metallicity.
[Mg/Fe] and [Ca/Fe] are around +0.3\,dex for both stars. 
The S/N ratios obtained allowed us to derive only upper limits for the Li abundance for the two stars. 
To obtain upper limits on Li abundance, we estimated the minimum measurable equivalent width (EW) 
at $1 \sigma$ detection for the \ion{Li}{i} doublet using the Cayrel 
formula\footnote{Cayrel's formula estimates the error on the EW of an isolated line ($\mathrm{
\delta W}$), and it is written in the form $\mathrm{\delta W \,\simeq\, 1.5\times (FWHM*\delta x)^{1/2}\,/\,(S/N)}$,  where FWHM is the full-width half maximum of the line, $\mathrm{\delta x}$ is the pixel size of the spectrum and S/N is the signal-to-noise in the continuum neighbouring the line.} \citep{Cayrel1988}. 
We then computed the curve of growth for the Li doublet by measuring the EW of the Li doublet in synthetic spectra for different values of A(Li).
The synthetic spectra were computed with the spectral-synthesis code \texttt{SYNTHE} \citep[see][]{2005MSAIS...8...14K,2004MSAIS...5...93S}, starting from \texttt{ATLAS\,9} 1D plane-parallel model atmosphere computed using an ODF by \citet{CastelliKurucz}, and atomic data including the hyperfine structure of the Li doublet from Kurucz's database\footnote{\url{http://kurucz.harvard.edu/atoms/0300/}}.

\subsubsection{GHS69}

   \begin{figure}
   \centering
   \includegraphics[width=0.75\hsize,clip=true,angle=-90]{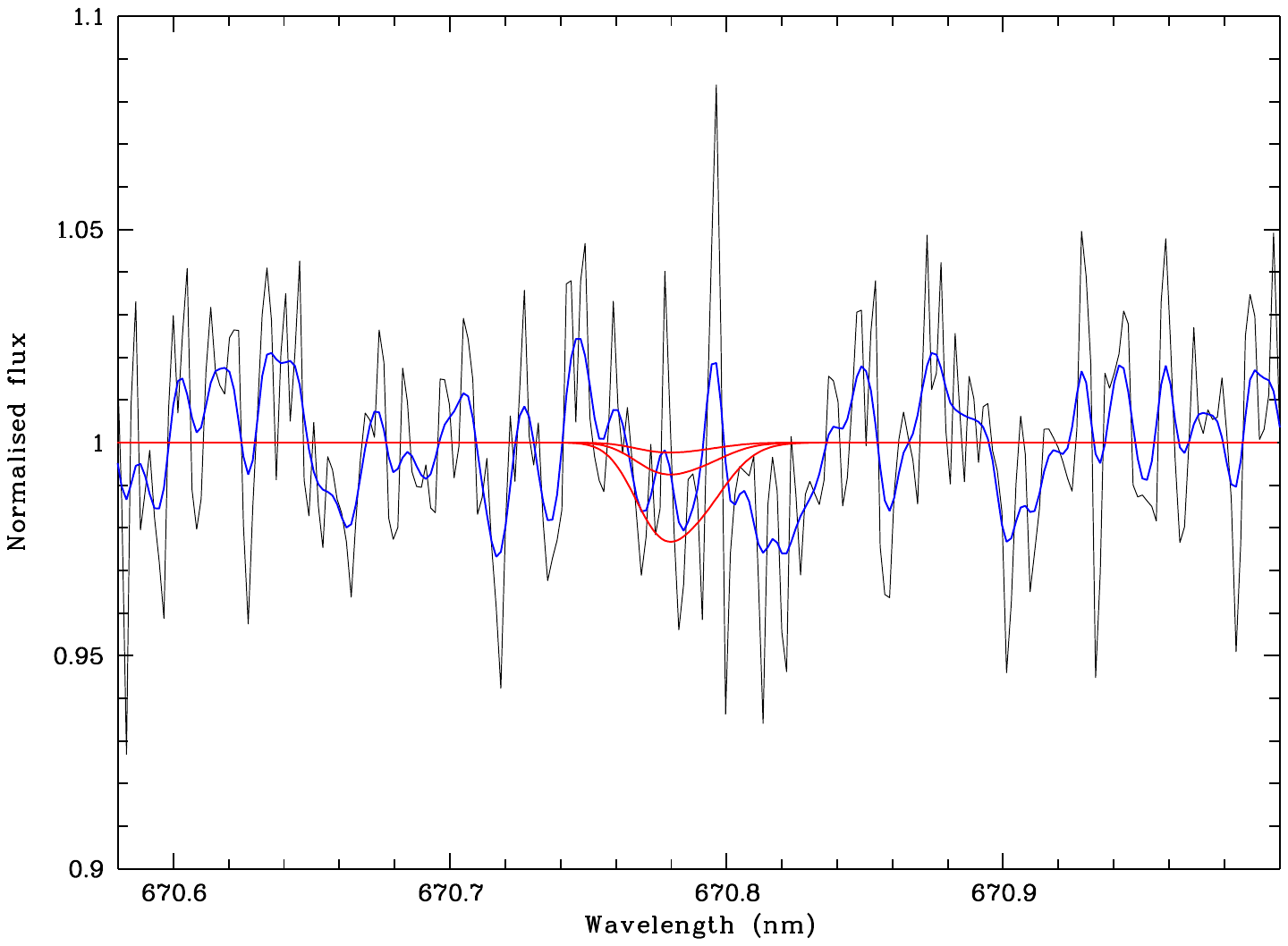}
      \caption{Spectra of GHS69 in the region of the \ion{Li}{i} 670.7\,nm doublet. Red lines represent synthetic spectra with Li abundances of A(Li)=1.0, 1.5, 2.0 dex. Blue line represents GHS69 spectrum with a broadening of 3 \kms.}
         \label{highspeed:li_6707_GHS69}
   \end{figure}

For  star GHS69,  a model atmosphere with \teff=6700\,K, \logg=3.8, $\xi$=1\,\kms, and 
[Fe/H]=--2.5, a S/N\,$\sim$\,42 would imply 
a minimum measurable EW for the Li doublet of 0.8 pm, which corresponds to a Li abundance of 2.0 dex, thus the $1 \sigma$ upper limit is A(Li)$< 2.0$. 
The spectrum of GHS69 (black) is compared to three synthetic spectra with A(Li)=1.0, 1.5, 2.0 dex in Fig.\,\ref{highspeed:li_6707_GHS69}. In blue, the GHS69 spectrum degraded with a broadening of 3 \kms\ is plotted for a better comparison.

\subsubsection{GHS70}

   \begin{figure}
   \centering
   \includegraphics[width=0.75\hsize,clip=true,angle=-90]{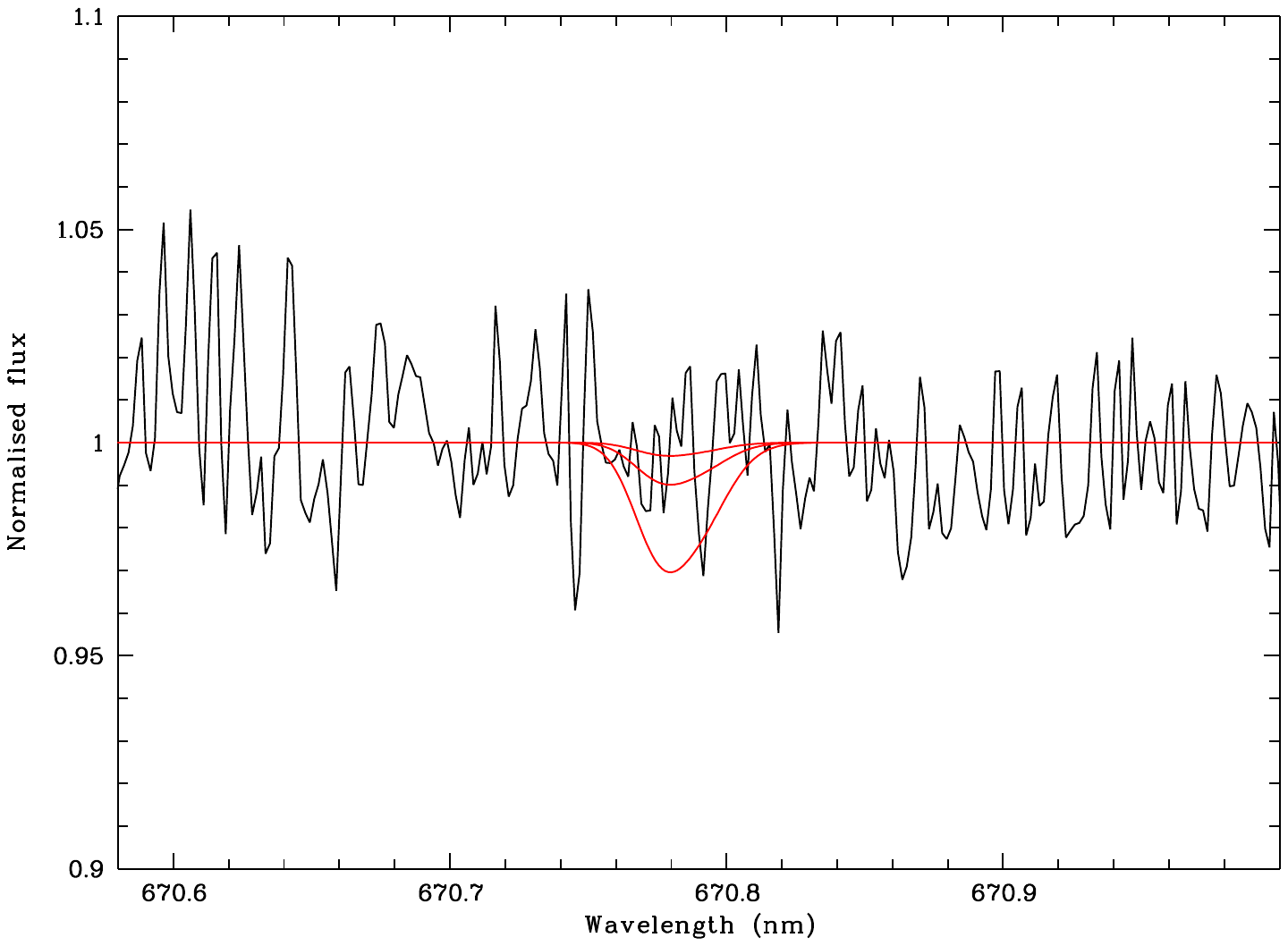}
      \caption{Spectra of GHS70 in the region of the \ion{Li}{i} 670.7\,nm doublet. Red lines represent synthetic spectra with Li abundances of A(Li)=1.0, 1.5, 2.0 dex. }
         \label{highspeed:li_6707_GHS70}
   \end{figure}

For the star GHS70, a model atmosphere with 
\teff=6500\,K, \logg=3.7, $\xi$=1 \kms, and [Fe/H]=--2.0, a 
S/N\,$\sim$\,55 would imply a minimum measurable EW for the Li doublet of 0.6 pm, which corresponds to a Li abundance of 1.8 dex, thus the 1 $\sigma$
upper limit is A(Li) $< 1.8$.
The spectrum of GHS70 (black) is compared to three
synthetic spectra with A(Li)=1.0, 1.5, 2.0 dex in Fig.\,\ref{highspeed:li_6707_GHS70}.

\section{Discussion}

\begin{figure}
\centering
\includegraphics[width=\hsize,clip=true]{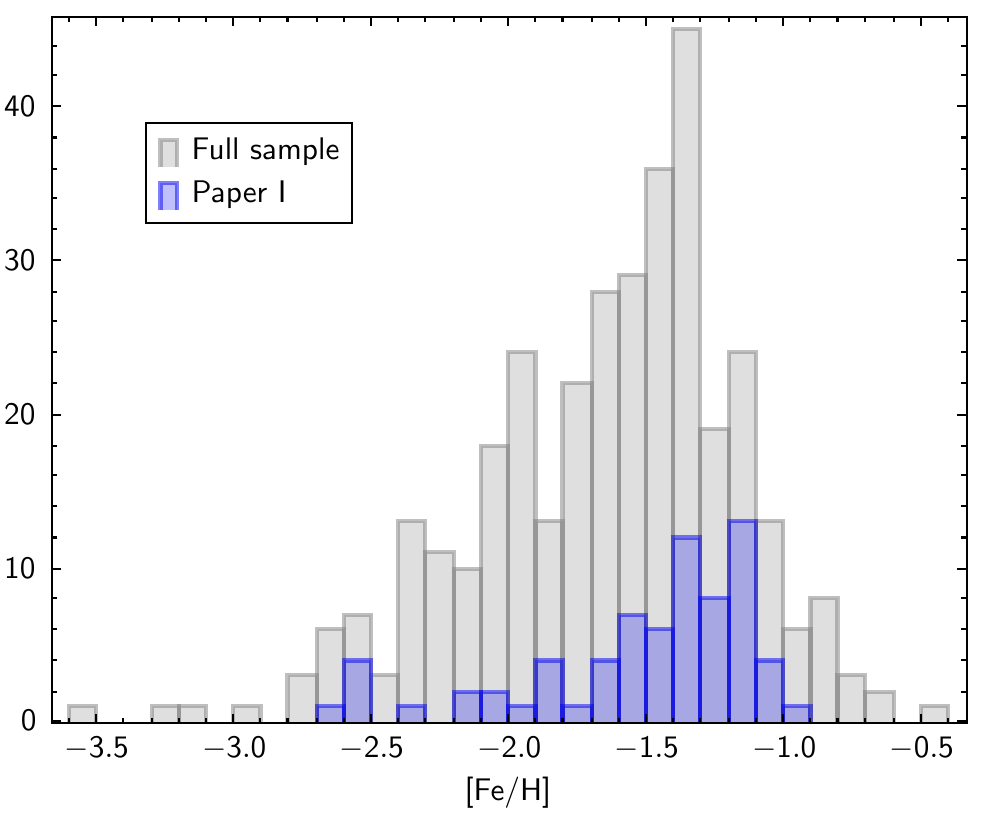}
\caption{Metallicity distribution of the observed sample, including the sample
of Paper\,I.}
\label{fig:histomet}
\end{figure}

\begin{figure*}
\centering
\resizebox{14cm}{!}{\includegraphics[clip=true]{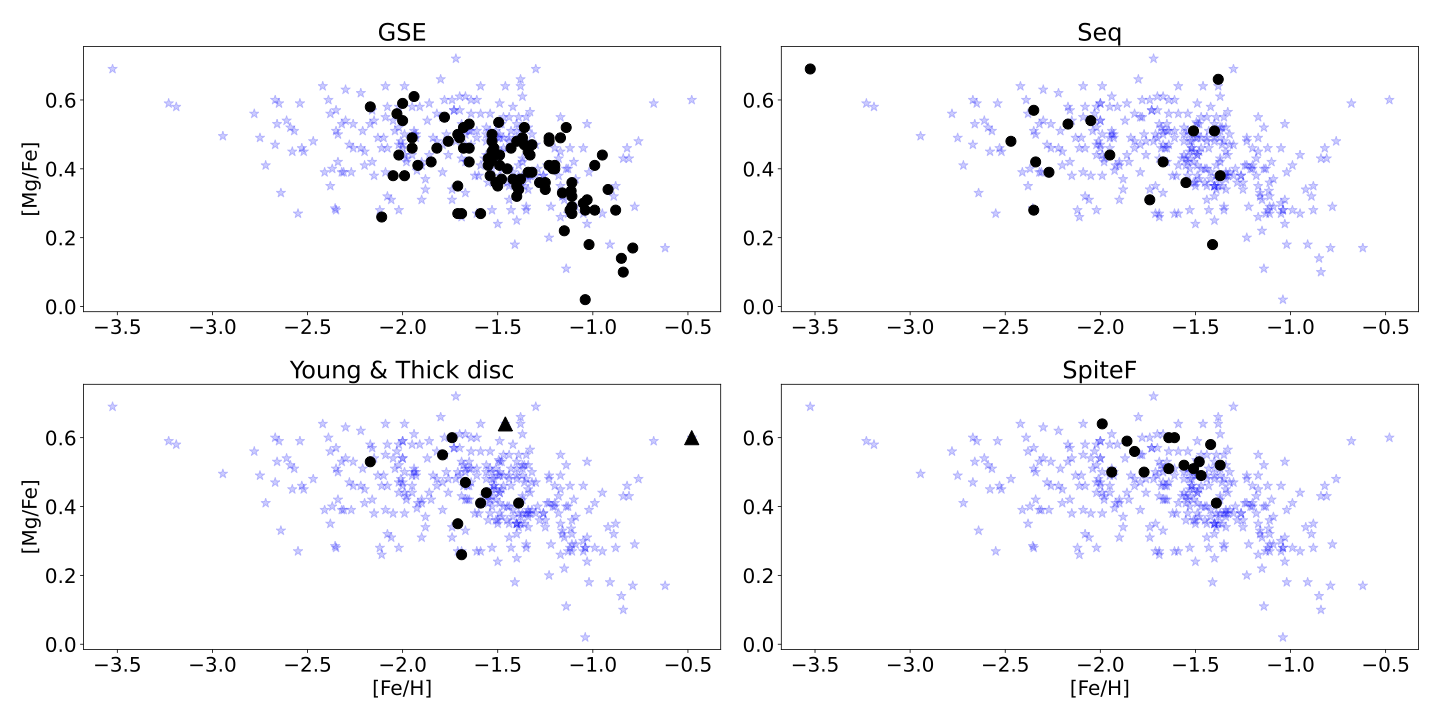}}
\caption{[Mg/Fe] { vs} [Fe/H] for the programme stars. In the various planes specific groups (filled black circles and black triangles) are plotted on top of all stars (cyan stars). Top panels: GSE (left) and Seq (right) candidates. 
Bottom panels: young (filled circles) and thick disc stars 
(filled triangles, left panel) and bulge (right).} \label{char3}
\end{figure*}
%
The basic result of Paper\,I was that the stars that are extreme in kinematics
are not necessarily extreme in chemistry. The whole sample (including the 72 stars of
Paper\,I) shows a marked peak around --1.4 and a decrease towards
lower metallicities, as shown in Fig.\,\ref{fig:histomet}.
Because the present sample is larger than that in Paper\,I, 
we have now 81 stars with [Fe/H]$\le -2.0$, 22 stars with [Fe/H]$<-2.5$
and four stars with [Fe/H]$<-3.0$. 

Our dataset allows chemical and kinematical data
to be combined to gain further insight into the origins of the different sets of stars. 
 In Fig.\,\ref{char3} we show our targets in the [Mg/Fe]  vs  [Fe/H] plane, highlighting GSE, Seq, young, thick disc and bulge stars, on top of the ensemble of the targets. The apparent bifurcation in the plane visible at about [Fe/H]$\simeq$-1.5\,dex is mostly due to stars belonging to GSE (top-left) and the bulge (bottom-right). 
Stars GHS036 \citep[see][]{ghs104}  and GHS110 (variable star, young and bulge) 
for which a full chemical abundance analysis was not performed, are not shown in the figure. 

Our sample has allowed us to highlight four sets of stars
that we consider particularly interesting:
{\sl (i)}  one unbound star; {\sl (ii)} the young stars; {\sl (iii)} a set of metal-poor bulge stars; and {\sl (iv)} candidate stars of the Aurora population, supposed to be formed
before the creation of the Galactic disc.
We shall discuss each of these in turn.

\subsection{On the origin GHS143 \label{origin_ghs143}}

\citet{koposov20} discovered a main-sequence 
A-type star (S5-HVS) with a total speed of about 1700 \kms\ that 
they interpret as ejected from the Galactic Centre. The mass of S5-HVS is estimated
to be 2.35 M\sun , that of GHS143 is considerably larger  (3.1 to 3.8 M\sun), consistent 
with the fact that it is evolved.
In order to gain insights into the origin of GHS143, we backwards integrated its orbit. At odds with S5-HVS1, GHS143 is currently approaching 
the Galactic plane and, thus, was not expelled by the Galactic Centre, as a consequence of an encounter with the central super massive black hole \citep[the Hills mechanism,][]{hills88}. We then searched for possible past association with Milky Way (MW) satellites and Galactic Globular clusters (GCs). We therefore back-integrated their orbits as well, considering the phase space positions from \citet[][]{pace22,baumgardt19,vasiliev21,baumgardt21} as already available in {\tt galpy}. 

The closest encounter occurred with the GC 
NGC\,6584, $\sim$1\, Ma ago at a distance of 2.5\,kpc.
The colour-magnitude diagram of NGC\,6584 \citep[][and \url{https://people.smp.uq.edu.au/HolgerBaumgardt/globular/}]{vasiliev21} shows an extended sequence of 
blue stragglers,
some of which could be more massive than twice the mass of the Turn-Off stars.
The metallicity of the star is --1.74 while that
of NGC\,6584 is -1.50 \citep{2012AJ....144..126D} thus the two are compatible within errors.
The closest encounter with a dwarf galaxy occurred with the Sagittarius dwarf spheroidal (Sgr dSph) $\sim$8-10\, Ma ago, at a distance of 9.8\,kpc. 
 If we consider 1000 random realisations of the initial parameters, the closest encounters are obtained with the same objects at median distances of 3.0\,kpc (standard deviation, std, 1.8\,kpc) and 9.8\,kpc (std=1.9\,kpc) for NGC\,6584 and Sgr, respectively.  
The encounters are compatible with NGC\,6584 within 2 std and with Sgr within slightly over 2 std and 2 half light radii \citep[r$_h$=2.59\,kpc, ][]{mc12}.

Finally, we used the {\tt GALSTREAM} library \citep[][and references therein]{mateu23}, to search for association with any of the many known stellar streams identified in the last period in the galaxy. Again, we could not find any clear association, with the closest stream being that associated to the GC NGC\,6362 at a distance of 5.3\,kpc.
We notice that candidate hyper/high-velocity stars likely originating from the Large Magellanic Cloud and the Sgr dSph were reported \citep[][]{erkal19,huang21,li22, li22a}.
Even though the high mass and corresponding young age of this star may suggest its origin in the ejection of a runaway disc star,  
its low metallicity ([Fe/H]=-1.74\,dex) opens also the possibility that GHS143 may be a debris of a low-mass, dwarf galaxy \citep[see][and references therein]{boubert18}. Recently, \citet[][]{li22a} identified a number of late-type, metal-poor candidates hyper-velocity stars and proposed ejection from dwarf galaxies or globular clusters as their likely origin.

One can invoke a binary system in which the massive companion explodes as supernova resulting in the disruption
of the system and the secondary star leaving with a high kinetic energy \citep{perets12} or 
a type\,Ia supernova in which the donor star becomes unbound after the thermonuclear detonation of the white dwarf \citep{geier15}.
It is also interesting to consider the possibility of an asymmetric supernova explosion as an accelerating mechanism for
a companion star of a Supernova  \citep[hereafter SN][]{tauris15}.
However at the present time we see no way of discriminating among these possible mechanisms.

\subsection{Young stars}

The set of ten young stars, whose 
estimated ages range from  300\,Ma to 2.5\,Ga
and metallicities from --1.3 to --2.2, as  
summarised  in Table\,\ref{masses}, is
unexpected. Although in Paper\,I we had in fact suggested the presence of some young stars, 
here there are many more. 
It is interesting to note that also in the sample of nearby high-velocity stars selected from Gaia\,DR2
by \citet{hattori18} there are several stars compatible with a young age (see their Figure\,2).
One could wonder whether the apparently young age could be a problem of overestimated
reddening. 
From Fig.\,\ref{cmd} it is easy to estimate that the $E(G_{BP}-G_{RP})$
should be smaller by about 0.2\,mag, in order for the young stars to fall
on an old red-giant branch. This is impossible, since the highest colour excess
for these stars is
$E(G_{BP}-G_{RP})=0.122$ for GHS145, six out of the ten young stars have
a colour excess smaller than 0.08.  
Rare objects, like post-AGB stars can have colours in the range covered by our young stars, 
however, such stars are typically 0.5 to 2 magnitudes brighter than
our brightest young star (GHS143), according to the catalogue
of  \citet{vickers}, that includes most of the known post-AGB stars.
The presence of debris discs or dust-shells around the stars
would make them redder, not bluer, so also these objects can be discarded.
We believe that the only two classes of stars that can occupy this region of the colour-magnitude diagram
are young stars of mass in excess of 1.5 solar masses or evolved blue stragglers.

\subsubsection{Young or rejuvenated}

The issue is to decide if these are truly young stars
or they are `rejuvenated' old stars, that is, evolved blue stragglers.  
A look at the masses in Table\,I suggests that only a few could be evolved blue stragglers,
in fact most of them are likely more massive than 2 M\,\sun.
The most credited channels to rejuvenate stars and create blue stragglers
involve binary stars. Either mass transfer in a binary system
\citep{1964MNRAS.128..147M} or merging of two stars in a binary \citep[e.g.][]{1976ApJ...209..734Z}.
It is possible to create a blue straggler also by direct collision of two stars, previously unbound
\citep[see e.g.][]{1976ApL....17...87H}. Among the many papers on the topic we refer the reader
to \citet{1993ASPC...53....3L}, \citet{2000AJ....120.1014P} and \citet{2005AJ....129..466C}, that we found very illuminating.
Whichever of the above mechanisms is invoked  it is not possible
to create a star whose mass is larger than the sum of the masses of the two stars involved.

If we consider the star formation history of \citet{2016A&A...589A..66H}, it is clear that outside the Galactic disc,
we do not expect stars younger than about  8\,Ga. Now in an isochrone of 8  Ga and metallicity --1.5
the most massive stars are about 0.9M\,\sun.  
This implies that stars of mass larger than 1.8\,M\sun\ cannot be formed by merging or colliding, or through mass transfer,
of two stars of  8\,Ga or older. The upper limit is even lower if you consider that some of
the mass must be lost in the process of blue straggler formation. 
Blue stragglers that have masses larger than twice the mass of Turn-off stars are
known both in open clusters \citep[e.g. M\,67][]{1992IAUS..151..473M} and 
globular clusters \citep[e.g. NGC\,6397][]{2002ASPC..263..157S}. In these cases
the evolution of a triple system through a common envelope phase
of the inner couple is invoked \citep[see][]{1980IAUS...88..145M}.
However, these massive blue stragglers are rare. 
\citet{2014ApJ...783...34F} determined 
pulsational  masses for blue stragglers in the
Globular Cluster NGC\,6541 (metallicity --1.76, age 13.25  Ga thus compatible with our halo
population) and  found  
masses in the range 1.0 -1.1 $ M_\odot$. 
\citet{raso19} found in 47\,Tuc five stars with mass estimate, derived
from the spectral energy distribution,  larger than twice
the TO mass. However the errors on these estimates
are large enough that they are all consistent, within  $1\,\sigma$,
with a mass smaller than twice the TO mass.

\subsubsection{Mergers of three stars as a possible explaination of  the observations}

In order to claim that all the young stars are evolved blue stragglers  it is necessary
to postulate that most of them descend from the common envelope evolution
of a triple system and in the case of GHS143 it is not sufficient but we would need a quadruple system. 
The fraction of binary and multiple systems is discussed in
\citet{arenou11}. For solar type stars the generally assumed fraction of triple
systems is 8.4\% , based on the investigation of \citet{1991A&A...248..485D}.
This is an upper limit to the number of triple systems that may form
a massive blue straggler, since the architecture of the system must be such
that the inner couple undergoes the common envelope phase
and the semi-major axis of the orbit of the third
star is small enough that it will experience friction and ultimately
spiral into this common envelope.
Recently \citet{moe19} have argued that the fraction of close binary stars, defined as those
having a semi-major axis $ a \la 10$\,au, increases with 
decreasing metallicity.
This result relies on the completness correction that the authors
apply to various samples, the trend in the uncorrected data is not detectable (see their
figure 3).  These authors find that for solar type stars with
[Fe/H] $\le -1.0$, the fraction of close binaries is about 50\%
and the fraction of triples plus quadruples is of the order of 35\%.
This number is much larger than the 8.4\% cited above. For the sake of discussion
let us consider 35\% the fraction of multiple systems potentially
suitable to be the progenitors
of a massive blue stragglers.
In our view the crucial question to ask is what is the fraction of these systems
that will result in the fusion of at least three stars.
Some insight may come from the study of \citet{toonen22} who perform extensive
simulations on the fate of destabilised triple systems. One striking
result is that although collisions in destabilised triple systems are fairly common,
for only at most 2.4\% of them the collision includes the third star.
Most collisions occur only in the inner couple of the hierarchical system.
If we add the estimate of the fraction of destabilised triple systems 
that is at most 4\%, we obtain a fraction of massive blue straggles
$F_{MBS} = 0.35\times 0.04 \times 0.024 = 0.000336 \approx 0.3\%$.
In our view even this large fraction of triples would still imply a tiny
fraction of triple fusions.
We therefore do not expect more than a few per cent of the stars
to be blue stragglers descendants from triple systems.

\subsubsection{Where do they come from?}

We back integrated the orbits of all the young stars for  1 Ga and did the same for all the known
Globular Clusters, dwarf galaxies and stellar streams, to look for close encounters.
If a close encounter with a GC can be identified, this would support a BSS nature
for the given star.
GHS143 had a wide encounter with NGC\,6584 1\, Ma ago and is further discussed
in Sect.\,\ref{origin_ghs143}. 
Of the remaining stars seven appear to have had an encounter, within  several kpc
about 550  Ma ago with Tuc III. These distances may appear large, however our estimated
errors on the minimum distance are of the order of 1.5 to 2 kpc, therefore most of these encounters
are significant at less than $3 \sigma$.
Although suggestive we discard a possible origin in Tuc III for any of these stars for two reasons:
{\em i)} the metallicity of Tuc III is --2.4 with a small metallicity dispersion, less than 0.1\,dex 
\citep{2017ApJ...838...11S}, while our stars are about 1\,dex more metal-rich, on average; the only star
that has a metallicity compatible with Tuc III is GHS120 ([Fe/H]=--2.17);
{\em ii)} the colour magnitude diagram of Tuc III \citep[see Figure 1 of][]{2017ApJ...838...11S},
does not show any radial velocity members that can be interpreted as a young population,
blue stragglers or evolved blue stragglers.
Two more stars, GHS209 and GHS212, had the closest approach with a dwarf galaxy with Bootes III, the minimum
distances are 5.3 kpc and 13.2 kpc. Although the metallicity of both stars is about --1.7, thus compatible
with the metallicity of Bootes III \citep[--2.1][]{2009ApJ...702L...9C}, within errors, the large
minimum distance makes this origin not very probable, although it cannot be ruled out.
There are very few confirmed members of Bootes III, thus it is unclear if the colour-magnitude
diagram of this galaxy supports the existence of a young population, blue stragglers or evolved
blue stragglers.

For what concerns close encounters with Globular Clusters and known stellar streams, the situation is complex. 
For each star we have a large number (up to 92) of close encounters (at less than 1 kpc) with GCs and over ten with streams.
For each star there are several GCs that have compatible metallicity and possess a sizeable 
blue straggler population, thus making the association plausible. 
For streams, we often do not have the information on the metallicity, thus it is difficult
to assess the likelihood of the association.
This state of affairs does not allow to draw clear conclusions on the possible association of any
of these stars with either GCs or known stellar streams.
We stress that in any case the orbits of our young stars do not coincide with that of any
dwarf galaxy, GC or stellar stream. Thus even if an association existed we would still
have to think of a mechanism that `kicks' the star out of its Galactic orbit placing
it in the high speed state in which we observe it. 

In a series of  papers \citet{2021ApJ...922...93H,hammer23} argued that the orbital energy of most Milky Way dwarf
galaxies is too high for them to be long lived satellites and suggest a recent first infall of $\la 2 $\,Ga for most of them. 
If the galaxies were gas rich at the time of the infall, the interaction
with the hot gas of the Milky Way halo would strip them of the gas and, likely,
trigger a starburst. Carina is known to posses a young population
\citep{monelli03,weisz14}, as well as Fornax \citep{deboer12}.
It  seems that a recent starburst is possible among dwarf galaxies,
even though their dominant population is old.
We add that \citet{rvs2} in a sample of high radial velocity stars
have found two stars apparently younger
than 1\,Ga and masses larger than 1.8\, solar masses.
These objects appear to be similar to the young stars found in this paper.
In spite of the fact 
that when we backwards integrate the orbits of our young stars,
we find no close encounter with any known dwarf galaxy or stellar stream
we still think it plausible that they were born in a dwarf galaxy
that had a sturburst during its infall in the Milky Way.
The fact that its remnant/stream has not yet been identified suggests
that these galaxies were very small  in mass. 
In fact the fewer stars in the galaxy, the fewer stars in the stream,
thus the stream is more difficult to detect observationally.
In fact, in our view, this is the only way to explain  young metal-poor
stars in the Galactic halo, since the halo does not contain metal-poor gas
of high enough density to support a recent star formation event.

\subsection{
Distinguishing an evolved blue straggler from a young star}

Is it possible to distinguish between an evolved blue straggler and a genuine young star
of the same mass? We suggest that it may be possible
by looking at the stellar rotation.
The masses of Blue Stragglers cover the range from late F type
to early B type \citep[see e.g.][for a mapping of masses to spectral types]{1981Ap&SS..80..353S}.
The best comparison to blue stragglers are the stars studied in \citet{2012A&A...537A.120Z}.
In this exhaustive study one finds clearly that for masses up to
$\le 2.5\,M_\odot$ there is a lack of slow rotators (defined
as $v\sin i \le 100\,{\rm kms^{-1}}$), and the distribution has a wide and
flat peak between 110 and 220 $\rm kms^{-1}$ of $v\sin i$.
At larger masses the distribution is bimodal, and a peak with slow rotators
appears, yet it  comprises only 20\% of the stars.
When these stars evolve to the RGB or even to the red clump, 
their rotation slows down, but one can still find rotational 
velocities in excess of 20 $\rm kms^{-1}$ \citep{2021A&A...656A.155L}.

The rotation of blue stragglers in Globular Clusters
is extensively discussed in \citet{Mucciarelli14},
that also contains an exhaustive set of references.
To summarise the discussion: most Globular Cluster blue stragglers
have projected rotational velocities in the range
30--40 $\rm kms^{-1}$, the exceptions are M4 and 
$\omega$ Cen, that have a small, but significant
population of stars that have projected rotational velocities
above 40\,$\rm kms^{-1}$.
In fact, \citet{Ferraro23} have shown that blue stragglers
with projected rotational velocities above 40\,\kms\
are only found in loose globular cluster, suggesting that
these are blue stragglers formed through mass accretion
in binary system. This because mass accretion transforms
part of the orbital angular momentum into rotational angular
momentum of the accreting object.
On the other hand \citet{Ferraro23} also argue that, in spite
of the fact that breaking mechanisms are not fully understood,
these objects slow down in less than one  Ga, based on the
observed correlation between rotational velocities and
ages \citep{2018ApJ...869L..29L} among blue stragglers in Open
Clusters and in the field.
They also argue that collisional blue stragglers probably slow down
even faster, since no fast rotating blue stragglers are observed
in dense environments where collisions are expected to dominate.
It is reasonable to expect that a sample
of evolved blue stragglers will have, on average, lower rotational
velocities than `normal' stars of the same mass. This because 
at the
beginning of the spin-down, caused by the envelope expansion, as
the star leaves the Main Sequence, the blue stragglers should arrive
with lower projected rotational velocities, than young stars of
the same mass.
The paucity of fast rotators seems  
to hold also for field blue stragglers, in 
\citet{2000AJ....120.1014P} the bulk of the stars 
has $v\sin i \le 40\rm kms^{-1}$ and the highest measured
projected rotational velocity is 160  $\rm kms^{-1}$.
We are aware that we are here ignoring the difference
in metallicity and age between the Pop I stars
in the \cite{2012A&A...537A.120Z} sample and the Pop II
stars in GCs, however only among Pop I stars we can find
a sample of stars of masses comparable to those of blue straggler stars.
We are making the assumption that for the rotational history
of a star the mass is the most important quantity. 

From what above said,  we expect that
evolved blue stragglers should, on average, have lower
rotational velocities than young stars that occupy
the same place in the colour-magnitude diagram.
This is also supported by the observation
of an evolved blue straggler  in 47 Tuc \citep{Ferraro16}.
It is important to understand that this is a statistical
criterion, it cannot be applied to a single star, but only to a sample of stars.
We thus encourage observations at high spectral resolution
of these young stars in order to probe the distribution of their rotational velocities.

It is interesting in this context to look at the two stars observed with UVES.
In Paper\,I we did not estimate ages and masses for these two stars.
Since they are both subgiants there is no ambiguity as to their mass and age estimate.
GHS69 has a mass of 0.78\,$M\odot$ and an age of 8.6\,Ga;
GHS70 has a mass of 0.81\,$M\odot$ and an age of 8.0\,Ga.
Thus they appear young, but not nearly as young as the more evolved
stars discussed above.
Moreover their masses are fully compatible with a blue straggler status,
as hinted by the upper limits on lithium.

\subsection{SpiteF: Metal-poor bulge stars, relics of an accretion event}
\label{sec:spitef}

The bulge stars we detected reach down to the very metal-poor regime ($\rm [Fe/H]\simeq-2$\,dex).
They thus belong 
to the metal-poor tail of the bulge metallicity distribution function \citep[][]{argos13,gonzalez15},
a rare bulge population. Surveys dedicated to the search for 
the most metal-poor stars in the bulge were able 
to detect  stars
down to the extremely metal-poor regime ([Fe/H]$<$-3\,dex, \citealt{embla16,combs22,pigs23}).
The [Mg/Fe] of our bulge stars is remarkably uniform and enhanced by about 0.5\,dex and with a very small
dispersion (0.06\,dex).  The sample of \citet{embla16}
has 29\% of the stars (four out of 14) that overlap with the metallicity range of the bulge stars
in our sample, and their [Mg/Fe] has a significantly larger dispersion than ours (0.1\,dex),
which can hardly be attributed to their errors, since they use spectra of
much higher resolution than ours.
\citet{combs22},
also cover the metallicity range of our sample and show a larger dispersion in [Mg/Fe].
In fact \citet{combs22} state: ``the inner bulge and halo distributions are 
not significantly different in [Ca/Fe] or [Mg/Fe], as they both have large scatter.''.
To be noted that the `inner bulge' of \citet{combs22} is defined as stars
with $r_{ap} < 3.5$ kpc, thus consistent with our definition of stars confined to the bulge.
An alternative definition of the SpiteF structure is discussed in appendix\,\ref{sel_spitef}.
It is true that the metal-poor bulge sample of \citet{lucey19} shows a negligible scatter in
in [Mg/Fe] and other $\alpha$ elements. One should however keep in mind that the sample
of \citet{lucey19} has been selected using a not clearly defined combination of medium
resolution ($R\approx 11\,000$) spectroscopy, centred on the
infra-red \ion{Ca}{ii} triplet  and SkyMapper photometry \citep{casagrande19},
that includes an intermediate-band filtered centred on the UV \ion{Ca}{ii} H\&K lines.
It is thus possible that their selection function implied a low scatter in $\alpha$ elements.
It is clear that all the above-discussed samples are heavily biased.
\citet{embla16,lucey19} and \citet{combs22} are biased on metallicities and, possibly,
on $\alpha$-to-iron ratios, our sample is biased on transverse velocities.
What is of essence here, however, is that our sample that is  unbiased with respect
to chemical composition turns out to be metal-poor and with a small scatter in [Mg/Fe].
The converse is not true, the chemically biased samples do not all show high transverse
velocties.
The presence of young stars in the bulge has been noted in the past
\citep{2013A&A...549A.147B,ness14,2021NatAs...5..311F}.
However, at odds with the two young stars in our sample, the  young stars detected so far in the bulge are metal-rich ([Fe/H]$>-0.5$\,dex).

The ratio of $\alpha$ elements to iron is often used to distinguish between
stars that have been formed in external dwarf galaxies and then accreted
by the Milky Way and stars that were formed in the Milky Way. This is
discussed in Sect.\,\ref{halo_aurora} with respect to  the stars
classified as halo. It should however be kept in mind
that while many dwarf galaxies show a sequence of low [Mg/Fe]
for their higher metallicity stars, with respect to Milky Way halo stars, at lower
metallicities the sequences merge to a plateau and 
the [Mg/Fe] criterion becomes not informative. 
This is due to the fact that at very low metallicities, in any galaxy, 
Mg and Fe are only produced by massive stars that end their lives as 
core-collapse supernovae. Only when Type Ia supernovae begin to explode, 
producing large amounts of Fe, but little or no Mg and other $\alpha$ 
elements, the [Mg/Fe] starts to decrease.
Since Type Ia supernovae are the result of the evolution of binary 
systems formed by less massive stars which have a longer lifetime, they  
begin to explode later when the metallicity of the galaxy has already 
increased, due to the enrichment from core-collapse supernovae alone.
This is obvious for Ultra Faint dwarf galaxies \citep[see figure 1 of][]{francois2016},
but  also for classical dwarf galaxies \citep[see figure 11 of][]{tolstoy09}.
As discussed in  Sect.\,\ref{halo_aurora} the use of [Mg/Fe] criterion at low metallicities
can lead to serious contamination.

It is noticeable in Fig.\,\ref{char2} 
that our bulge stars lie on a well defined sequence with a positive 
slope in the Z  vs  R plane (top-left panel), 
at R$<$3\,kpc and $-2<$Z$<-0.5$\,kpc. This group of stars was 
noticed already during the target selection phase. 
A random sample of bulge stars should occupy a range
of Z values at any given R. 
This issue is further elaborated upon in Appendix\,\ref{e2}
where we compare our sample to that of \citet{rix22}.
Figures \,\ref{fig_lb_zr_spitef} and 
\ref{fig_lb_zr_spitef_hs} show in fact that selecting from the \citet{rix22}
sample either in metallicity or transverse speed, 
we end up with a wide range in Z for any given R.
Of the 16 bulge stars in our sample 14 are on retrograde
orbits. \citet{2022Univ....8..206K} wrote an extensive review of
bulge kinematics, based on RR Lyr stars. The latter are metal poor
and in fact of the same metallicity range of our sample.
The dominant structure of the bulge is a triaxial bar
\citep[e.g.][and references therein]{2013MNRAS.435.1874W},
that is characterised by cylindrical rotation 
\citep{2009ApJ...702L.153H,2017A&A...599A..12Z}.
A structure of this kind can form from a disc
that undergoes instability \citep{2011ApJ...734L..20M}.
The RR Lyr rotate slower than the bar, but there is no consensus
on whether the RR Lyr trace the bar or if they have a more spherical
distribution \citep[][and references therein]{2022Univ....8..206K}.
\citet{2021NatAs...5..311F}, from the properties of two bulge
Globular Clusters, Terzan\,5 and Liller\,1,  that host a young stellar
population, argued in favour of  a hierarchical assembly
of the bulge. The two clusters could be the relics of nuclear
star clusters in the merging galaxy.
This view is opposite to that of \citet{2019A&A...626A..16R}, who argued,
based on the [Mg/Fe] distribution of bulge stars, that accretion cannot have played
a major role in the formation of the bulge, considering the chemical
composition of present-day dwarf galaxies in the Milky Way vicinity.

The Pristine Inner Galaxy Survey \citep{2020MNRAS.491L..11A} was
specifically targeted at metal-poor stars in the bulge and inner disc,
and determined rotation curves for these populations, that are
very similar for different metallicity bins.
So far 
the extensive kinematic surveys of the bulge
have relied only radial velocities \citep[see e.g.][]{argos13,2020MNRAS.491L..11A}.
Under these conditions only the  rotational velocity
projected along the line of sight 
can be deduced.  In all these studies a dispersion around the mean
rotational velocity is observed, and this includes also a retrograde population.
In the present study, however, we make use of full three dimensional space velocities.
In particular if the bulge does contain a  pressure-supported component, as hinted at by
the kinematics of RR Lyr stars, in a radial velocity investigation, it would be seen 
as a dispersion around the mean  projected rotational component deduced from 
radial velocities.

We tentatively associate the metal-poor bulge population to an accretion
event that remained confined to the bulge. We call this population/accretion event
SpiteF, in memory of Fran{\c c}ois Spite, who began this investigation with us
and recently passed away.
There are three  reasons for which we believe this is an accretion event 
and not a selection effect.
\begin{enumerate}
\item We selected the stars only on r$_{ap}$, yet all our
bulge stars are metal-poor and, as above-mentioned, 
such stars are rare. 
\item
The stars show constant  [Mg/Fe] and [Ca/Fe]
ratios with a small dispersion that can be fully attributed to observational
errors. As above-mentioned other bulge samples in the same 
metallicity range show a larger dispersion in [Mg/Fe].
\item
The stars occupy a range in Z for
any given R and a narrow range in Y for any given X. 
\end{enumerate}
A random selection of bulge stars would not have these characteristics. 
The SpiteF contains two  young stars (GHS108  whose estimated mass is
1.9 M\sun\ and GHS110  whose estimated mass is in the range 2.0--2.8 M\sun,
see Table\,\ref{masses}), if we assume
the recent star formation to have occurred at the time of accretion, 
when the merging galaxy was still gas-rich, this implies that the accretion event is very recent,
at most 1\,Ga ago (see ages in Table\,\ref{masses}). 
Whether it is dynamically possible that a recent accretion
event penetrates to the bulge, rather than remaining in the halo is
a difficult question to answer and is further discussed in Appendix\,\ref{e3}.

We used the GUM
to get some insight in our selection biases. 
We are aware that GUM is based on an older version of the Besan{\c c}on model \citep{robin03} and
that our knowledge on the bulge structure has considerably evolved since. 
Yet we believe this comparison can give useful indications.

We queried the GUM, with the same query that we used 
for the Gaia source catalogue. The sample is made of 6685 stars.
In GUM stars are labelled as thin disc, thick disc, spheroid and bulge.
The selected sample (6685 stars) is dominated by the bulge (70\%) and the spheroid (halo in our definitions, 28\%) with negligible contributions of thick disc (15 stars)
and thin disc (90 stars). This can be expected, from our selection on transverse
velocity. We analysed this sample using galpy, similarly as we did with the observed stars. 

The GUM bulge stars are mostly confined to galactic latitudes and longitudes comprised between -15$<$b$<$+15, l$>340$ and l<30, with a density which is decreasing moving away from b=0. All but 34 of our stars, are located outside of this area, covering a region dominated by halo stars, consistently with our classification. The halo sample (including stars in the l,b region dominated by bulge stars) is composed by 1898 stars, 1439 (76\%) of which in retrograde motion (L$_Z<$0). The largest density of stars in both prograde and retrograde motion is also observed around the Galactic bulge, at -30$<$b$<$+30, l$>320$ and l<40, while almost only stars in retrograde motion are found outside of this area. Therefore, our sample is, actually, expected to be dominated by stars in retrograde motions.   

Thirty-four of our stars are located in the region -14$<$b$<$-8 and 350$<$l$<$1, which is dominated by the bulge in  the GUM model, with 271 (80\%) versus 67 (20\%) halo stars. halo stars are in almost equal number in pro/retrograde motion (33/34), while bulge stars are predominantly on retrograde orbits (157/114, 58\% versus 42\%).  However, bulge stars in the GUM models in pro/retrograde motions occupy different spatial regions. This is especially clear when looking at the stars in cartesian galactocentric coordinates X, Y, Z, our stars being concentrated in the region of the planes strongly dominated by retrograde stars. According to the GUM model, it is, therefore, again expected that also this particular group of stars is predominantly in retrograde orbits. Indeed, only 8 out the 34 stars  of our sample in this region are prograde (23\%), the others being in retrograde motion.

Fifteen, out of these 34 stars, belong to the sample that we labelled as `bulge' stars, the sixteenth being GHS108, which is somewhat separated by the rest of the sample at b$<$-18. These stars were identified as those having apocentric distances (r$_{ap}$) from the galactic centre lower than 3.5 kpc. This conditions is verified after integration under a potential which combines a rotating bar to the standard MWPotential2014 potential in galpy. However, they were identified also considering the MWPotential2014\footnote{Using the MWPotential2014, we identified 18 bulge stars, two of them not confirmed after integration in the potential including a rotating bar}. The remaining of the 34 stars in this spatial region, do not 
respect this condition, and are classified as halo. 

Stars confined to r$_{ap}<$3.5\,kpc represent a 4.6\% of our sample (16/348), and a 0.8\% and a 0.1\% of the entire GUM bulge and complete GUM samples (37/4682, 3/1898), respectively. If we limit the request to the l,b region previously indicated, none is left in the GUM sample. This supports the fact that the bulge population that we have discovered is real, and not a result of our selection function.

GUM does not have built in sub-structures like GSE or Sequoia, it
is however interesting to point out that 7\% of the GUM sample would 
be classified as Sequoia, to be compared with the 5\% of our sample.
This shows that one may conclude that a given dynamic group exists, even
when none exist in the underlying universe, and casts some doubt on the
reality of the stars we labelled as Sequoia as belonging to a real structure. 

We also find 1081 stars that we would classify as GSE (16\%) to be compared
with 26\% in the observed sample. This suggests that GSE is certainly real,
but our selection criterion implies some contamination. The contamination
of a pure dynamical selection for GSE has been discussed, for instance  by \citet{topos6}.

\subsection{Halo stars and Aurora candidates \label{halo_aurora}}

We classify all the stars that are not in GSE, 
Sequoia, bulge or Thick Disc as halo stars and these are the majority of the stars,
comprising 224 unique stars.
The discovery of GSE \citep{belokurov18,haywood18,helmi18}
allowed to realise that much of the samples of halo stars used in the
past were heavily contaminated by GSE stars. At the same time it was clear
that the collision of the GSE progenitor with the Milky Way
would also result in the scattering of stars formed in the disc into halo
orbits \citep[see e.g.][]{2009ApJ...702.1058Z,2017A&A...604A.106J}.
For this reason in the recent literature people call {\em in situ} stars,
both those formed during the gas collapse and those formed in the disc
and scattered into halo orbits by the collision with the GSE progenitor
or other major mergers.
This usage brings about confusion with the older literature, in the following
we shall, nevertheless, stick to it.
We can eliminate from our halo sample the stars that we classified
as GSE, keeping in mind that this dynamical selection still
has a 20\% contamination \citep[see e.g.][]{topos6} and conversely
we can expect the non-GSE sample to still contain GSE stars
\citep[see e.g.][]{amarante22,orkney23}.
Yet it is even more tricky to separate the sample
between {\em accreted} and {\em in situ}.
At  metallicities above --1.0 the Mg  abundance or generally
the abundance of $\alpha$ elements can be used 
to identify the accreted component 
\citep{1998A&A...338..161F,2003A&A...406..131G} at lower metallicities this is not
so simple.
\citet{2015MNRAS.453..758H} argued that the [Al/Fe] ratio is a much more powerful
indicator to select accreted stars. Theoretically this can be understood
as a combination of 
metallicity dependent  Al yields \citep{1995ApJS..101..181W,2013ARA&A..51..457N} and
the lowering of the ratio when the Fe production
of Type\,Ia SNe kicks in. For Mg instead only the latter mechanism is useful
to identify stars that were formed in an environment that was characterised by
low or bursting star formation.
We stress that these chemical criteria work only if the accreted component
comes from dwarf galaxies, if the accreted stars come from a more
massive galaxy, or anyway, by a galaxy characterised by continuous and
vigorously star formation, these will not work.

\citet{BV2022} used the [Al/Fe] ratio to create a high purity sample
of halo {\em in situ} stars. The study of this sample in the Lz-E plane
reveals that in the lowest metallicity bin ($\rm -1.7\le [Fe/H]< -1.3$)
there is no presence of a fast rotating disc, that appears instead in
the metallicity bin $\rm -1.3\le [Fe/H]< -0.9$.
They call this phenomenon the spin-up and identify it as the formation
of the disc, that, supported by galaxy formation simulations, they estimate
to have formed  10--11 Ga ago for a duration of 1--2  Ga.
Moreover they identify the lowest metallicity population, characterised
by high energy and a wide range in angular momentum, as that
formed in the turbulent and chaotic phase of the collapse, literally the dawn
of Galactic star formation.

\begin{figure}
\resizebox{7.7cm}{!}{\includegraphics[clip=true]{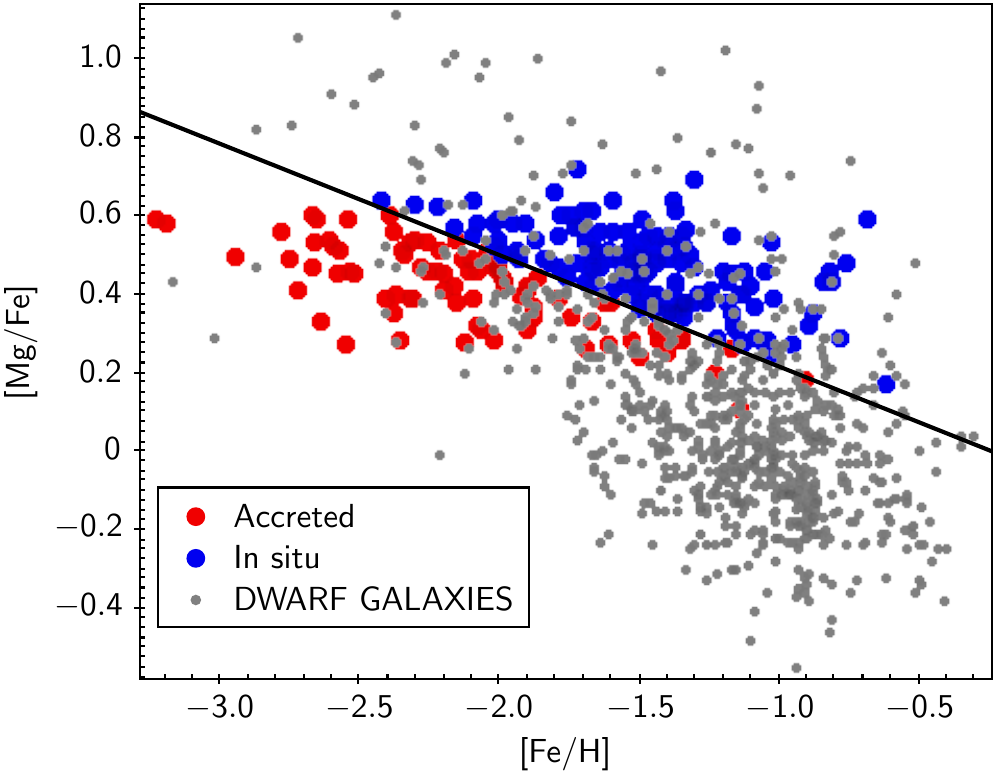}}
\caption{Our programme halo stars  split
 between accreted (red dots) and in situ (blue dots), using the [Mg/Fe] 
criterion of \citet{BV2022}. The grey dots are from Local Group dwarf galaxies, taken
from the SAGA database \citep{SAGA1,SAGA2}. }
\label{aurora_mg}
\end{figure}

In our sample we do not have the luxury to use [Al/Fe], however we use the criterion
on [Mg/Fe] shown in figure 2 of \citet{BV2022}. In Figure\,\ref{aurora_mg}
we show our stars,    that are not in GSE, Sequoia, bulge or Thick Disc,
divided into accreted and in situ. For reference we show also the points
from Local Group dwarfs galaxies, retrieved from the SAGA database \citep{SAGA1,SAGA2}.
This shows that the criterion is fairly good, since most of the points of LG galaxies
lie in the accreted region, however the in situ sample is likely to have some contamination.
According to this criterion the Aurora candidate stars are 102. 
\citet{BV2022}, further
introduced a cut on the energies and distances of the stars, 
namely: $-0.75\times 10^5 {\rm km^2/s^2} < E < -0.4 \times 105 {\rm km^2/s^2}$\,\footnote{ 
We note the typo in \citet{BV2022}, page 692, $10^5$ 
should be $10^{-5}$} and 
$d\le 15$ kpc. 
We do not trim down the number of our Aurora candidates to take also these criteria
into account. 
We also note that in subsequent papers of ther same \citep{BK2023,BK2023_arXiv}
authors the definition of Aurora is revised, as is the Milky Way
potential adopted to compute the integrals of motion. 
Clearly the field is still evolving and we are not in a position
to comment on this evolution.

We stress that our are candidates and higher resolution spectroscopic observations, including Al abundances, 
are encouraged to establish the possible membership to Aurora. 
The full chemical and kinematical data is available in a Table at CDS, 
thus the researchers interested in performing follow-up observations can pick their targets according to the criteria they consider more appropriate.
From the limited chemical inventory at our disposal we cannot see any significative difference 
between the Aurora candidates and the other stars.

\begin{figure*}
\centering
\resizebox{17cm}{!}{\includegraphics[clip=true]{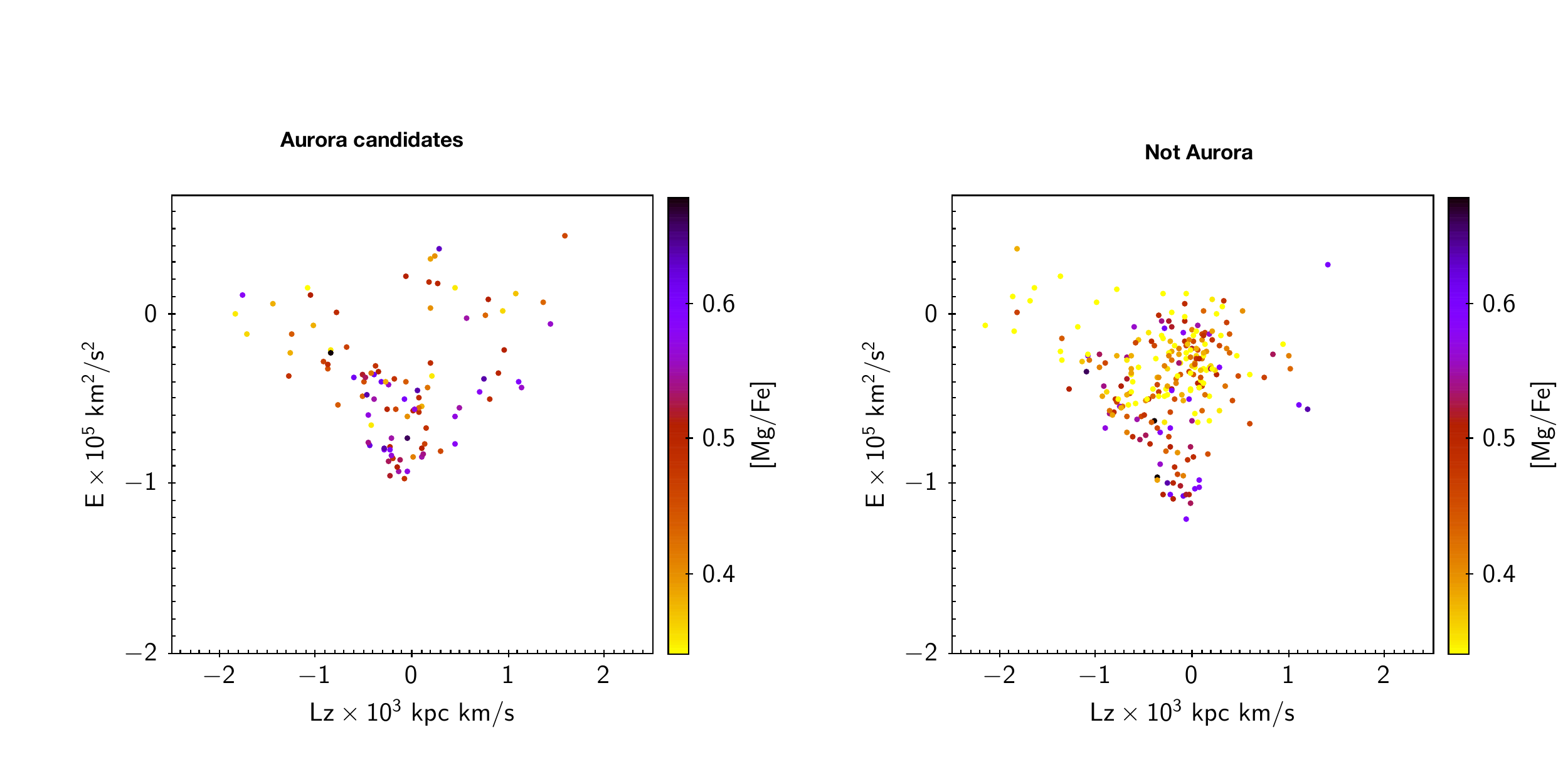}}
\caption{Candidate  Aurora stars in our sample, in the Lz--E plane (left panel), and the
other stars (right panel).}
\label{aurora}
\end{figure*}

In Fig.\,\ref{aurora} we show the sample of our {\em Aurora} candidates
in the Lz--E plane. These candidates are simply our halo in situ sample, 
with [Fe/H]$< -1.3$.
The morphology is very similar to that of figure 1 of \citet{BV2022}, bottom row,
last panel to the right.

According to our criteria, out of the ten young stars, 
six are classified as halo, of these five belong to the Aurora population and only one to the accreted
population. Young age is incompatible with an origin in the first 2\,Ga of the Milky Way formation,
as is expected for Aurora.
This implies that the chemical criterion based on [Mg/Fe] 
introduces substantial contamination between the two populations,
at least for data of this precision. To estimate the level of contamination
one needs [Al/Fe] data.
One of the issues that requires further theoretical and observational investigation
on the Aurora stars, is their metallicity distribution. In our sample the bulk of the stars
lie in the interval $-1.8\le \rm [Fe/H] \le -1.4$ decreasing rapidly at lower metallicity, the most metal-poor
stars being at --2.4.
Is this result simply driven by small number statistics? Why do we not find extremely metal-poor stars
in Aurora? Are all extremely metal-poor stars formed in dwarf galaxies and subsequently accreted?
If so  what were the sources that pre-enriched the gas out of which Aurora stars were formed?
Where they halo stars, or was it the pollution from SNe in the dwarf galaxies that created
a metallicity  floor in the halo stars, before it began star formation?
Assembling larger and more accurate samples of Aurora stars may answer some of these questions.

\section{Conclusions}

Our selection on a high transverse speed is certainly at the origin of the large number of stars on
retrograde orbits.   
The main conclusion of our investigation is that the stars selected with a high transverse velocity are not
a homogeneous group. When we began this investigation, we suspected that we might detect traces 
of past accretion events among them. This is true since we detected stars from the GSE structure and from the
Sequoia structure, as well as from the SpiteF, tentatively identified in this paper.
However, the majority of stars cannot be identified with accretion events and must be considered
as halo stars. 
By using a chemical selection on [Mg/Fe], we could highlight 102 candidates of the Aurora population as
defined by \citet{BV2022}. 

We encourage further investigation of the tentative SpiteF structure. We have provided circumstantial evidence that it is associated with a recent accretion event and argue that it is unlikely that it is just a result of our selection bias. We however underline that this possibility cannot be totally ruled out yet. 
The existence of late accreting galaxies with a recent burst of star formation is in line with the scenario described in \citet{2021ApJ...922...93H,hammer23}.

Unbound stars, such as GHS143, are extremely rare. 
In Paper\,I we have signalled the
possible presence of unbound stars, which turned out to be bound instead after the revision
of parallaxes in Gaia DR3. What is even more surprising is that this star is falling
into the Galactic potential and not escaping from it, as is the case for other
unbound stars previously found. The most promising encounter with a stellar system is that
with the globular cluster NGC\,6582, albeit at two standard deviations. The corollary
is that one should look for the presence of an intermediate mass black hole at the centre
of NGC\,6582, since this would provide the mechanism to supply GHS143 with its current
kinetic energy. 
Other acceleration mechanisms, such as ejection from a binary system,
in which the primary companion has undergone a SN explosion are possible.
The formation of young stars in a globular cluster
is something difficult to envisage, due to the lack of gas, but we shall keep
the cases of Terzan\,5 and Liller\,1 in mind \citep{2021NatAs...5..311F}.

The finding of what is likely a metal-poor young population has been somewhat of a surprise.
In Paper\,I we were more inclined to interpret these stars as blue stragglers, and certainly
some of them are. However, in this investigation we have concentrated on the masses of these stars and found
that they are exceedingly large; blue stragglers typically have masses around 1\,M\sun\ and below
1.3\,M\sun. We recall that when pulsational masses were determined
for blue stragglers, they were found to be in good agreement with the evolutionary masses
that were derived from isochrones, as we did, \citep{2014ApJ...783...34F}.
Thus there is no reason to believe that our mass estimates are not accurate.
Clearly, at this stage we cannot exclude that all of these stars are evolved blue straggler descendants
from triple systems, but we find this hypothesis contrived, and in contrast with the
very few massive blue stragglers found so far in the literature.
We have proposed a test to distinguish between truly young stars and evolved blue stragglers:
the former should, on average, have higher rotational velocities than the latter.
This requires higher resolution follow-up observations of these stars.
The colour-magnitude diagram of the high-speed stars indicates that there are many
other stars that are consistent with a young or evolved blue straggler status. 
These stars are a prime target to further investigate this issue.

The UVES observations of two of the apparently young stars investigated in Paper\,I imply
that the 670.7\,nm \ion{Li}{i} doublet is not detected and the upper limits imply an abundance that 
is below the Spite plateau \citep{spite82a,spite82b,sbordone10}.
Although this supports the blue straggler nature of these stars, the result is not conclusive.  
For both stars the upper limit should be pushed down below A(Li)$\approx 1.0$ with higher quality spectra to be able
to reach a stronger conclusion.


\begin{acknowledgements}
This paper is dedicated to Fran{\c c}ois Spite, who began this investigation with us
and passed away on July 21st 2022.  
We are grateful to R. Lallement for estimating the reddening from the DIBs of GHS076.
We express our gratitude to Paola di Matteo and Misha Haywood for their comments 
on earlier versions of this paper.
We gratefully acknowledge support from the French National Research Agency (ANR) funded project `Pristine' (ANR-18-CE31-0017).
This work has made use of data from the European Space Agency (ESA) mission
{\it Gaia} (\url{https://www.cosmos.esa.int/gaia}), processed by the {\it Gaia}
Data Processing and Analysis Consortium (DPAC,
\url{https://www.cosmos.esa.int/web/gaia/dpac/consortium}). Funding for the DPAC
has been provided by national institutions, in particular the institutions
participating in the {\it Gaia} Multilateral Agreement.
This research has made use of the SIMBAD database, operated at CDS, Strasbourg, France.
\end{acknowledgements}

   \bibliographystyle{aa} 

   \bibliography{biblio} 

\clearpage
\onecolumn

\begin{appendix}

\onecolumn
\section{Target stars}
\label{starnam}
\begin{longtable}{rr}
\caption{\label{tablenam} Names and Gaia DR3 identifiers for our target stars} \\
\hline\hline
{Name} & {Gaia DR3 id} \\
\hline
\endfirsthead
\caption{continued.}\\
\hline
{Name} & {Gaia DR3 id} \\
\hline
\endhead
\hline
\endfoot
\hline
\endlastfoot
  GHS01 & 6408116258177723776\\
  GHS02 & 6406908375935227136\\
  GHS03 & 6568152615142338944\\
  GHS04 & 6467021184886451712\\
  GHS05 & 6402407078410758144\\
  GHS06 & 6459038215072888576\\
  GHS07 & 6578634740526049024\\
  GHS08 & 6396836437107984000\\
  GHS09 & 6792492500909072768\\
  GHS10 & 6375153449333298048\\
  GHS11 & 6480138079431700480\\
  GHS12 & 6480968592964063616\\
  GHS13 & 6469971896141617408\\
  GHS14 & 6779859249744647168\\
  GHS15 & 6450745698376857600\\
  GHS16 & 6453807460302929280\\
  GHS17 & 6697340214885797248\\
  GHS18 & 6472285955798151168\\
  GHS19 & 6443749093572296320\\
  GHS20 & 6686309364478769152\\
  GHS21 & 6455330593146017664\\
  GHS22 & 6425686935027126016\\
  GHS23 & 6678886836357605760\\
  GHS24 & 6454181878372335104\\
  GHS25 & 6424093364721856640\\
  GHS26 & 6673777852499865472\\
  GHS27 & 4659670836211433728\\
  GHS28 & 2892473389378566656\\
  GHS29 & 5188812082642658944\\
  GHS30 & 5401875170994688896\\
  GHS31 & 5388804142407125248\\
  GHS32 & 3481141194650183936\\
  GHS33 & 5371147153902398080\\
  GHS34 & 6186384413992963072\\
  GHS35 & 6186522402702308992\\
  GHS36 & 6070459535828774400\\
  GHS37 & 5792409434759815680\\
  GHS38 & 5819863862157033728\\
  GHS39 & 6494419743340318848\\
  GHS40 & 6378867354796867584\\
  GHS41 & 6500170326593156352\\
  GHS42 & 6344288714832612224\\
  GHS43 & 6390856571321213568\\
  GHS44 & 6490034581730775296\\
  GHS45 & 6490954013971262976\\
  GHS46 & 4629181692264635520\\
  GHS47 & 4817932482581995776\\
  GHS48 & 4849168336616387712\\
  GHS49 & 4945774589328875520\\
  GHS50 & 4697867771333023744\\
  GHS51 & 5008808800675101056\\
  GHS52 & 5008468123868998400\\
  GHS53 & 4719106247173470592\\
  GHS54 & 5120933594860876544\\
  GHS55 & 4709272180814135936\\
  GHS56 & 4687368809680370176\\
  GHS57 & 4684725274488105728\\
  GHS58 & 4924385446036517760\\
  GHS59 & 4901276357319875072\\
  GHS60 & 4901665206478819200\\
  GHS61 & 4703413673624094208\\
  GHS62 & 4977325865764082688\\
  GHS63 & 2321153334969276160\\
  GHS64 & 6567028295783009664\\
  GHS65 & 6578468095795178112\\
  GHS66 & 6376033402233095040\\
  GHS67 & 6431108077108143488\\
  GHS68 & 6363531542708636032\\
  GHS69 & 5189373658205822848\\
  GHS70 & 5791687571014695168\\
  GHS71 & 4907295702445579776\\
  GHS72 & 4983217530100818176\\
  GHS073 & 5246774448712573312\\
  GHS074 & 5464821314034339456\\
  GHS075 & 5400199515274230912\\
  GHS076 & 5335896261418144512\\
  GHS077 & 3479111049508586240\\
  GHS078 & 3467123383468755328\\
  GHS079 & 6185397739746042240\\
  GHS080 & 6158820275960197376\\
  GHS081 & 6168833906311688704\\
  GHS082 & 6188285877618232192\\
  GHS083 & 6188952593982134400\\
  GHS084 & 6164028215864795520\\
  GHS085 & 6190776275751144320\\
  GHS086 & 5769999326196730240\\
  GHS087 & 6124121132097402368\\
  GHS088 & 6172555134696611072\\
  GHS089 & 6221350429945324032\\
  GHS090 & 6106866049448870400\\
  GHS091 & 5906261080680763648\\
  GHS092 & 6216808072534120576\\
  GHS093 & 5795956218759895296\\
  GHS094 & 6199825801908510080\\
  GHS095 & 6001009845939999104\\
  GHS096 & 5820475465505773312\\
  GHS097 & 5823880137628573952\\
  GHS098 & 5807115608941452032\\
  GHS099 & 5809181660013572992\\
  GHS100 & 5914788652362791168\\
  GHS101 & 5818235966468374272\\
  GHS102 & 5802555247021285760\\
  GHS103 & 5817399547356081792\\
  GHS104 & 5811622919419811712\\
  GHS105 & 5813048986006795392\\
  GHS106 & 5919704946794195712\\
  GHS107 & 5919241704497101824\\
  GHS108 & 6743400234444296320\\
  GHS109 & 6438773631658600832\\
  GHS110 & 6734274043115986176\\
  GHS111 & 6708752832046080896\\
  GHS112 & 6709429688837208704\\
  GHS113 & 6446167263239773696\\
  GHS114 & 6660061925981458816\\
  GHS115 & 6724884733535094016\\
  GHS215 & 6548706893010307712\\
  GHS116 & 6759345867468008320\\
  GHS117 & 6723858816145372544\\
  GHS118 & 6639733192933584640\\
  GHS119 & 6725398102402401536\\
  GHS120 & 6654836978006631936\\
  GHS121 & 6657610152491402240\\
  GHS122 & 6703622308933487744\\
  GHS123 & 6710061495703908608\\
  GHS124 & 6644138936026516864\\
  GHS125 & 6644182229296625408\\
  GHS126 & 6719559180233706368\\
  GHS127 & 6658033498828025600\\
  GHS128 & 6732823997746014592\\
  GHS129 & 6711011572528262272\\
  GHS130 & 6727332315145507840\\
  GHS131 & 6730967884676130048\\
  GHS132 & 6661006990584788224\\
  GHS133 & 6437230084835781248\\
  GHS134 & 6435544984482961408\\
  GHS135 & 6662545722747834624\\
  GHS136 & 6633963230792340352\\
  GHS137 & 6715529298307602176\\
  GHS138 & 6433337199495213056\\
  GHS139 & 6715711301840606208\\
  GHS140 & 6661669583779428480\\
  GHS141 & 6632232152810751488\\
  GHS142 & 6730547355843299840\\
  GHS143 & 6632370485122299776\\
  GHS144 & 6715108524650751104\\
  GHS145 & 6431315919166662400\\
  GHS146 & 6438403096241999744\\
  GHS147 & 6436884219708078976\\
  GHS148 & 6432994632905279616\\
  GHS149 & 4044446251671465600\\
  GHS150 & 6726211496149367168\\
  GHS151 & 6726207068062588160\\
  GHS152 & 6680627019666959232\\
  GHS153 & 6475332462000476800\\
  GHS154 & 6424891713242837632\\
  GHS155 & 6658676438250022656\\
  GHS156 & 6373751572009351808\\
  GHS157 & 6445943065946638848\\
  GHS158 & 6659358788294532096\\
  GHS159 & 6697640862596021504\\
  GHS160 & 6665227019292614400\\
  GHS161 & 6647138330729258240\\
  GHS162 & 6741699461753415680\\
  GHS163 & 6639271707287676160\\
  GHS164 & 6759529344169496960\\
  GHS165 & 6666726856229854976\\
  GHS166 & 6796363262874820480\\
  GHS167 & 6685546612647538560\\
  GHS169 & 6689078862471972224\\
  GHS170 & 6443729130564198144\\
  GHS171 & 6685927421627177600\\
  GHS172 & 6797044341609133184\\
  GHS173 & 6643886804265865344\\
  GHS174 & 6640961691018592128\\
  GHS175 & 6669295968226934144\\
  GHS176 & 6754545399700231936\\
  GHS177 & 6474154300931920128\\
  GHS178 & 6670239486643467392\\
  GHS179 & 6447711355521757824\\
  GHS180 & 6850286783435901568\\
  GHS181 & 6679621207045185280\\
  GHS182 & 6671398921654234624\\
  GHS184 & 6674250127104208768\\
  GHS185 & 6805170454092098688\\
  GHS186 & 6464518421544150016\\
  GHS188 & 6476230419403065728\\
  GHS189 & 6409485837349533440\\
  GHS190 & 6476479355707573376\\
  GHS191 & 6465543303819966720\\
  GHS192 & 6477482453910217728\\
  GHS193 & 6477737025212150528\\
  GHS194 & 6410024838564847744\\
  GHS195 & 6809524618920019840\\
  GHS196 & 6579524863908225408\\
  GHS197 & 6479895400896199168\\
  GHS198 & 6570773301106866048\\
  GHS199 & 6410742265605581056\\
  GHS200 & 6577763721159122816\\
  GHS201 & 6676759487516643584\\
  GHS202 & 6589825157556394880\\
  GHS203 & 6456587609813249536\\
  GHS204 & 6678553203298351616\\
  GHS205 & 6456746489243603968\\
  GHS206 & 6412415142484095488\\
  GHS207 & 6775157600584648832\\
  GHS208 & 6584883127668353024\\
  GHS209 & 6585219956184571648\\
  GHS210 & 6519464075600798080\\
  GHS212 & 6380245322040654976\\
  GHS213 & 6485376840021854848\\
  GHS214 & 6549893059898174336\\
  GHS216 & 6488964550758904448\\
  GHS217 & 6620131271429742464\\
  GHS218 & 6408639556993271168\\
  GHS219 & 6526793076516023808\\
  GHS220 & 6494632155242530176\\
  GHS221 & 6543884916047853312\\
  GHS222 & 6395827669549221376\\
  GHS223 & 6625197335678814208\\
  GHS224 & 6610555792166870528\\
  GHS225 & 6604051944665974656\\
  GHS226 & 4900325966956503040\\
  GHS227 & 4638383814314197248\\
  GHS228 & 4719222967204604544\\
  GHS229 & 4705166776194991488\\
  GHS230 & 5002359512143209600\\
  GHS231 & 4706695612753634048\\
  GHS232 & 4925067486843323264\\
  GHS233 & 5006315211381992448\\
  GHS234 & 4960618924017083008\\
  GHS235 & 4709117768150098176\\
  GHS236 & 4962433942836758528\\
  GHS237 & 2321267065703224320\\
  GHS238 & 4930039787661776896\\
  GHS239 & 4931202693006868736\\
  GHS240 & 4744503625745191808\\
  GHS241 & 4751801221857764096\\
  GHS242 & 4948075248690809600\\
  GHS243 & 4828700412269231872\\
  GHS247 & 4037996035430475904\\
  GHS248 & 6727695561968103680\\
  GHS249 & 6727647016003354112\\
  GHS251 & 6727262839751268608\\
  GHS252 & 6726960027365884928\\
  GHS253 & 6727000498823237376\\
  GHS254 & 6734285553628576512\\
  GHS255 & 4044903682871414912\\
  GHS256 & 4045089088020401536\\
  GHS257 & 6734881454571852672\\
  GHS258 & 4044716491016787200\\
  GHS259 & 6726771529840056192\\
  GHS260 & 6726100037506559104\\
  GHS261 & 6725786947220199040\\
  GHS262 & 4044669972293549184\\
  GHS263 & 6727838502804991104\\
  GHS264 & 6725823613397479936\\
  GHS265 & 6724945412835328768\\
  GHS266 & 6726512354342331520\\
  GHS267 & 4038117501457765376\\
  GHS268 & 4036049040854615040\\
  GHS269 & 4045078165976720640\\
  GHS270 & 6726511529708519040\\
  GHS371 & 4044697391366831488\\
  GHS372 & 6733322901158296192\\
  GHS373 & 6726816919056054016\\
  GHS271 & 2306798282955504896\\
  GHS272 & 4972994958180431360\\
  GHS273 & 4688513916687143040\\
  GHS274 & 4685472564442295680\\
  GHS275 & 4902917550222823424\\
  GHS276 & 5011444467485885056\\
  GHS277 & 4698962919274310528\\
  GHS278 & 4956345603356257792\\
  GHS279 & 4622163406464165376\\
  GHS280 & 5496239828735395712\\
  GHS281 & 5242575280666939648\\
  GHS282 & 5659111104633796224\\
  GHS283 & 5451988918123679616\\
  GHS284 & 5373547525219529984\\
  GHS285 & 5370387941123150336\\
  GHS286 & 3477848844520166528\\
  GHS287 & 5379743032720204160\\
  GHS288 & 3470573788395088256\\
  GHS289 & 3470203837092167552\\
  GHS290 & 6184615986979107072\\
  GHS291 & 6155020569933314816\\
  GHS292 & 5844585178525184000\\
  GHS293 & 6175800927381924608\\
  GHS294 & 6094901301356994944\\
  GHS295 & 6114612624260756096\\
  GHS296 & 6273531636891321984\\
  GHS297 & 6274166295618635008\\
  GHS301 & 6106034131463681152\\
  GHS302 & 5793152189225305600\\
  GHS303 & 5785885963551502720\\
  GHS311 & 6005027083467410688\\
  GHS312 & 5902898327464861696\\
  GHS315 & 5793965793469740544\\
  GHS316 & 6009153825488821248\\
  GHS319 & 6670601874506513152\\
  GHS320 & 6846937636656976128\\
  GHS321 & 6471981562876265344\\
  GHS322 & 6472175764117277952\\
  GHS323 & 6692433987842986624\\
  GHS324 & 6686342727784978816\\
  GHS325 & 6694064524930075136\\
  GHS326 & 6681027551137071744\\
  GHS327 & 6696370200815375744\\
  GHS328 & 6429536600114085120\\
  GHS329 & 6471075427855732352\\
  GHS330 & 6779593889485911424\\
  GHS331 & 6482816425335411328\\
  GHS332 & 6453858828111935488\\
  GHS333 & 6778364841643442816\\
  GHS334 & 6471283304276414720\\
  GHS335 & 6425109657062446464\\
  GHS336 & 6774572007563813888\\
  GHS337 & 6481694098841003520\\
  GHS338 & 6369854486186048896\\
  GHS339 & 6484277985525429504\\
  GHS340 & 6677197471101594112\\
  GHS341 & 6677416102116682496\\
  GHS342 & 6781914168256940544\\
  GHS343 & 6775382759950459776\\
  GHS344 & 6775379701933700736\\
  GHS345 & 6451791402653847936\\
  GHS346 & 6580668630879478272\\
  GHS347 & 6579947213812310016\\
  GHS348 & 6782909879114518016\\
  GHS349 & 6575448287109093888\\
  GHS350 & 6578906766575595776\\
  GHS351 & 6786759406762161408\\
  GHS352 & 6459085528432629504\\
  GHS353 & 6397098464472800128\\
  GHS354 & 6461290530282488320\\
  GHS355 & 6402073479711443712\\
  GHS356 & 6411301646442226816\\
  GHS357 & 6625938990631507456\\
  GHS358 & 6566282380222388096\\
  GHS359 & 6624839036622440576\\
  GHS360 & 6386080739486924544\\
  GHS361 & 6504959661804736384\\
  GHS362 & 6384892167417294336\\
  GHS363 & 6512326905107154048\\
  GHS364 & 6512769355458511232\\
  GHS365 & 6393487530847705344\\
  GHS366 & 2379631410648067968\\
  GHS367 & 6492089843841047808\\
  GHS368 & 6341831375063721088\\
  GHS369 & 2310689007929797760\\
\hline
\end{longtable}
%
\twocolumn

\section{ADQL query used}
\label{query}

The ADQL query used  to select high transverse  speed
stars in Fig.\,\ref{cmd_hs}
\begin{verbatim}
SELECT 
   *
   FROM gaiadr3.gaia_source
   WHERE
      4.740470446*pm/parallax >= 500
      AND
      phot_g_mean_mag between 7 and 13

\end{verbatim}

\section{Remarks on individual stars}
\label{remarks}

\subsection{GHS076 a highly reddened B type star}

\begin{figure}
\centering
\includegraphics{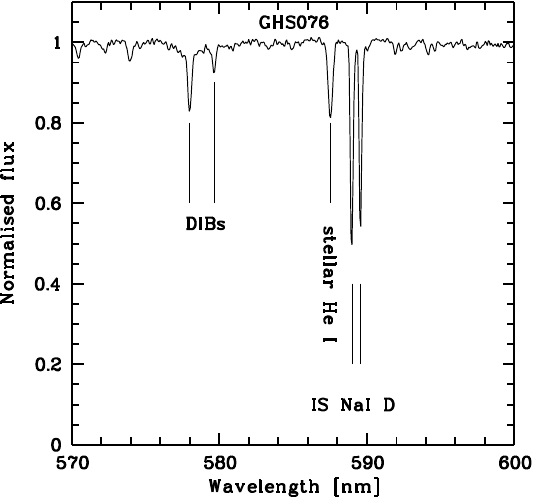}
\caption{Portion of the spectrum of GHS076 displaying stellar and interstellar features.}
\label{ghs076}
\end{figure}

Since the colour selection was done on the observed Gaia $G_{BP}-G_{RP}$ and no
cut on Galactic latitude, 
it is not surprising that among the selected targets  we found a highly reddened B star.
Star GHS076 (=Gaia DR3 5335896261418144512) is at low Galactic latitude ($b=1.40284$, $l=293.75475$)
and  is predicted by the Planck reddening maps
\citep{2019A&A...623A..21I} to have $A_V=3.3$, in Fig.\,\ref{ghs076} we show a portion
of the spectrum of GHS076 that covers the two  Diffuse Interstellar Bands
at 579.7\,nm and 578.0\,nm \citep[see e.g.][]{2018A&A...616A.143E}, the interstellar
\ion{Na}{I} D doublet and the stellar \ion{He}{I} 587.5\,nm line.
From the two DIBs shown in Fig.\,\ref{ghs076} one can estimate $A_V(578.0) = 3.2$
and $A_V(579.7) = 3.5$ (R. Lallement, priv. comm.) in substantial agreement
with the Planck-based estimate.
To determine the atmospheric parameters of this star we assumed a solar
metallicity (as appropriate for a disc young star) and computed ATLAS 9 models
\citep{2005MSAIS...8...14K}, in its Linux version \citep{2004MSAIS...5...93S,2005MSAIS...8...61S},
using the new opacity distribution functions of Mucciarelli \& Bonifacio (in preparation).

We computed Balmer lines H$_{\gamma}$ and H$_\beta$ in LTE using SYNTHE and \ion{He}{I} lines
in NLTE using the Kiel code \citet{1984A&A...130..319S}
and the same 36 level model atom used in \citet{2014AN....335...59C}, collisions with
hydrogen where computed with the \citet{1984A&A...130..319S} generalisation of the
\citet{1969ZPhy..225..483D} formalism assuming $S_H=1/3$.
Using a well established practice in the analysis of B-type stars, for each Balmer line we
computed profiles for several values of effective temperature and surface gravity. 
For each temperature we fitted the observed line profile to determine the surface gravity values.
This set of \teff\ and \glog\ values determines  a curve
in the \teff , \glog\ plane.
Likewise for the \ion{He}{I} lines we computed NLTE line profiles for several values of 
\teff\ and \glog . For each \ion{He}{I} line and each value of \glog\ we fitted \teff.  
Again
these points define a curve in the \teff , \glog\ plane.
The `best' parameters for the star are defined by the region that is enclosed by all these curves.
In our case this results in \teff = 21000\,K and \glog = 3.4 .
We can detect in our spectrum clearly several metallic lines like \ion{C}{II}, \ion{N}{II}, \ion{O}{II},
\ion{Mg}{II}, \ion{Al}{III}, \ion{S}{II}, however a quick check showed that an NLTE analysis is required
to determine the chemical abundances of this star and we are not equipped to do so for all these ions.
A higher resolution spectrum of  this highly reddened, and relatively bright ($G=13.15$) would certainly 
be valuable for the study of DIBs and interstellar features.

\subsection{Variable stars\label{var}}

Five stars in our sample are variables: GHS029, GHS046 GHS065, GHS110, GHS236.
The first two were already discussed as RR\,Lyr in Paper\,I. 

GHS065 was reported as one of the two CEMP stars in the sample. Gaia DR3
classifies it as $\delta$ Scuti/$\gamma$ Dor, however with a low probability (0.12).
We inspected the Gaia epoch photometry and the peak-to-peak magnitude variation
is of the order of 0.02 mag in all three bands. We believe this implies that the
photometric parameters derived in Paper\,I are correct.

GHS110 is classified as a Type II Cepheid, subclass W Vir, by Gaia DR3, albeit with a very low probability
of 0.05.   
The Gaia DR3 period is 8.5927 days, which is coherent
with the Lomb-Scargle \citep{1976Ap&SS..39..447L,1982ApJ...263..835S} 
periodogram that we computed for the 
$G$ band epoch photometry, note however that the false alarm probability 
for this peak is 0.01 casting some doubt on the reality of this period.
We tried to phase the epoch photometry with this period and could not find a curve that
can be reasonably fitted. We believe the star is probably
multi-periodic and longer and better sampled time series are required
to elucidate its nature. For the $G_{BP}-G_{RP}$ the situation is even worse,
however the peak-to-peak variation in colour is of only 0.07 mag,
i.e a change in \teff\ of the order of 150\,K.
In the impossibility of determining a reliable light curve to derive what
the appropriate colour was at the time of observation, we stick with the 
parameters derived from the mean photometry and consider an error on \teff\
pf 150\,K for deriving the stellar metallicity.

Star GHS236 is classified as RS CVn by Gaia DR3, with a low
probability of 0.26. Inspection of 
the spectrum does not show any sign of chromospheric
activity, distinctive of RS CVn, thus not supporting this classification.
The Lomb-Scargle periodogram shows a highly significative
peak corresponding to a period of 32.59 days. Such a period does not rule 
out an RS CVn, however the majority of these close binaries have
periods less than 20 days \citep[see e.g.][]{Martinez}.
Phasing the $G$ data with this period the light curve is quite reasonable,
the colour curve is much more scattered. However, for our purpose what 
is relevant is that the peak-to-peak variation in $G_{BP}-G_{RP}$ colour
is of the order of 0.04\,mag  i.e. roughly 100\,K in effective
temperature. This supports our atmospheric parameters deriving using the mean
colour.

\section{Clustering analysis}\label{clustering}

The motivation for performing a clustering analysis has been given in Sect.\,\ref{kin}. We did so on the stars of our sample using the HDBSCAN\footnote{\url{https://hdbscan.readthedocs.io/en/latest/index.html}} 
library and a Gaussian Mixture model\footnote{\url{https://scikit-learn.org/stable/modules/mixture. html}}.
After a number of attempts using different input variables, we decided to use E, L$_Z$ and J$_R$ only. The data was scaled using a `RobustScaler' within the scikit learn package.

\begin{figure*}
\centering
\includegraphics[width=\hsize,clip=true]{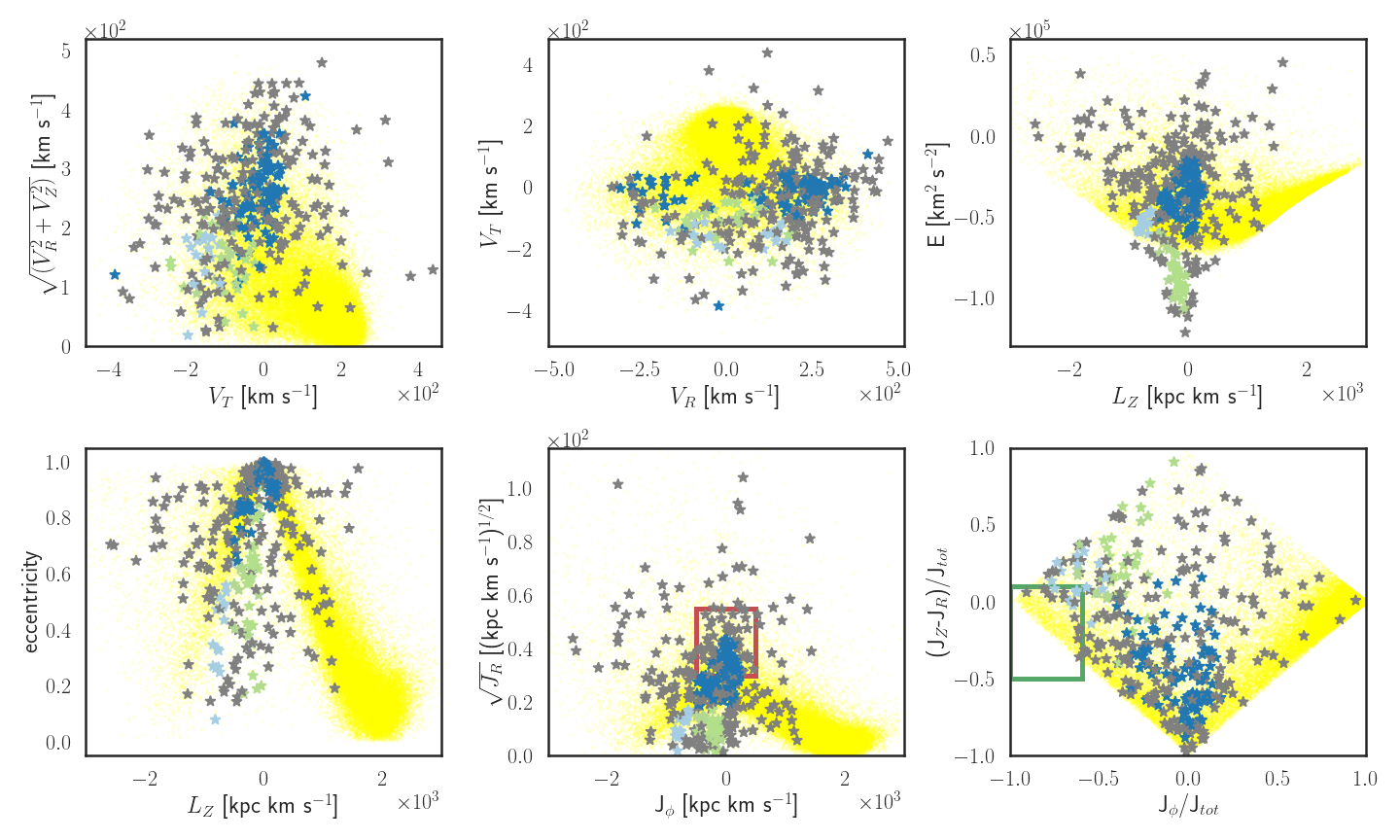}
\caption{Same planes as in fig.\,\ref{char1}. Programme targets are marked by filled stars: grey for unclustered, green blue and light blue for the three identified clusters. Blue and green clusters can be associated to stars classified as GSE and bulge in section\,\ref{kin}. Yellow points are stars from \citet[][]{topos6}. The areas used to select GSE and Seq stars in section\,\ref{kin} are still outlined here for reference but using red and green squares in the bottom middle and right panels, respectively.}
\label{char5}
\end{figure*}

\begin{figure}
\centering
\includegraphics[width=\hsize,clip=true]{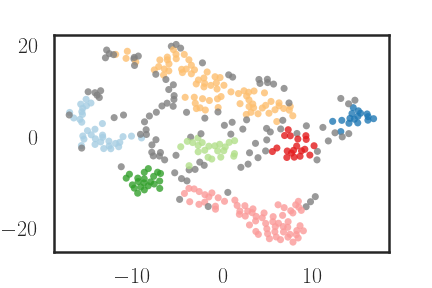}
\caption{Two-dimensional t-SNE projection of the E, L$_Z$ and J$_R$ distribution of the program stars. The 7 clusters identified by HDBSCAN are indicated with different colours. Grey filled circles are unclustered stars.}
\label{char8}
\end{figure}
\begin{figure*}
\centering
\includegraphics[width=\hsize,clip=true]{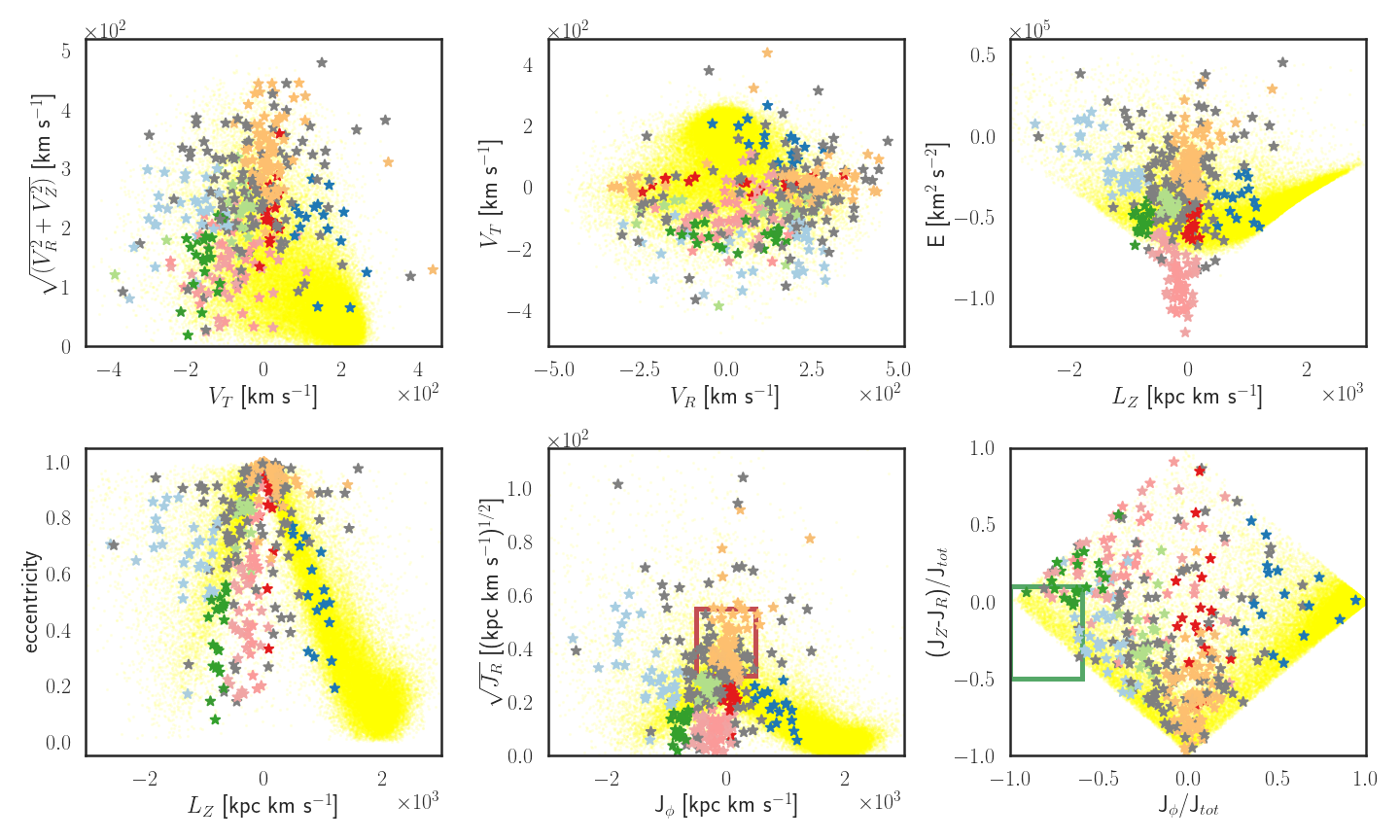}
\caption{Similar to fig.\,\ref{char5} but for the clusters identified using HDBSCAN using the t-SNE projection of the input parameters.}
\label{char10}
\end{figure*}
\begin{figure*}
\centering
\includegraphics[width=\hsize,clip=true]{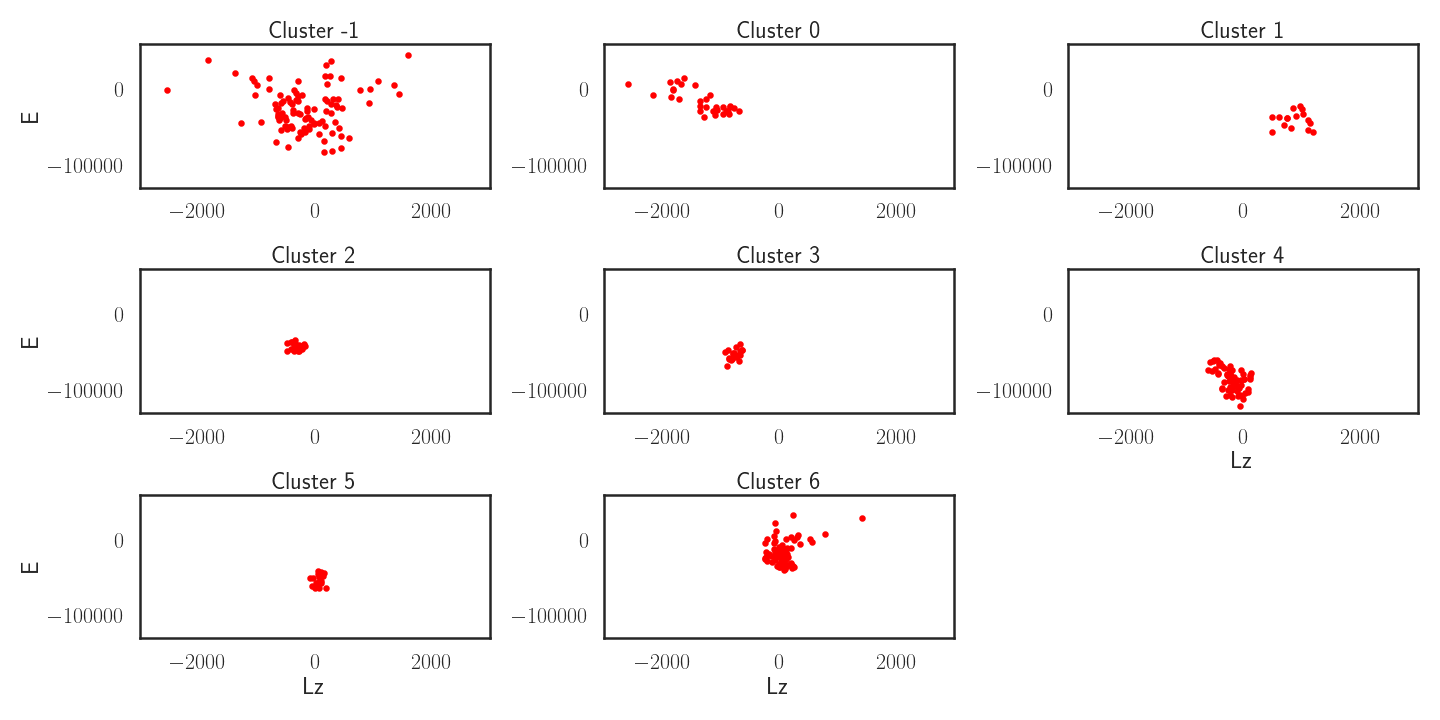}
\caption{E  vs  L$_Z$ plane. On each panel we plot stars belonging to one of the 7 clusters (the top label indicates the number of the cluster from 0 to 6) identified using HDBSCAN using the t-SNE projection of the input parameters. Unclustered stars (top-left panel) are labelled `Cluster -1'.}
\label{char12}
\end{figure*}
\begin{figure*}
\centering
\includegraphics[width=\hsize,clip=true]{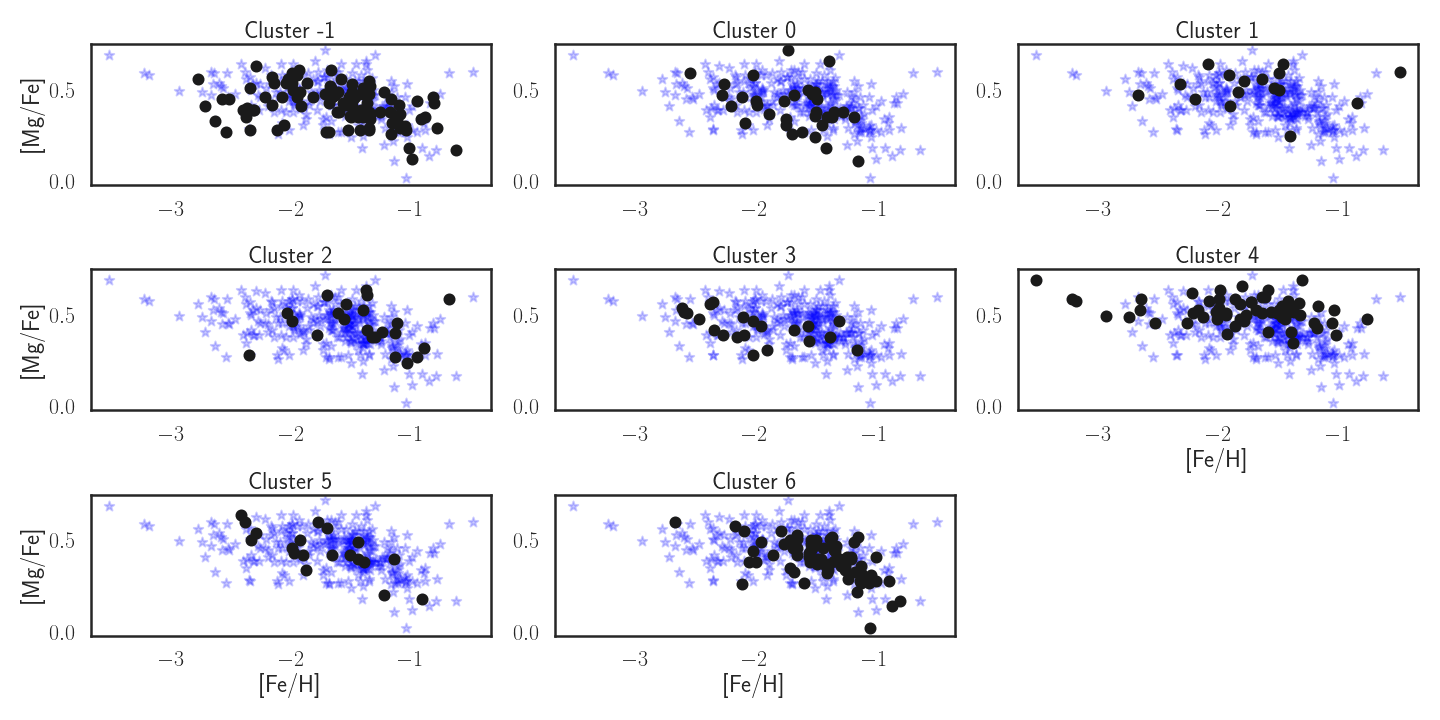}
\caption{Similar to fig.\,\ref{char3} but for the clusters identified using HDBSCAN using the t-SNE projection of the input parameters. Clusters 6 and 4 can be associated with GSE and bulge stars as defined in section\,\ref{kin} (see figs.\,\ref{char12} and \ref{char1}).}
\label{char13}
\end{figure*}

\begin{figure}
\centering
\includegraphics[width=\hsize,clip=true]{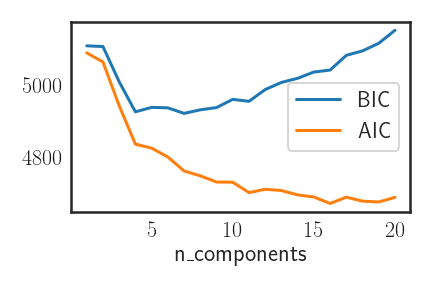}
\caption{AIC and BIC estimators for the GMM applied on the two-dimensional projection of the input parameters.}
\label{char14}
\end{figure}
\begin{figure}
\centering
\includegraphics[width=\hsize,clip=true]{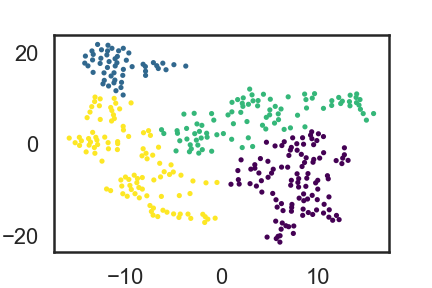}
\caption{Two-dimensional t-SNE projection of the input parameters. The stars belonging to a GMM with 4 components are indicated with different colours.}
\label{char15}
\end{figure}
\begin{figure*}
\centering
\includegraphics[width=\hsize,clip=true]{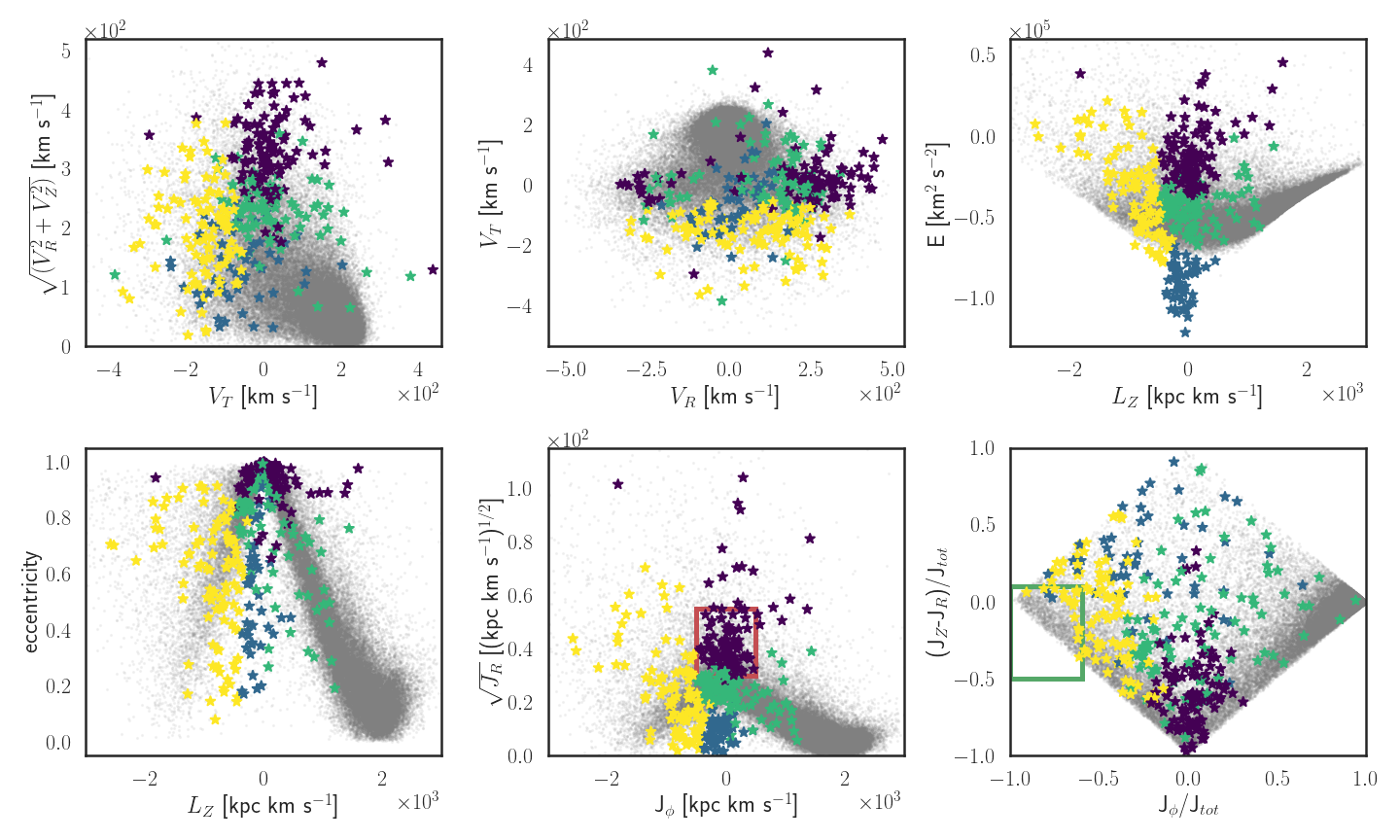}
\caption{Same as fig.\,\ref{char1}. Grey points are stars from \citet[][]{topos6}, stars belonging to the different group of the 4 components of the GMM have different colours.}
\label{char16}
\end{figure*}
\begin{figure*}
\centering
\includegraphics[width=\hsize,clip=true]{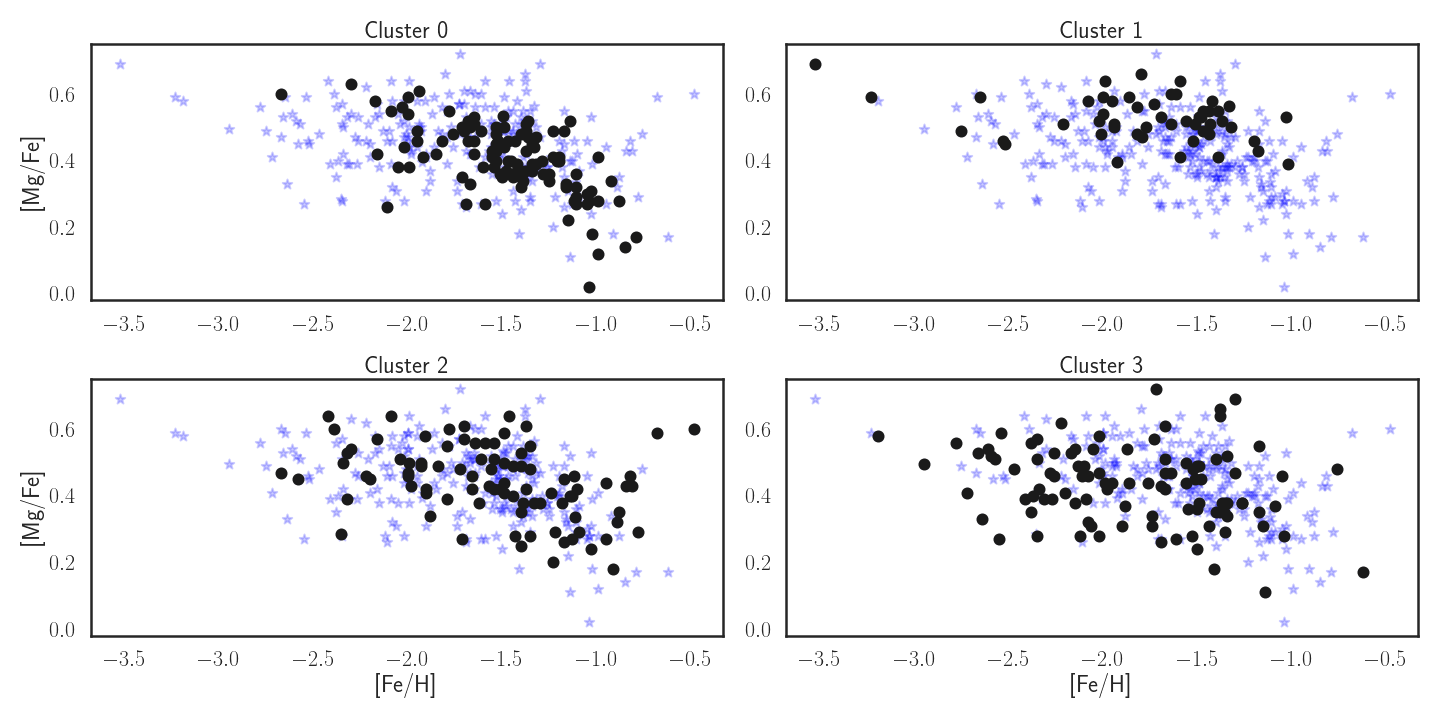}
\caption{
Same as Fig.\,\ref{char3}. Symbols are coloured as in Fig.\,\ref{char16}. 
Cluster 0 (bottom-right) and 1 (top-right) correspond to yellow and blue stars in
figure D.8 and can be associated to GSE and bulge stars as identified in section 4.
}
\label{char18}
\end{figure*}

\subsection{Minimum size 15 and minimum samples 10}

HDBSCAN was setup with min\_cluster\_size=15 and min\_samples=10. Under these conditions, HDBSCAN identifies 3 clusters, the most prominent of which can be identified with GSE (cluster "1", 90 members) and bulge (cluster "2", 34 members) stars, although the stars belonging to the groups are not exactly the same as those of Sect.,\ref{kin}. A third group of 17 stars was also identified. 
Figure \ref{char5} is analogous to figure \ref{char1}, but this time unclustered stars are presented in grey and stars from \citet[][]{topos6} in yellow, while coloured stars are stars belonging to the different clusters. 

\subsection{Dimensionality reduction}

If we perform a dimensionality reduction using the 
t-SNE\footnote{\url{https://scikit-learn.org/stable/modules/generated/sklearn.manifold.TSNE.html}} 
(T-distributed Stochastic Neighbour Embedding) library first and then apply HDBSCAN on the resulting projection, we end up with 7 clusters. Similarly, the two most prominent clusters, "6" and "4", with 72 and 62 members each, occupy in the planes of figure \ref{char1} regions similar to those of GSE and bulge stars but, again with larger scatter when we look at them in the [Mg/Fe]  vs  [Fe/H] plane for bulge stars. Here, however, clusters 5 (19 stars) and perhaps 2 (22 stars) and 0 (33 stars) seems also to present a coherent behaviour in this plane. Figures \ref{char8} presents the t-SNE projection resulting from the dimensionality reduction process, showing the different clusters identified by HDBSCAN with different colours. 
Figure \ref{char10} is analogous to figure \ref{char5}. Figure\,\ref{char12} presents each cluster separately in the E  vs  L$_Z$ plane to allow the association of a kinematic structure with the chemical behaviour presented in fig.\,\ref{char13}. It may be noticed here that the region defined by \citet[][]{feuillet21} to select likely GSE candidates is roughly coincident with the stars selected in cluster 6 (bottom-middle panel in Figs.\,\ref{char10} and \ref{char12}), apart for a few stars having higher values of J$_R$ in our case. 

\subsection{Gaussian mixture model}

We also applied a Gaussian mixture model (GMM) to the stars in our sample by using the same quantities and scaling and applying a t-SNE dimensionality reduction first. Using the t-SNE projection to train the model, the BIC estimator suggest a number of components between 4 and 6 to represent our sample (figure \ref{char14}). Figure \ref{char15} shows the t-SNE projection with stars assigned to different groups by applying a GMM with 4 components.  Again, figures\,\ref{char16} and  \ref{char18} are analogous of figures \ref{char1}, \ref{char2}, \ref{char3}. Clusters 0 and 1 can be identified with GSE and bulge stars, respectively. The GSE selection box results once again very similar to stars assigned to component 0 of the GMM, with a few stars at higher J$_R$. Also, once more clusters 0 and 1 have defined patterns in the [Mg/Fe]  vs  [Fe/H] plane but bulge stars appear also more dispersed with respect to the pattern described by the stars selected in section\,\ref{kin}. 

\subsection{Conclusions on the clustering analysis}

In summary, the criteria adopted in \citet[][]{feuillet21} seems to be broadly consistent with the major group identified in our clustering analysis with HDBSCAN and the Gaussian Mixture model using E, L$_Z$ and J$_R$ as input variable, especially when a dimensionality reduction is applied first using the t-SNR algorithm. It is remarkable that a detection this clear can be made with a clustering analysis on a sample of about 350 stars. 

The second most prominent group is composed by stars which have low orbital energy and found in the inner part of the Galaxy. These two groups are identified as well in all the experiments we made. We associate this group to the stars confined to the bulge that we identified in section \ref{kin}. However, while the reality of the group seems unquestionable, stars belonging to this group present a larger scatter in the [Fe/H]  vs  [Mg/Fe] plane when selected according to our clustering analysis with respect to the group identified in section \ref{kin}.
This is the main reason why we decided to 
define the SpiteF structure requiring $r_{ap} < 3.5$ kpc.
This choice is, to some extent arbitrary, however it is also consistent with
the notion that the inner bulge is more spherical and different from
the bar and pseudo-bulge \citep[see e.g.][and references therein]{lucey19,reggiani2020}.

Finally, HDBSCAN when coupled with a dimensionality reductioN technique, identify up to 7 clusters. Besides, GSE and bulge stars, at least one but up to 3 more clusters seem to present a chemical signature in the  [Fe/H]  vs  [Mg/Fe] plane.

\clearpage

\section{SpiteF: Further discussion}
\label{spietef_app}

In this section we provide a more detailed discussion of
SpiteF, its selection and its properties.

\subsection{Selection of stars belonging to the bulge and SpiteF}
\label{sel_spitef}
\begin{figure*}
\centering
\includegraphics[width=\hsize,clip=true]{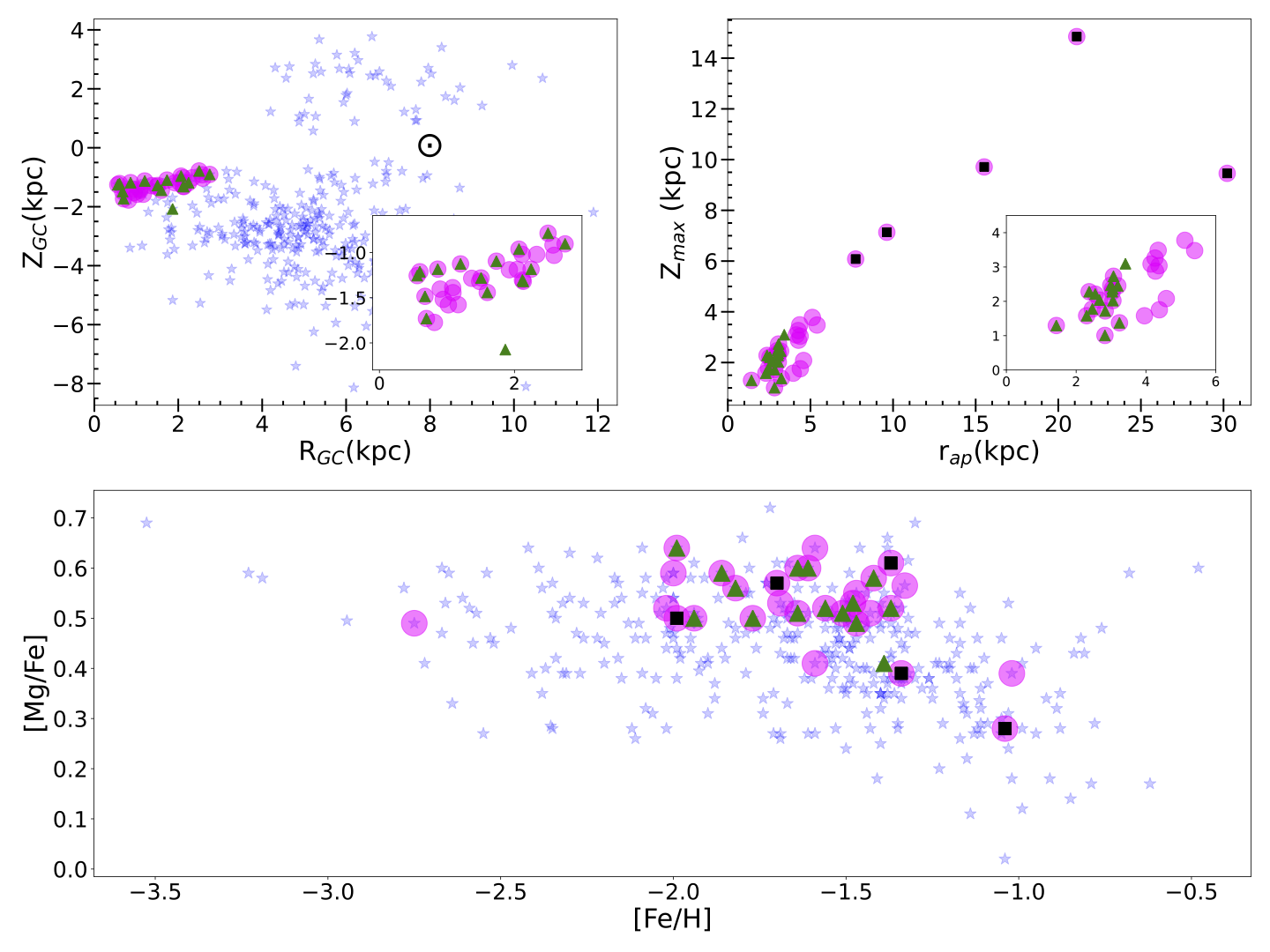}
\caption{
Possible alternative selection of the SpiteF structure.
Top left: in the Z$_{GC}$ vs  R$_{GC}$ plane (magenta large filled circles).
The SpiteF stars as selected in Sect.\,\ref{sec:spitef}  are identified as green 
triangles. the other stars in our sample are shown as blue stars.
Top right: the same in the $Z_{max}$  vs  $r_{ap}$ plane.
Bottom panel: The same in the [Mg/Fe]  vs  [Fe/H] plane.  
Stars with r$_{ap}$>7\,kpc are shown as black filled squares in the top-right and bottom panels.
}
\label{bulge_spitef}
\end{figure*}

The operational definition of the SpiteF structure is certainly arbitrary to some degree. 
In appendix\,\ref{clustering}, we saw that the second most prominent cluster among the stars of our sample is a group of stars at low energies which includes the SpiteF. 

The SpiteF was initially noticed as a group of stars with a well defined trend in the   Z$_{GC}$ vs  R$_{GC}$ plane. 
A possible selection of stars belonging to this structure is presented in the top-left panel of Fig.\ref{bulge_spitef} (magenta large filled circles). The SpiteF stars are also shown as green triangles, while the remaining stars in our sample are shown as blue stars. The inset shows a zoom-in  the region of the selected stars. Fifteen of the sixteen SpiteF stars belong also to the stars we select in this way. This larger sample is composed of 30 stars and we  call them 'RZ' in the following. 

The two different selections (RZ and SpiteF) are presented in the top-right panel in the Z$_{max}$  vs  r$_{ap}$ plane. A rotating disc was included in the potential in order to evaluate these quantities. By definition, SpiteF stars are confined to r$_{ap}<$3.5\,kpc. A group of five stars (filled black squares) extends to large values (r$_{ap}>$7\,kpc and Z$_{maz}>$6\,kpc). Excluding these five stars, the RZ is a sample of 25 stars confined to r$_{ap}<$5.4\,kpc and Z$_{max}<$3.8\,kpc, as can be better seen in the inset. Fifteen of these 25 stars also belong to the SpiteF. These RZ stars extend, thus, to larger Z$_{max}$ values than the SpiteF, which is confined to Z$_{max}<$3\,kpc. 

The bottom panel of the figure, presents the RZ and SpiteF stars in  the [Fe/H]  vs  [Mg/Fe] plane. Stars extending at large r$_{ap}$ and Z$_{maz}$ are also identified (black filled squares). 
The remaining stars in the sample are also shown as blue stars. As can be seen, 7 our of the 10 stars additional stars confined in the inner region of the Galaxy, have abundances compatible with the SpiteF. Of the remaining 3 stars, two present a lower [Mg/Fe] ratio, and one has a lower metallicity ([Fe/H]=-2.75). Three of the stars extending to large values of r$_{ap}$ and Z$_{maz}$ are also compatible with SpiteF, while two are not.  

In summary, the 25 stars in the RZ sample confined to the inner Galaxy provide a possible alternative definition for the structure we identified and labelled as SpiteF. All but one of the SpiteF stars are in fact contained in the RZ sample. The kinematics and chemistry of RZ and SpiteF are compatible between each other. 

Note that, \citet[][]{lucey21} adopted the same criterion we used to define stars belonging to the SpiteF (r$_{ap}<$3.5\,kpc), 
to discriminate stars truly confined to the bulge from halo interlopers. 
They indicate this as the distance usully adopted to define the Galactic bulge in the literature \citep[e.g.][]{ness13,2020MNRAS.491L..11A,kunder20}.

\subsection{SpiteF and the poor old heart of the Milky Way\label{e2}}
\begin{figure*}
\centering
\resizebox{8cm}{!}{\includegraphics{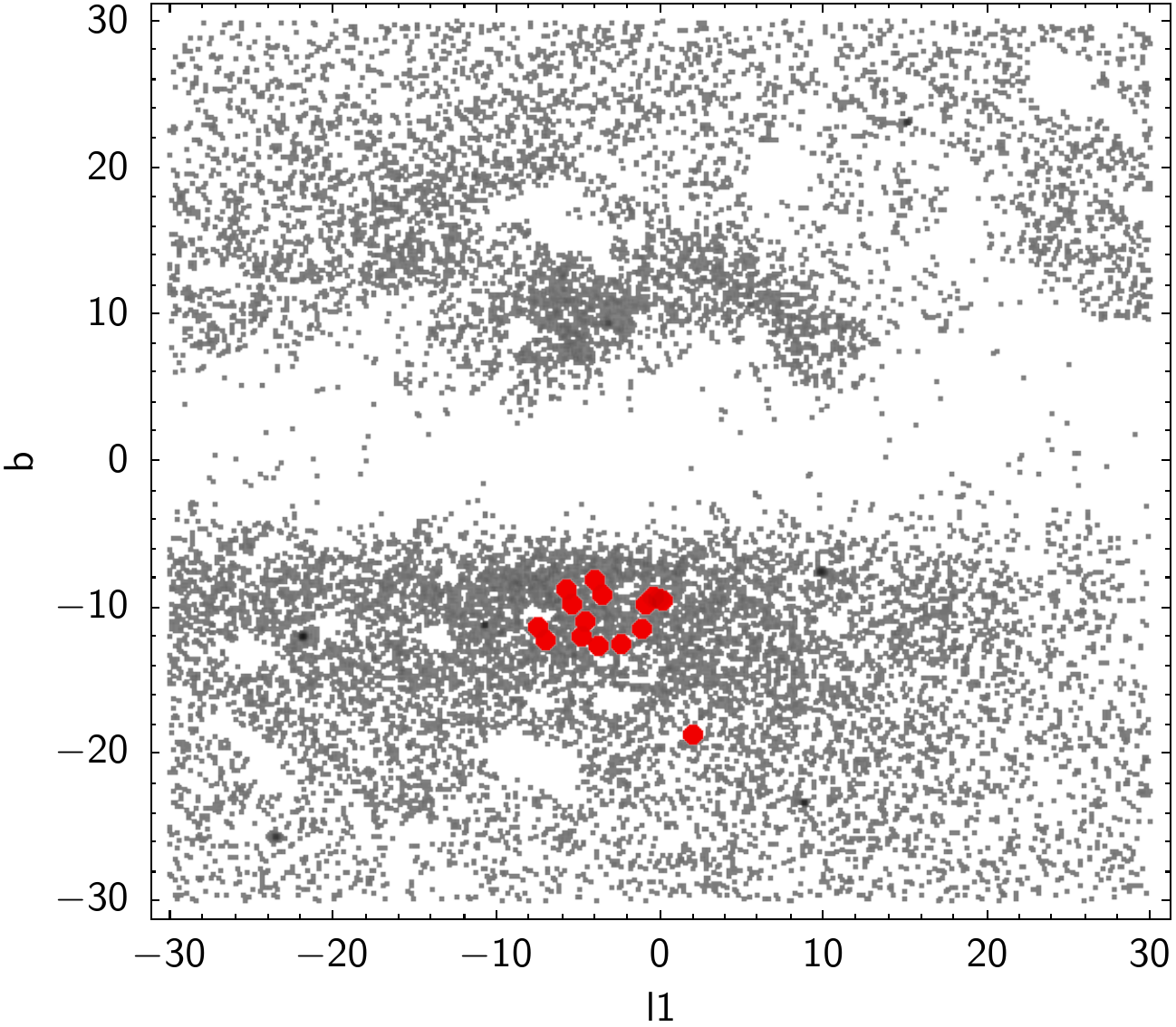}}
\resizebox{8cm}{!}{\includegraphics{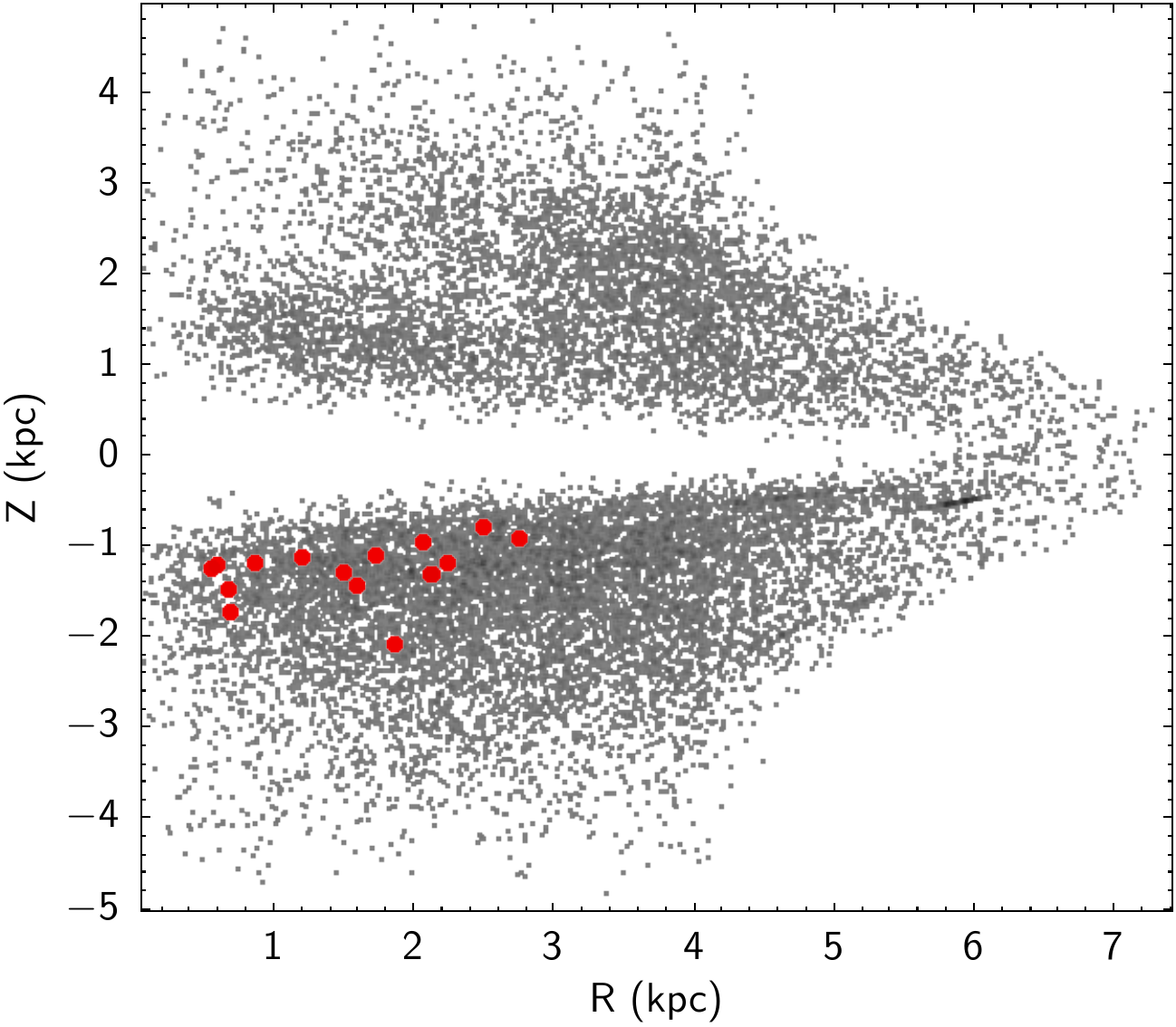}}
\caption{ Stars with [M/H]$< -1.5$ from the sample of \citet{rix22} (grey dots)
compared with the SpiteF stars of our stars in the planes $(l1,b)$ and
$(R,Z)$. To be consistent with figure 3 of \citet{rix22}, in the left panel
we plot $l1=\{l\ if \ l \le 180. ; l-360.\ if\ l > 180\}$.}
\label{fig_lb_zr_spitef}
\end{figure*}

\begin{figure*}
\centering
\resizebox{8cm}{!}{\includegraphics{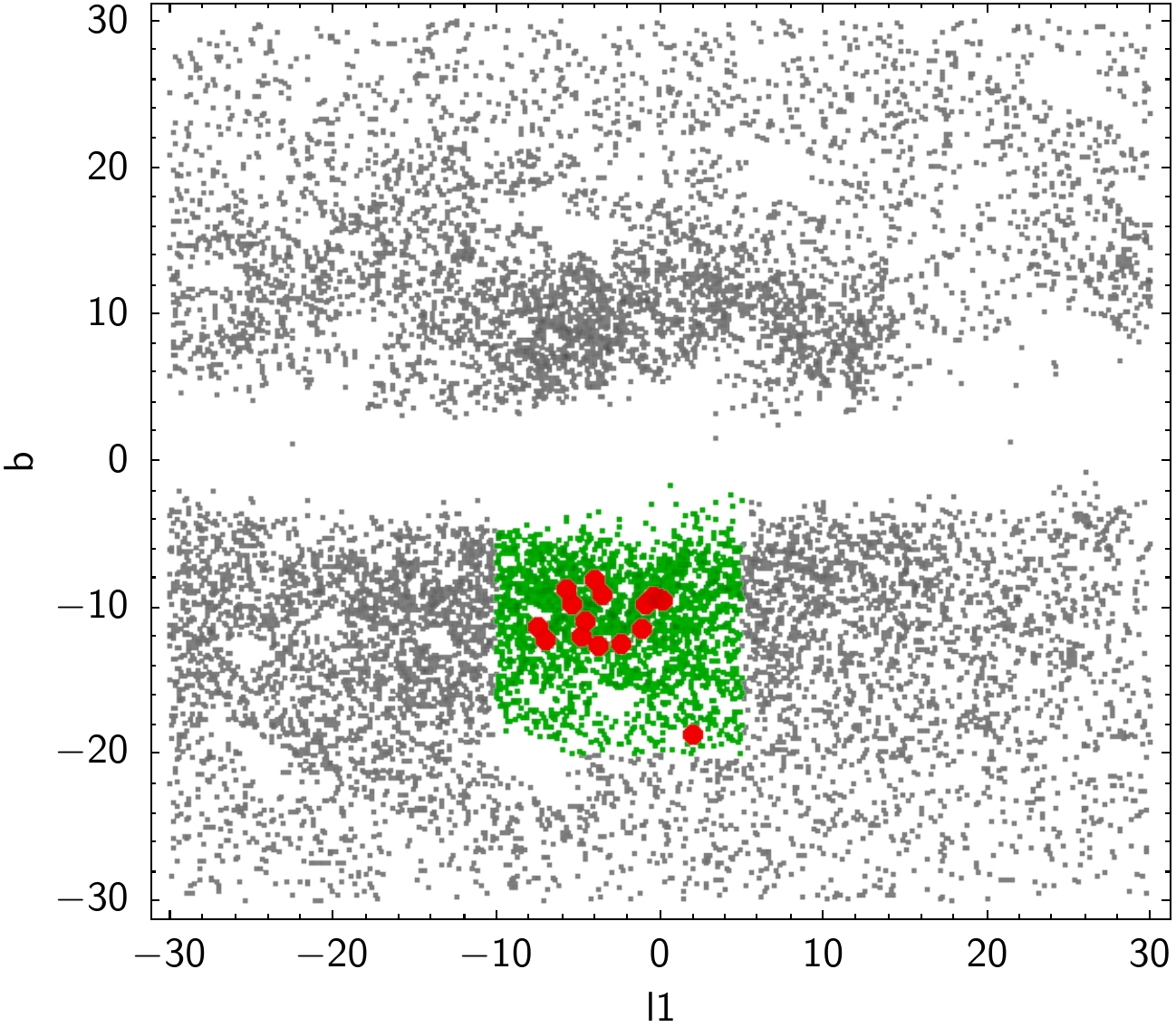}}
\resizebox{8cm}{!}{\includegraphics{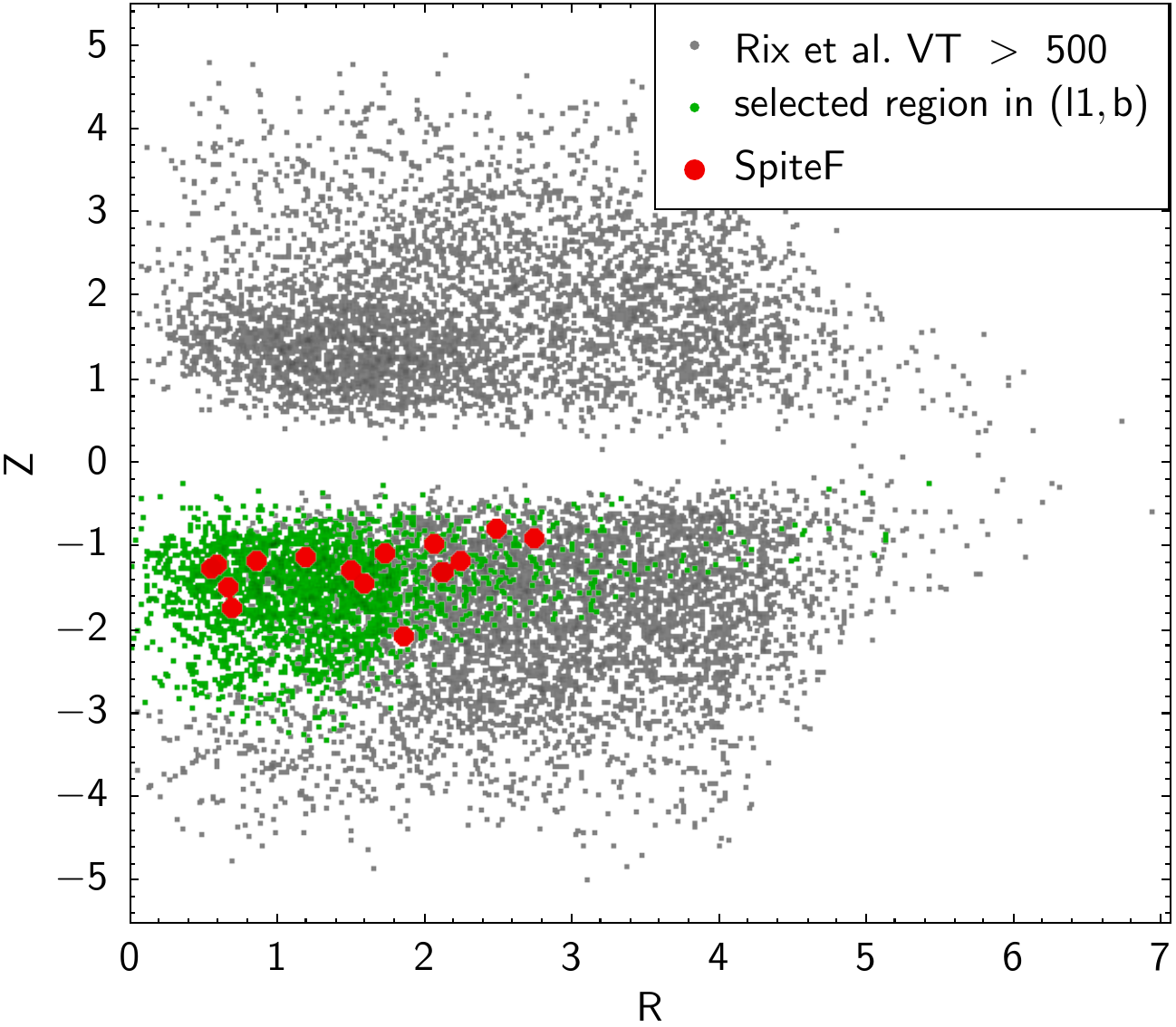}}
\caption{ Stars with transverse speed faster than 500 \kms\ 
from the sample of \citet{rix22} (grey dots)
compared with the SpiteF stars of our stars in the planes $(l1,b)$ and
$(R,Z)$. The green dots are  the subset of high transverse velocity stars 
from the sample of \citet{rix22}, selected in the $(l1,b)$ region
occupied by our SpiteF stars.
}
\label{fig_lb_zr_spitef_hs}
\end{figure*}

\citet{rix22} identified a metal-poor structure in the central part of the Milky Way,
that they interpret as the result of the coalescence of a few massive progenitor
galaxies and they refer to this structure as `proto-galaxy' or, more poetically,
in their title as `Poor Old Heart of the Milky Way', for short we shall refer to it
as POH in the following.
We would like to understand if SpiteF could simply be a subset of POH or, more in general,
if there is any connection between the two structures.

SpiteF has two young stars, while POH is interpreted by \citet{rix22} as an old
structure, this would allow us to conclude that the two structures have no connection.
Let us, for the sake of discussion, set aside this issue, one can always argue the
apparently young stars are just evolved blue stragglers.
SpiteF and POH overlap both in the $(l,b)$ an in the $Z,R$ planes, as shown in Fig.\,\ref{fig_lb_zr_spitef}.
However, to conclude that SpiteF is somewhat associated to the POH they should share further properties
than their spatial coincidence.

In order to make more meaningful the comparison we select from
the sample of \citet{rix22} stars with transverse velocity larger than 500\,\kms.
This is a sample of 11\,743 stars and is shown in Fig.\,\ref{fig_lb_zr_spitef_hs}.
In the left panel we also select a subsample of 2\,178 stars (shown as green dots) that occupy
the same region in $(l1,b)$ as our SpiteF panel. In the right panel
we show the same sets of stars in the $(R,Z)$ plane. It is  obvious that
the sample of high transverse speed stars that occupies
the same $(l1,b)$ region as SpiteF shows, for any given $R$, a wider
range in $Z$, as we already said in Sect.\,\ref{sec:spitef}.

\begin{figure}
\centering
\resizebox{8cm}{!}{\includegraphics{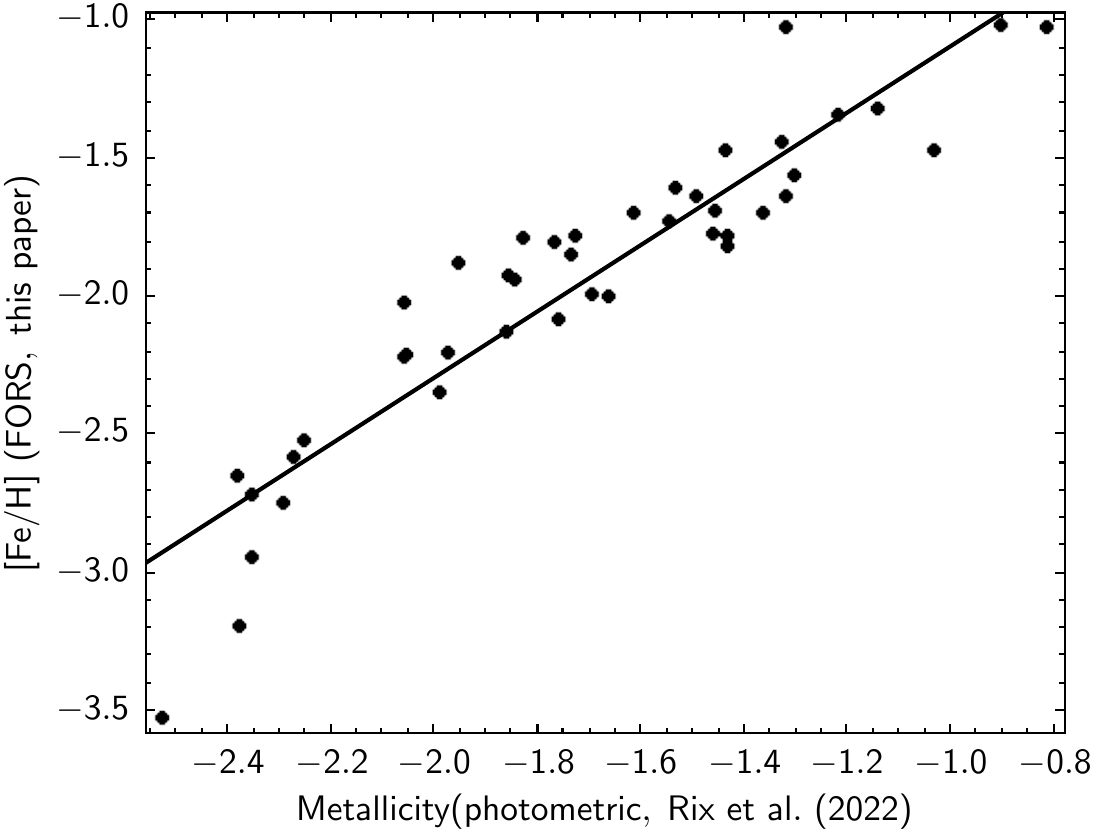}}
\caption{ Comparison of the spectroscopic and photometric metallicities for 43
stars in common between this paper and \citet{rix22}. The straight
line is a linear fit.}
\label{rix_fors}
\end{figure}

As a preliminary step we try to put the metallicities of \citet{rix22}
on the same scale as ours. In Fig.\,\ref{rix_fors} we compare 
the photometric metallicities of \citet{rix22} with our spectroscopic
metallicities for 43 stars in common. The correlation coefficient is about
0.93. A linear fit yields a slope of 1.199877 and an intercept
of 0.10886344, the root mean square error around this relation
is 0.2\,dex, fully compatible with the errors involved.
It is nevertheless noticeable how the correlation becomes
more dispersed below photometric metallicity of --2.0. 
This is expected, on the one hand the photometric sensitivity
to metallicity diminishes below this value, even for narrow
bands, on the other hand the calibration set
used by \citet{rix22} has very few calibrators at low metallicity
(see their figure 1).
Let us assume this linear correlation
as suitable to put the metallicities of \citet{rix22} on the same scale
as ours.

\begin{figure}
\centering
\resizebox{8cm}{!}{\includegraphics{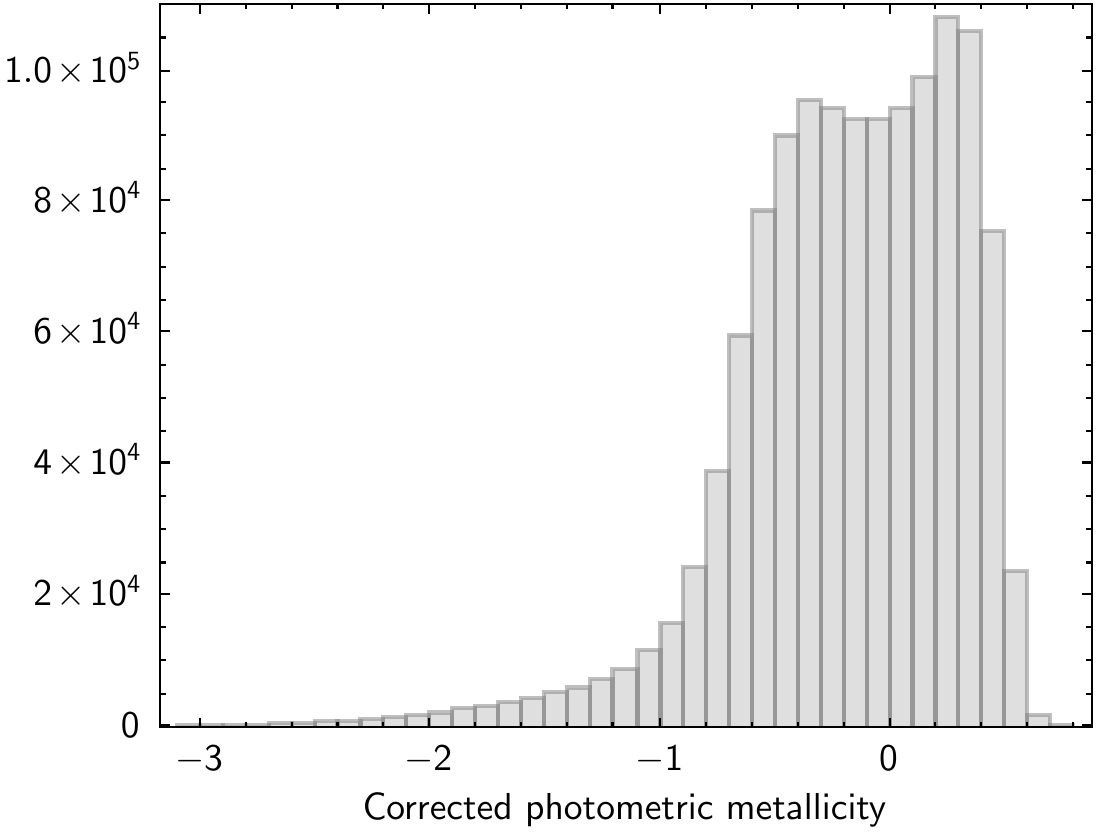}}
\caption{ Histogram of the corrected metallicity for the sample
of 1\,250\,764 stars for which \citet{rix22} computed orbits.}
\label{rix_histo}
\end{figure}

\begin{figure}
\centering
\resizebox{8cm}{!}{\includegraphics{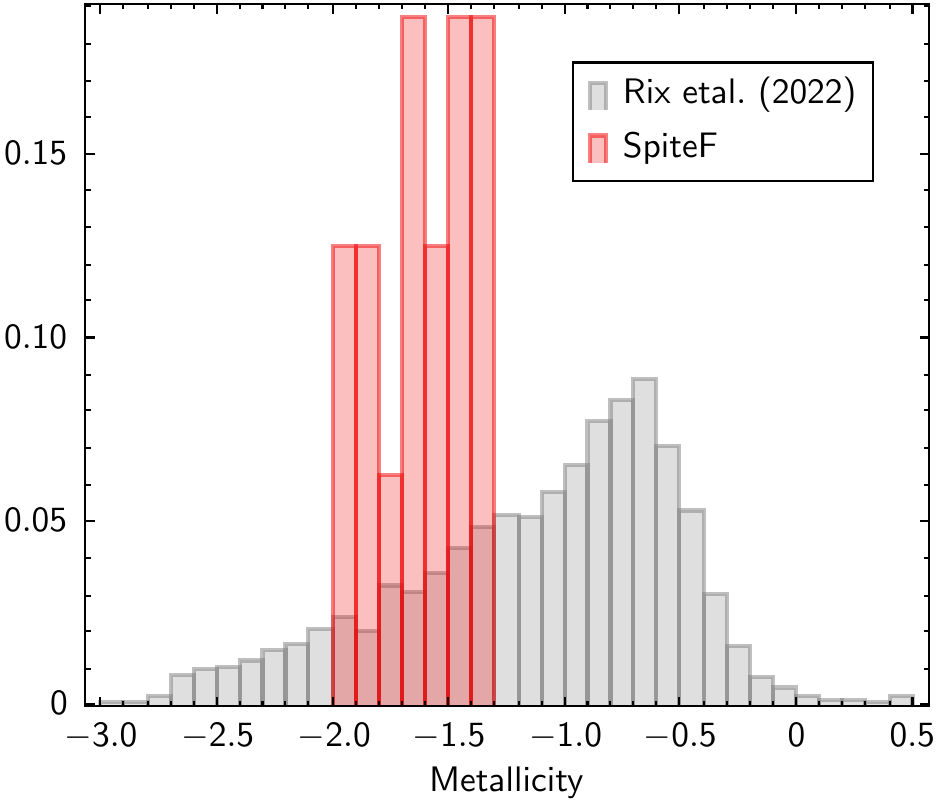}}
\caption{Histogram of the corrected metallicity for the sample
of 2\,178 high transverse speed stars of \citet{rix22} selected in $(l1,b)$
as shown in Fig.\,\ref{fig_lb_zr_spitef_hs}. Our sample of SpiteF stars are also shown. Both histograms have been normalised to the total number of stars.}
\label{histo_blob}
\end{figure}

In Fig.\,\ref{rix_histo} we show the histogram
of the corrected metallicities for the whole sample 
for which \citet{rix22} computed orbits.
In Fig.\,\ref{histo_blob} we show the normalised  metallicity histogram
for the high transverse speed stars of \citet{rix22} that occupy
the same $(l1,b)$ region as SpiteF and identified in 
Fig.\,\ref{fig_lb_zr_spitef_hs} as green dots.
In the same plot we also show the normalised metallicity
histogram of SpiteF. The difference in the metallicity
distribution of the two sets of stars is obvious, in spite of the
fact that they all have transverse speed in excess of 500\,\kms
and occupy the same region in $(l1,b)$.

\begin{figure}
\centering
\resizebox{8cm}{!}{\includegraphics{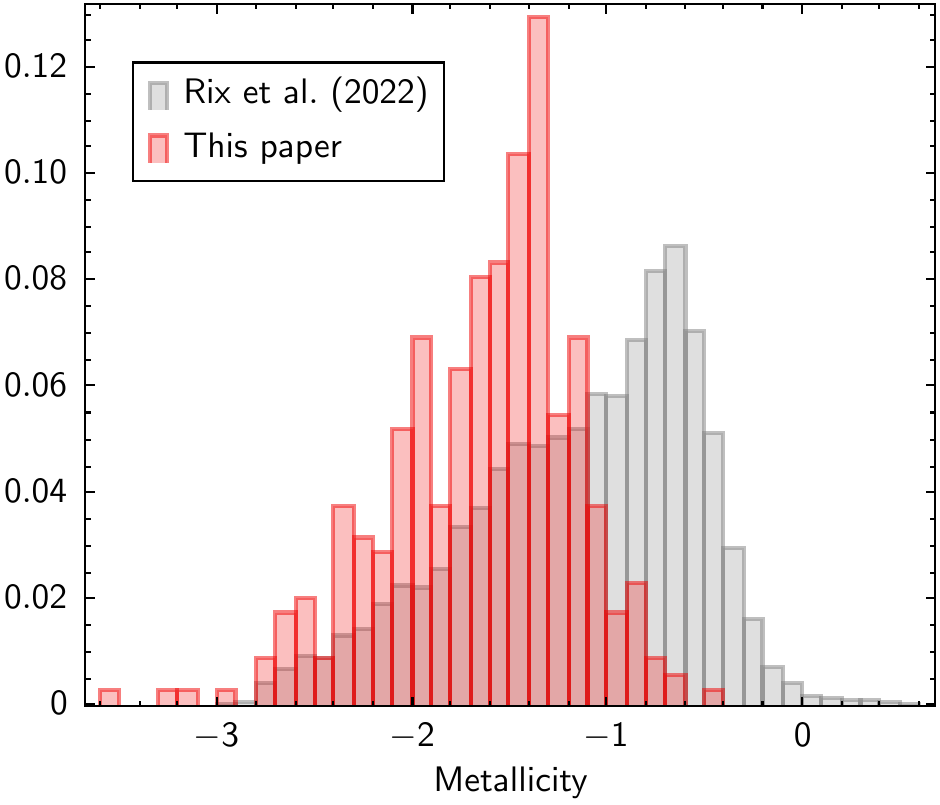}}
\caption{ Normalised histogram of the corrected metallicity for the sample
of 11\,743 stars from the sample of \citet{rix22} with transverse velocities larger
than 500\,\kms (grey). Superimposed is the metallicity histogram of our sample
of stars (red). Both histograms have been normalised to the total number of stars.}
\label{histo_vtgt500}
\end{figure}

In Fig.\,\ref{histo_vtgt500} we show the corrected metallicity normalised histogram
for the entire sample of the high transverse speed stars of \citet{rix22}.
On the same plot we also show the metallicity distribution of
our stars. It is clear that our sample peaks at much lower
metallicities than that of \citet{rix22}.

\begin{figure}
\centering
\resizebox{8cm}{!}{\includegraphics{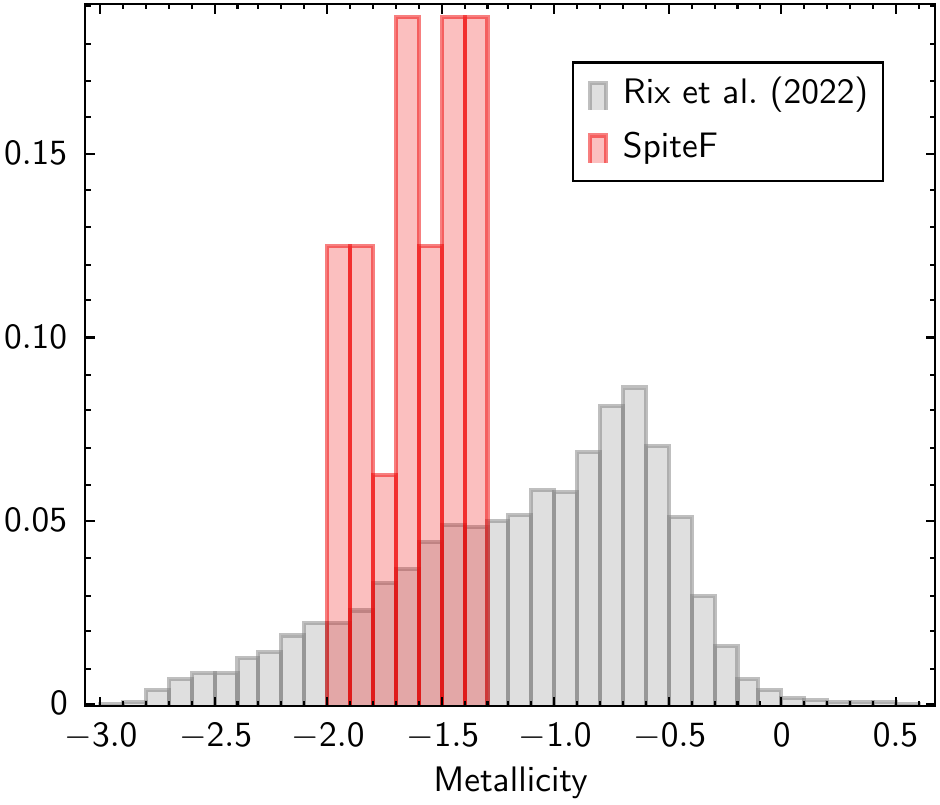}}
\caption{ 
Same as Fig.\,\ref{histo_vtgt500} except that now only
our sample of SpiteF stars is shown.
Both histograms have been normalised to the total number of stars.}
\label{histo_vtgt500_spitef}
\end{figure}

Let us now restrict the comparison to the sample of stars
that we identified as SpiteF, as shown in Fig\,\ref{histo_vtgt500_spitef}.
It is obvious that the two metallicity distributions are very different.
SpiteF spans a much smaller range in metallicity than the high transverse
velocity stars selected from the sample of \citet{rix22}.
The sample of \citet{rix22} has few but well understood biases: it comprises
only stars in a rectangular region of $60^\circ$ side centred on
the Galactic centre, it only selects giant stars, and 
$1.0 \le (G_{BP}-G_{RP}) \le 3.5$.

Quite likely the colour selection introduces some metallicity bias,
our sample contains bluer stars and we do not observe any of the
M giants that are part of the \citet{rix22} sample.
The difference in colour selection may be at the root of the different
metallicity distribution of our sample and that of \citet{rix22}, even
when restricted to the high transverse speed stars (Fig.\,\ref{histo_vtgt500_spitef}).
A more detailed investigation of the biases is hampered by the inhomogeneous selection
of our sample. 
Even keeping in mind this underlying
difference between the two samples, in
view of Fig.\,\ref{histo_blob} and Fig.\,\ref{histo_vtgt500_spitef} it does not
seem likely that SpiteF is just a random selection of the
population of high transverse velocity stars found in the central
part of the Galaxy.

\subsection{Is a recent accretion of a massive galaxy possible?\label{e3}}

To fully answer this question it is necessary
to conduct detailed dynamical simulations, including the
presence of gas, both in the accreted satellite and in the
Milky Way.
One argument in favour of a recent accretion is the fact
that stellar debris that has apocentre distances
smaller than 3.5\,kpc, and therefore
relatively short periods, will mix, in phase
space, on time-scales of a few Ga.
This statement is supported, for example,
by the analysis on tidal streams of dwarf galaxies and
globular clusters by \citet{amorisco},
see specifically his equation 16. 
Thus the fact that we still see a dynamical
coherence among the SpiteF stars implies that
the accretion is relatively recent.
On the other hand recent accretion events 
mostly deposit stars in the outer regions of the halo and not in the inner Galaxy. 
Dynamical friction can make the stars sink to the central regions
of the galaxy, but the process takes time and it may be incompatible
with the young age of the stars we observe in SpiteF \citep[][]{vassiliev22}.
We stress that to make the analysis realistic the effect of gas should be taken
into account, which is not the case for the analysis of \citet{vassiliev22}.
Awaiting for specific analysis we can look at the work of 
\citet{eliche2006} who
investigated the possible role of mergers in the growth
of bulges.  
It should be stressed that this work is mainly concerned with large accreting
satellites and classical bulges, so it may not be directly
relevant to the case of SpiteF.
However among their results, the one that may be
relevant for the present discussion is that the survival
of the merging galaxy at the centre of the bulge of the primary
galaxy depends on the structural properties of the merging
galaxy. Structural properties that \citet{eliche2006} quantify
through the index of the Tully-Fisher relation, $\alpha_{TF}$\footnote{
Initially introduced by \citet{TF} as a method
to estimate distances of galaxies from the widths of 
\ion{H}{I} 21\,cm lines, in its modern acception
the Tully-Fisher relation can be expressed
as $M \propto V_{rot}^{\alpha_{TF}}$,
where $M$ is the mass of the galaxy and $V_{rot}$ its rotational
velocity.  
}. 
In particular they find that if $\alpha_{TF} \ge 3.5$
the galaxy partially survives at the centre of the bulge, 
while for  $\alpha_{TF} = 3.0$ 
the merging galaxy totally disrupts. 
The time scales are also interesting, in their suite
of simulations \citet{eliche2006} find a time for full
merger of 1.3\,Gyr for a merging galaxy with
$\alpha_{TF}=4.0$ and mass 1/2 of the mass
of the bulge of the primary galaxy, corresponding to 
1/6 of the mass of primary galaxy.
While we do not claim that these simulations may 
explain our observations, we believe they show that
a recent merger with a massive galaxy with partial
survival of the merging galaxy at the centre of the
primary galaxy is not implausible. Clearly detailed
{\em ad hoc} simulations should be performed to further
explore the issue, and see other secondary
effects, such as the localisation of the disrupted debris.
We close this subsection by recalling that for external
galaxies there is a general consensus that the bulges grow
also through the contributions of mergers
\citep[see e.g.][and references therein]{sachdeva2017}.
Most of the growth is due to secular evolution
of disc stars being accreted by the bulge, but 
this process may also be triggered by accretion events.

\section{Limitations of \mygi\ for ionised species \label{mygi_ion}}

\begin{figure}
\centering
\resizebox{7.7cm}{!}{\includegraphics[clip=true]{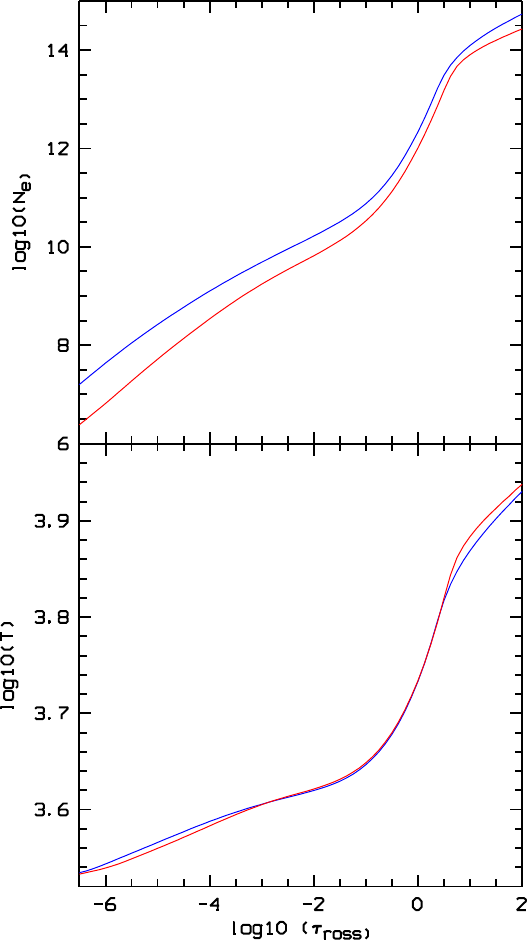}}
\caption{Structure in temperature and electronic density for two ATLAS 12 models of our grid.
\teff = 5000\,K, \glog = 2.0, $\xi=2$\,\kms and $\rm [\alpha/Fe]=+0.4]$ and metallicity
--1.0 (red) and --2.0 (blue).}
\label{modstr}
\end{figure}

\mygi\ uses grids of synthetic spectra that are computed from models of different metallicity. This implies that when \mygi\ fits a spectral feature it interpolates among profiles computed with {\em different} models.
This is different than what is usually done in abundance analysis, where 
profiles, or equivalent widths, are computed from a given model atmosphere and different abundances of the element under study. While being inconsistent (the model atmosphere had been computed with a fixed abundance of any element) this approach can be justified on the grounds 
that the element under study is a trace element and the effect of its abundance on the structure of the model atmosphere is negligible. Of course this condition may fail for some specific elements. The classical example is Mg in cool models around solar metallicity. Since Mg is the dominant electron donor in these models, changing its abundance can have a non-negligible effect on the model structure, mainly through the H- number density and associated opacity.
The \mygi\ approach is also inconsistent, yet in \citet{mygi14} we have shown that for lines of neutral species, it provides essentially the same results as the more traditional approach.

This is not true, in general, for ionised species, or for any feature that has a strong dependence on the electron density, like, for instance the [O I] line at 630\,nm.
The reason is that two models that differ in metallicity by a significant amount, let us say one order of magnitude to fix the ideas, have a  similar T(tau) structure but a very different electron density structure. This is illustrated in Fig.\,\ref{modstr}.
While for neutral trace species, like \ion{Fe}{i} the T(tau) is the most relevant characteristic of the model in computing the line profile, for dominant ionised species, like \ion{ Ba}{ii}, the run of electron density is also very important in determining the computed profile.
We made some experiments with \ion{Ba}{ii} features and, as a rule of thumb \mygi\ gives unreliable quantitative results when |[Ba/Fe]|>0.3. Qualitatively the result is correct, but the \mygi\ abundance cannot be trusted. By extension we assume this limit for any electron density sensitive feature.

\end{appendix}

\end{document}